\documentclass[prb,twocolumn,showpacs,superscriptaddress,floatfix]{revtex4}
\usepackage{graphicx,amsfonts,amssymb,amsmath,hyperref}

\newif\ifhyper
\hypertrue
\ifhyper
\hypersetup{
  citecolor = {green},
  colorlinks = {true}, 
  urlcolor = {blue} 
}
\fi

\begin{document}
\title{Quantum phase transitions in fully connected spin models : \\ an entanglement perspective}

\author{Michele Filippone}
\email{filippone@lpa.ens.fr}
\affiliation{Laboratoire Pierre Aigrain, CNRS UMR 8551,  \'Ecole Normale Sup\'erieure, 24 rue Lhomond, 75231 Paris Cedex 05, France}

\author{S\'ebastien Dusuel}
\email{sdusuel@gmail.com}
\affiliation{Lyc\'ee Saint-Louis, 44 Boulevard Saint-Michel, 75006 Paris, France}

\author{Julien Vidal}
\email{vidal@lptmc.jussieu.fr}
\affiliation{Laboratoire de Physique Th\'eorique de la Mati\`ere Condens\'ee,
CNRS UMR 7600, Universit\'e Pierre et Marie Curie, 4 Place Jussieu, 75252 Paris Cedex 05, France}


\begin{abstract}
We consider a set of fully connected spins models that display first- or second-order transitions and for which we compute the ground-state entanglement in the thermodynamical limit. We analyze several entanglement measures (concurrence, R\'enyi entropy, and negativity), and show that, in general, discontinuous transitions lead to a jump of these quantities at the transition point. Interestingly, we also find examples where this is not the case.

\end{abstract}


\maketitle

\section{Introduction}
\label{sec:intro}

During the last decade, the relationship between quantum phase transitions and
entanglement has become an important research domain \cite{Amico08}.
Although it is natural to expect some deep changes in the ground state of a
system at a transition point, the real problem is to measure these variations
or, in other words, to characterize the quantum state structure. In most cases,
the study of an order parameter or the behavior of correlation functions is
sufficient to detect and analyze a phase transition but one may wonder whether
more ``intrinsic" measures could be helpful. 
Following pioneering works in one-dimensional spin models
\cite{Osborne02,Osterloh02,Vidal03_1}, many studies have been devoted to this
problem (see Ref.~\onlinecite{Amico08} for a review), but only a few allow
for an exact solution in the thermodynamic limit which is a key ingredient to
characterize a phase transition.

The goal of this paper is to propose a class of simple, fully connected (collective) models in which
ground-state entanglement properties can be studied in details. These models,
in which degrees of freedom (spins 1/2) mutually interact, can be seen as
generalizations of the Lipkin-Meshkov-Glick model
\cite{Lipkin65,Meshkov65,Glick65} for which most entanglement features are
now well known \cite{Vidal04_1,
Vidal04_2,Vidal04_3,Dusuel04_3,Latorre05_2,Dusuel05_2,Unanyan05_1,
Barthel06_2,Vidal06_3,Vidal07,Morrison08_2,Cui08,Orus08_2,Caneva08,Ma09,Wichterich09_2}.
The main reason for introducing these collective systems is that they not only
allow us to study a second-order phase transition as in the Lipkin-Meshkov-Glick
model, but also allow the study of first-order transitions.
As surprising as it may seem, although ``collective'' may be thought to lead
to a pure mean-field behavior, we will see that the  is deeply
entangled.
Furthermore, although it might be naively expected that entanglement measures will
simply display jumps at first-order transitions, we will show that this does not
hold for one of the models, the spectral properties of which show some similarities with
those of a system exhibiting a second-order transition.

The study of these collective systems is also motivated by the fact that they
are much simpler to analyze than their nearest-neighbor counterparts on a
finite-dimensional lattice.
Indeed, most entanglement measures rely on a multi-partition of the microscopic
degrees of freedom. For instance, the concurrence \cite{Wootters98} is obtained
by separating a system of $N$ spins into two parts (of sizes $N-2$ and $2$), the
R\'enyi entropy is obtained by splitting it into two parts of arbitrary sizes ($N-L$
and $L$),  a tri-partition is required to compute the negativity
\cite{Vidal02} of a mixed state, etc. Thus, the main problem often consists of
computing reduced density matrices for a given partition. In the models studied
below, this crucial step can be achieved since the original spin problem can be
mapped onto a quadratic bosonic Hamiltonian. 

The structure of this paper is the following. In Sec.~\ref{sec:models_qpt}, we
introduce a family of models and  compute their low-energy spectrum
(ground-state energy and gap). This allows us to determine their phase diagram
and characterize the quantum phase transitions. Analytical expressions are
obtained in the thermodynamical limit and compared with exact diagonalization
results. In Sec.~\ref{sec:entanglement_measures}, we discuss the ground-state
entanglement by focusing on three different measures~: the concurrence, the
R\'enyi entropy, and the negativity, which rely on a one-mode, two-mode, and
three-mode description of the bosonic Hamiltonian, respectively. Once again, exact
results in the thermodynamical limit are compared to numerical data for a
representative set of parameters.

\section{Models and quantum phase transitions}
\label{sec:models_qpt}

\subsection{Hamiltonians}
\label{sec:sub:hamiltonians}

We consider a system made of $N$ spins 1/2 whose Hamiltonian reads
%
\begin{equation}
	\label{eq:ham}
	H = - N \left[\cos\omega \left(\frac{S_x}{S}\right)^m
	+ K_{m,n}\sin\omega \left(\frac{S_z}{S}\right)^n\right].
\end{equation}
%
In the above equation, $S_\alpha=\sum_{j=1}^N\sigma_j^\alpha/2$ are total spin
operators along the $\alpha=x$, $y$, $z$ direction, with $\sigma_j^\alpha$ being the
usual Pauli matrix at site $j$ and $S=N/2$ denotes the maximum spin value. 
This class of models is defined by the non-negative integer parameters $(m, n)$.
In what follows, we shall refer to a model with given values of $m$ and $n$ as
the $(m,n)$ model.

Without loss of generality, we shall restrict ourselves to $m\geqslant n
\geqslant 1$. We furthermore exclude the trivial case $m=n=1$, since
it describes a large spin in a magnetic field, and only displays a crossover,
but no quantum phase transition. The $(2,1)$ model is a collective version
of the transverse-field Ising model, known as the Lipkin-Meshkov-Glick model
\cite{Lipkin65,Meshkov65,Glick65}. The $(m>2,1)$ models are multispin
generalizations of such a model. The $(2,2)$ model can be seen as a collective
version of the quantum compass model \cite{Kugel82}. Note that, in two
dimensions, the latter is dual to the Xu-Moore model
\cite{Xu04,Xu05,Nussinov05_1}, but the collective version of the Xu-Moore model,
namely, the $(4,1)$ model, is not dual to the $(2,2)$ model (in fact, as will be
seen below, the latter two models have rather different properties).

As we shall see, all models under consideration exhibit a quantum phase
transition when the control parameter $\omega\in[0,\pi/2]$ is varied. This
quantum phase transition occurs at $\omega=\pi/4$, provided one imposes
$K_{m,n}$ to take the following value \cite{Maritan84}:
%
\begin{eqnarray}
	\label{eq:K21}
	K_{2,1} &=& 2,\\
	\label{eq:Km1}
	K_{m>2,1} &=& \frac{m^{m/2}(m-2)^{m/2-1}}{(m-1)^{m-1}},\\
	\label{eq:Kmn}
	K_{m\geqslant2,n\geqslant 2} &=& 1.
\end{eqnarray}
%
These values can be found easily (see Sec.~\ref{sec:sub:sub:GSE} for details).

Finally, let us note that all Hamiltonians preserve the magnitude of the total
spin, \textit{i.~e.}, $[H,\boldsymbol{S}^2]=0$. When $m$ ($n$) is even, the
Hamiltonian furthermore has a spin-flip symmetry since it commutes with
$\prod_j\sigma_j^z$ ($\prod_j\sigma_j^x$).

\subsection{Quantum phase transitions}
\label{sec:sub:qpt}

\subsubsection{Numerical spectra}
\label{sec:sub:sub:numerical_spectra}

Physically, the quantum phase transition stems from the competition between the
ferromagnetic $m$-spin interaction in the $x$ direction and the ferromagnetic
$n$-spin interaction in the $z$ direction if $n>1$, or magnetic field in the $z$
direction if $n=1$.

\begin{figure}[t]
	\includegraphics[width=0.49\columnwidth ]{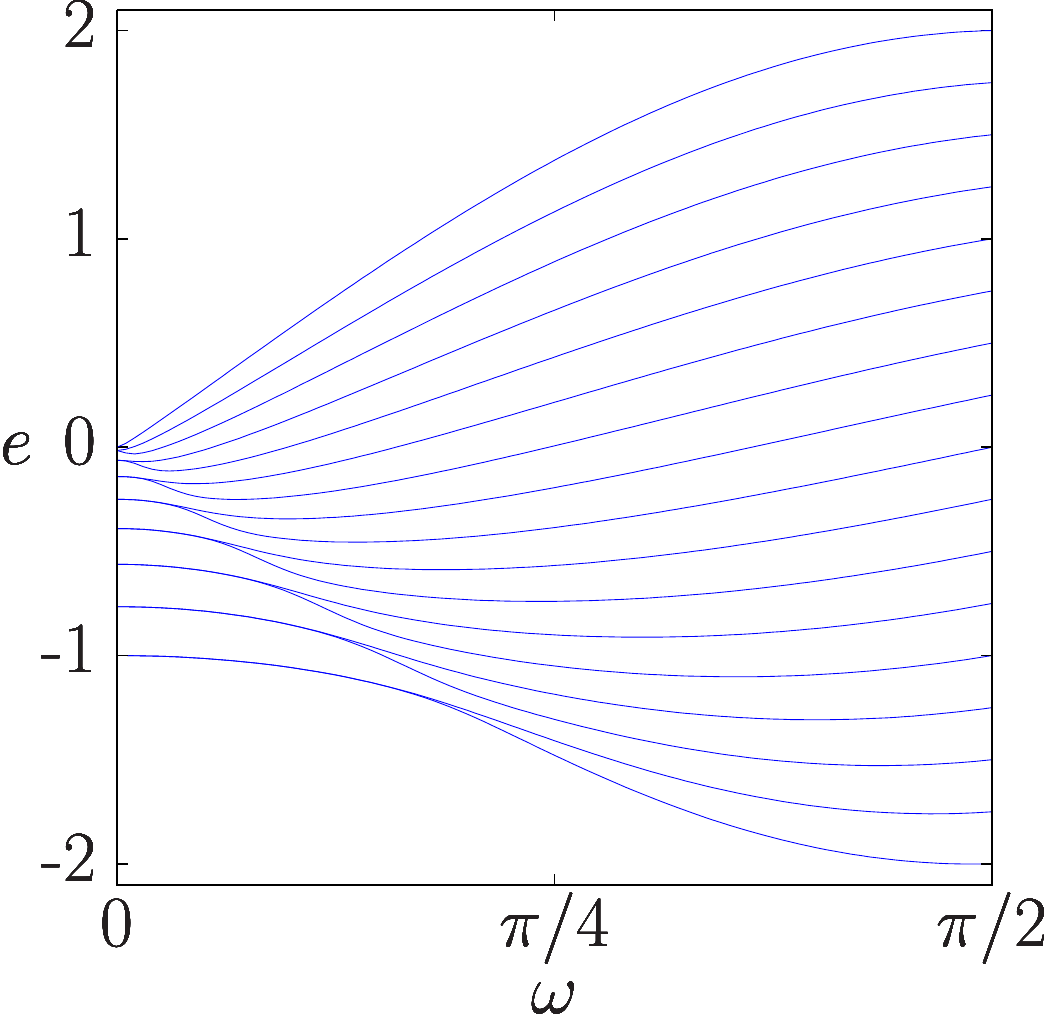}
	\includegraphics[width=0.49\columnwidth ]{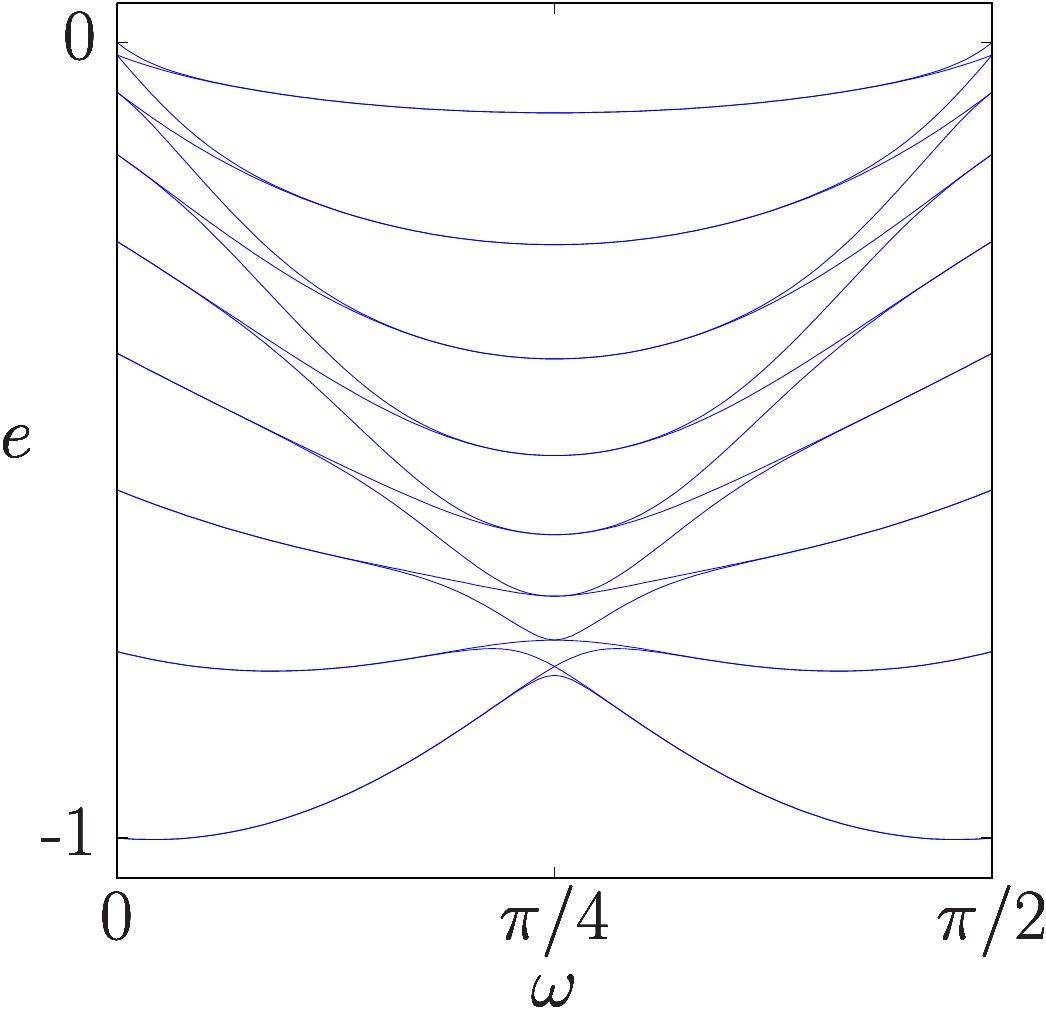}
	\vspace{0.15cm}

	\includegraphics[width=0.49\columnwidth ]{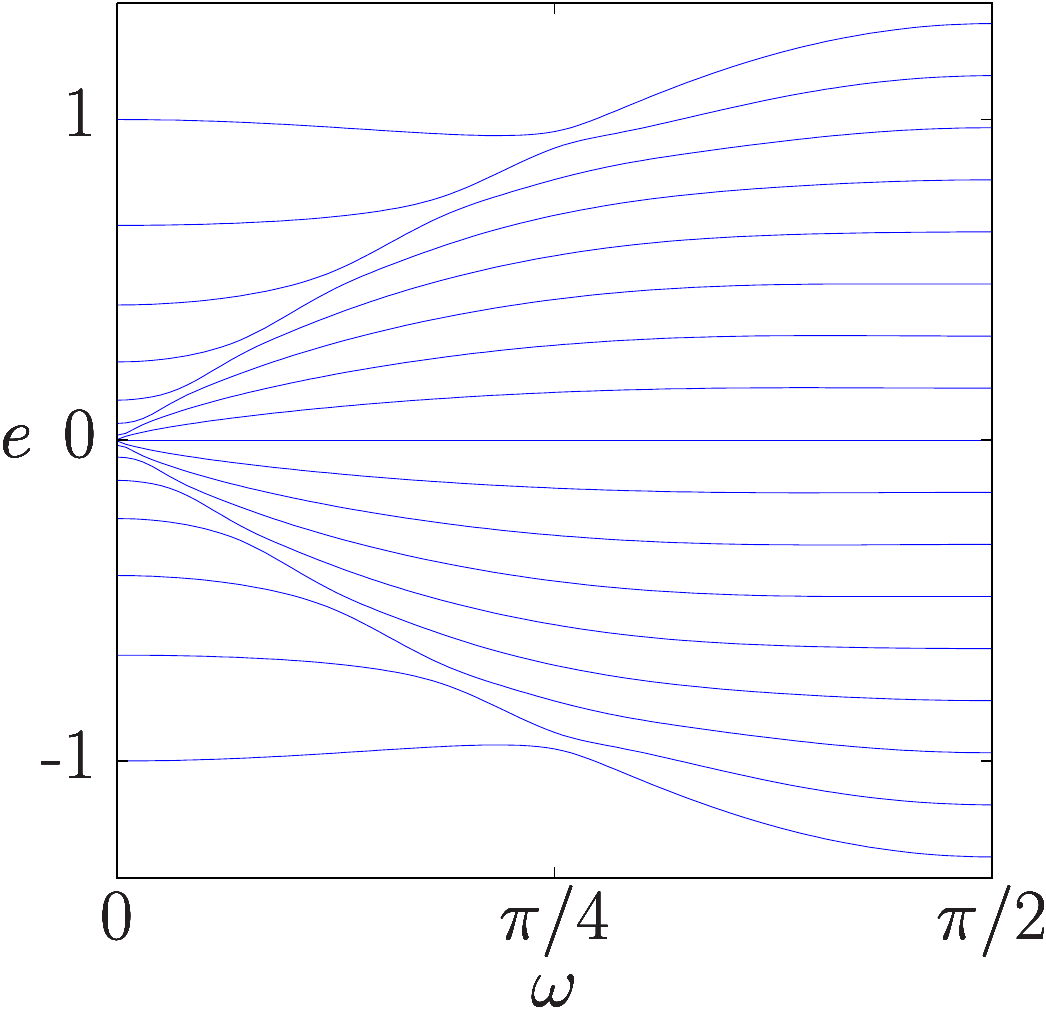}
	\includegraphics[width=0.49\columnwidth ]{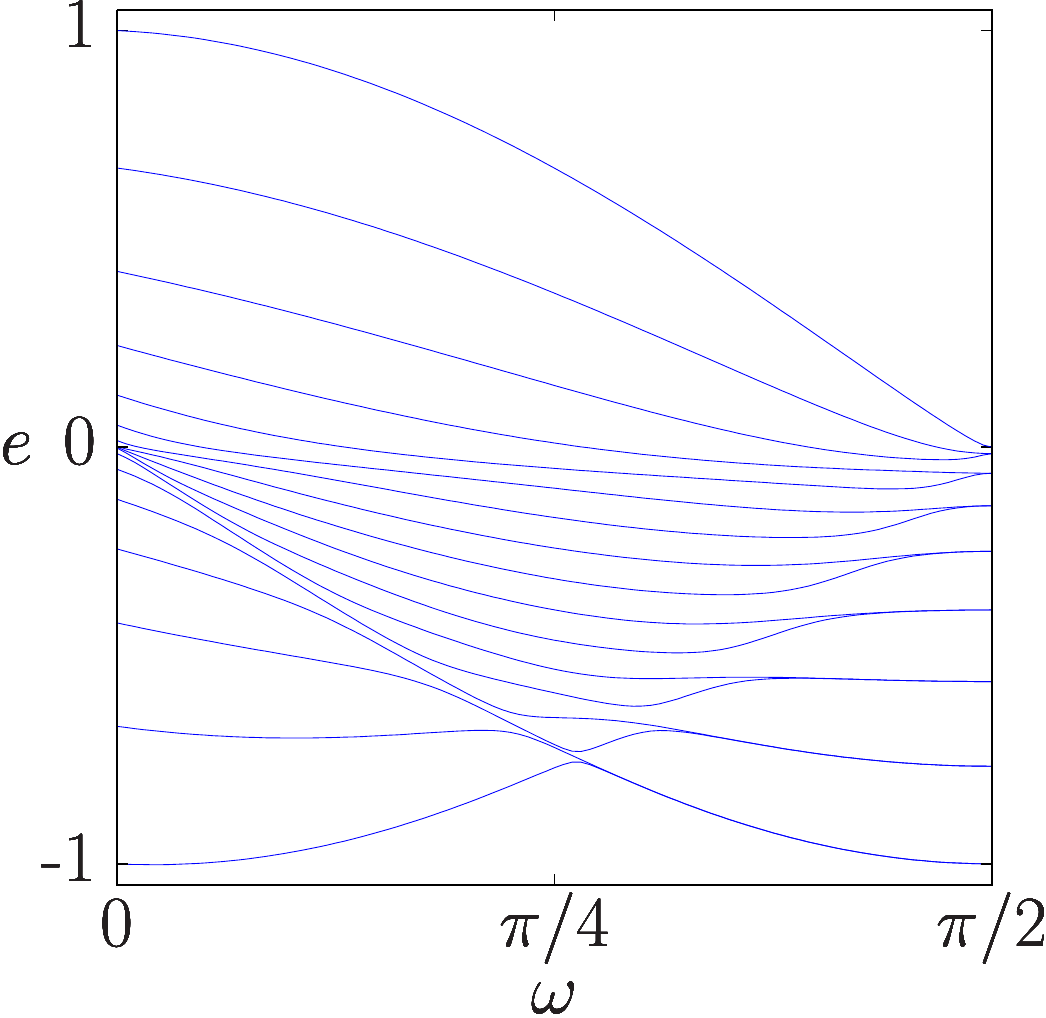}
	\vspace{0.15cm}

	\includegraphics[width=0.49\columnwidth ]{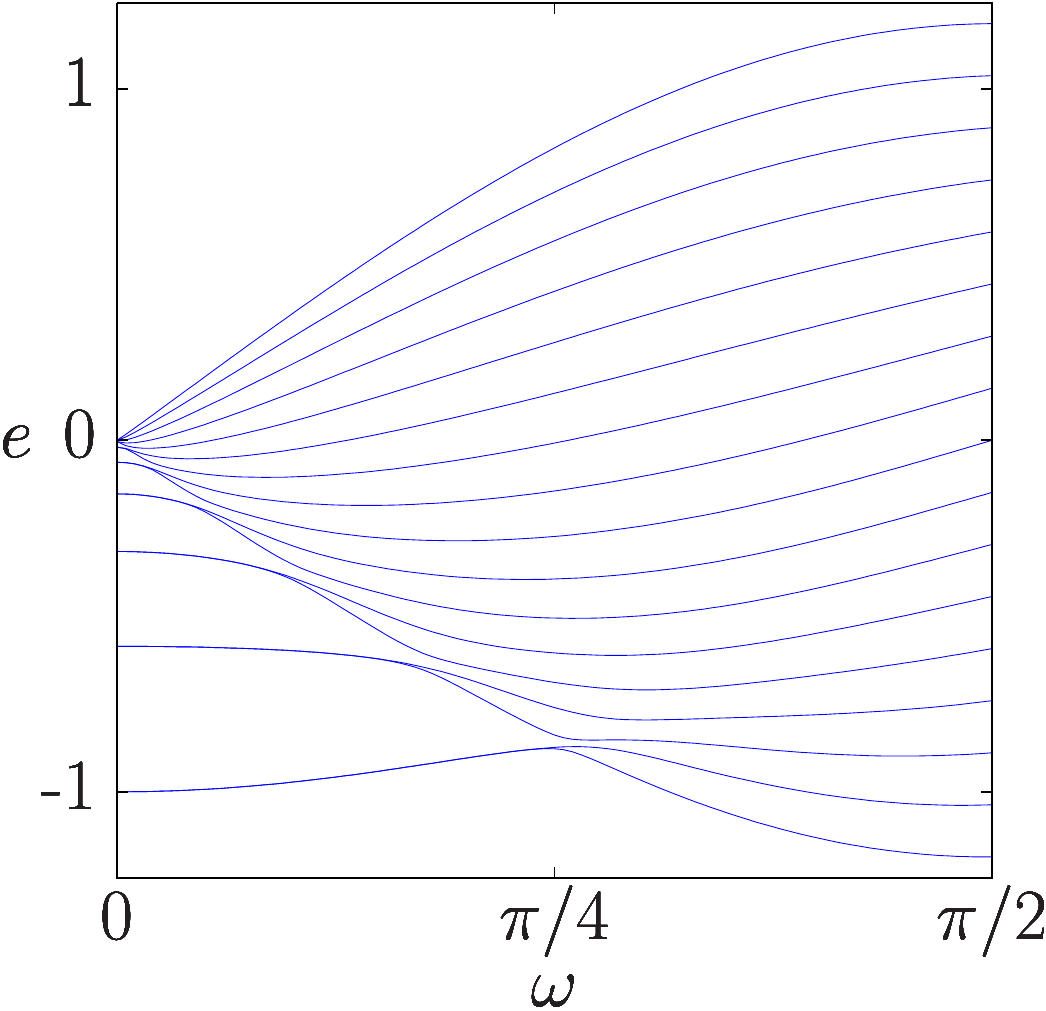}
	\includegraphics[width=0.49\columnwidth ]{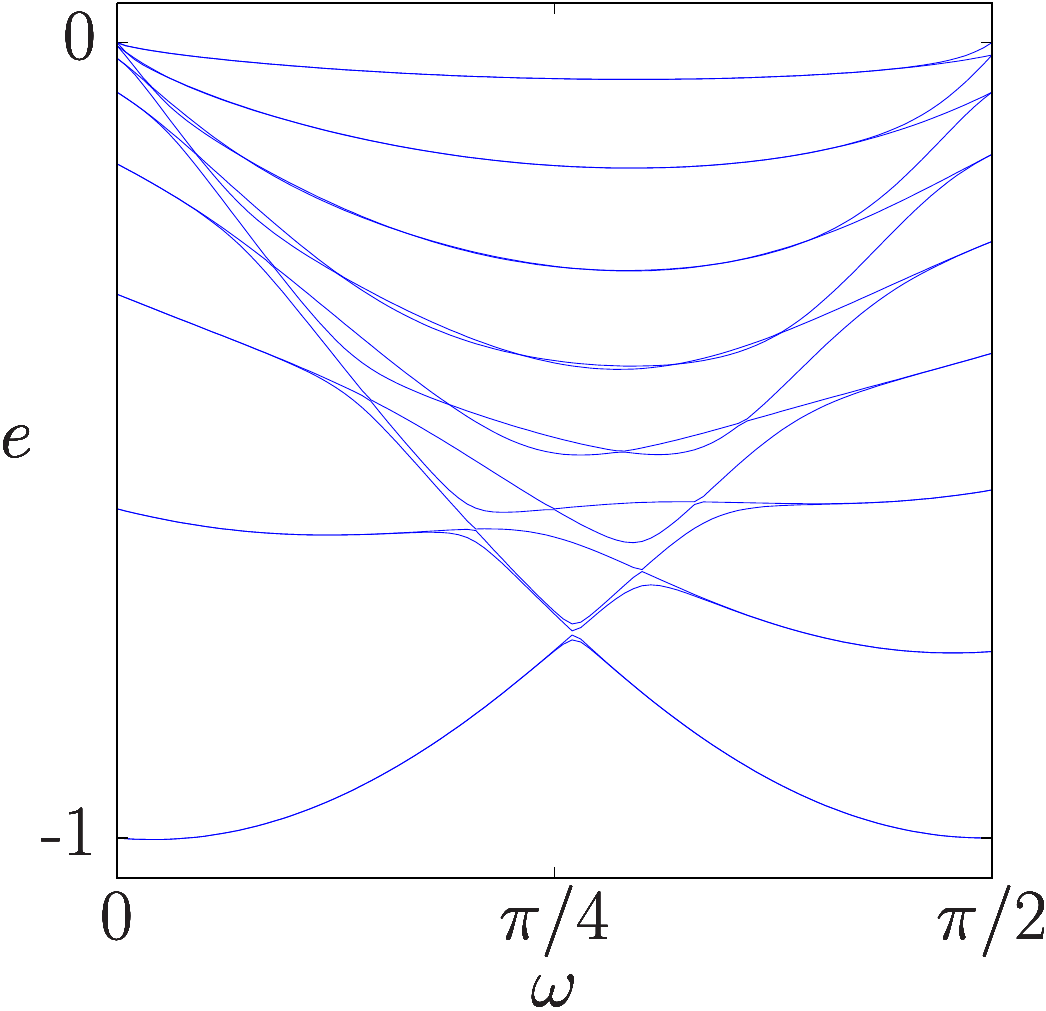}

    \caption{(Color online) Spectra (energies per spin) of six models as a
    function of the control parameter $\omega$, for a system of $N=16$ spins,
    in the maximum spin sector $S=N/2$.
    Left (right): $n=1$ ($n=2$). From top to bottom : $m=2$, $m=3$, and $m=4$.}
	\label{fig:spectrum_N16}
\end{figure}

\begin{figure}[t]
	\includegraphics[width=0.49\columnwidth ]{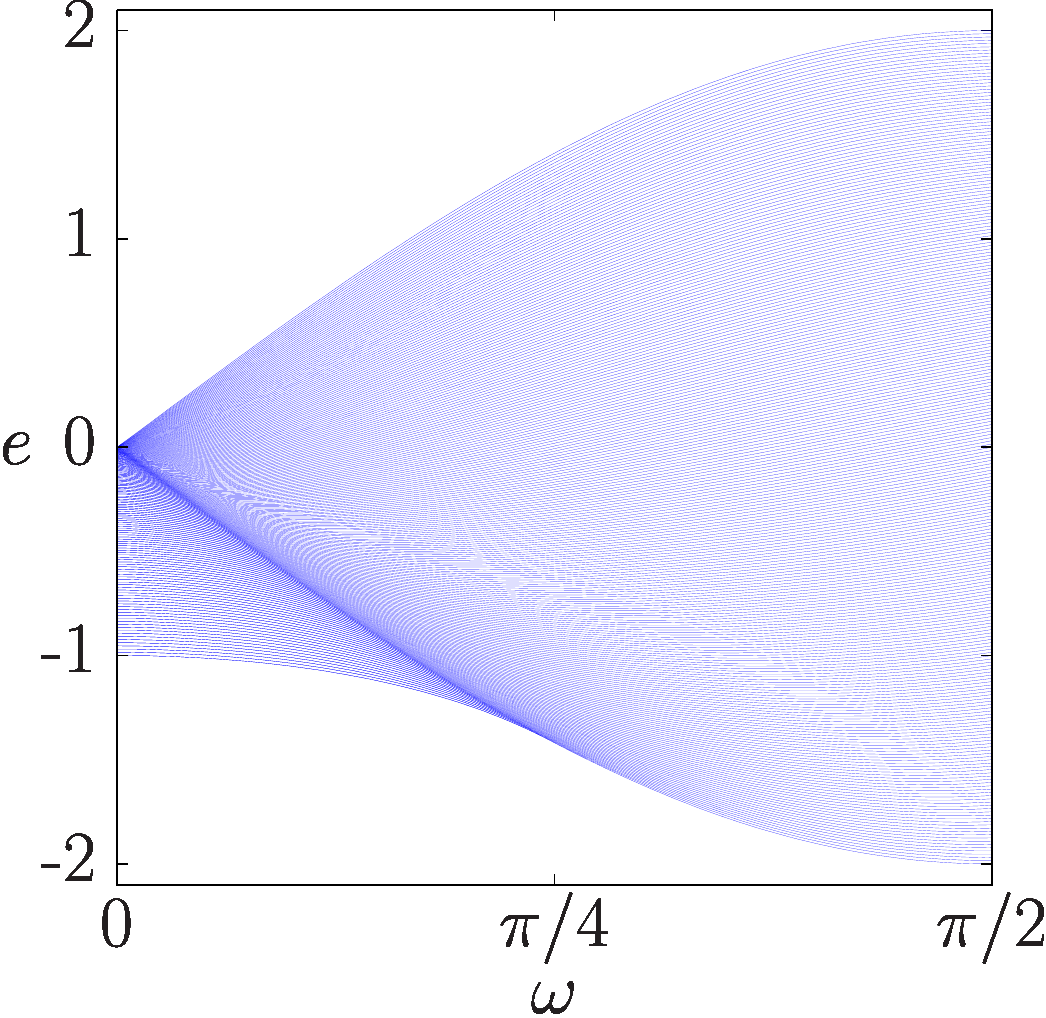}
	\includegraphics[width=0.49\columnwidth ]{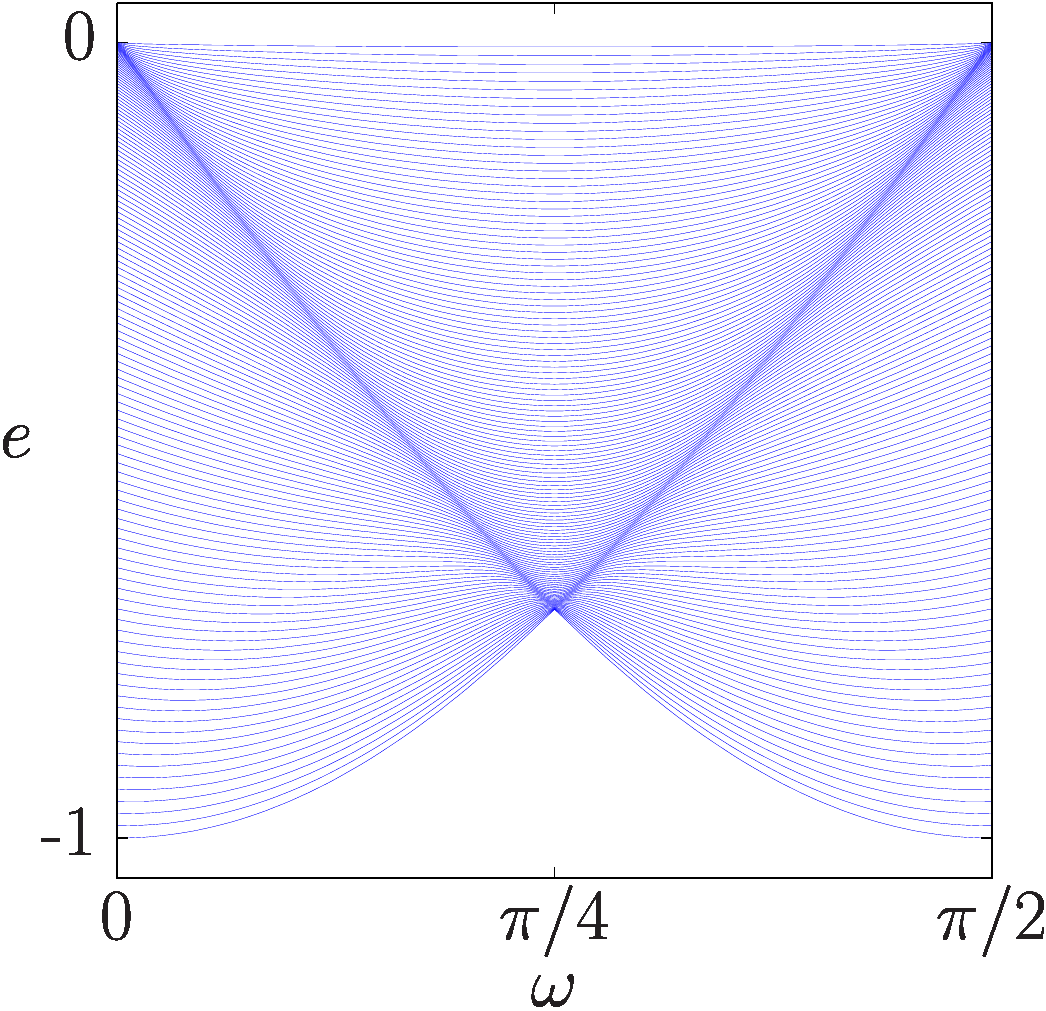}
	\vspace{0.15cm}

	\includegraphics[width=0.49\columnwidth ]{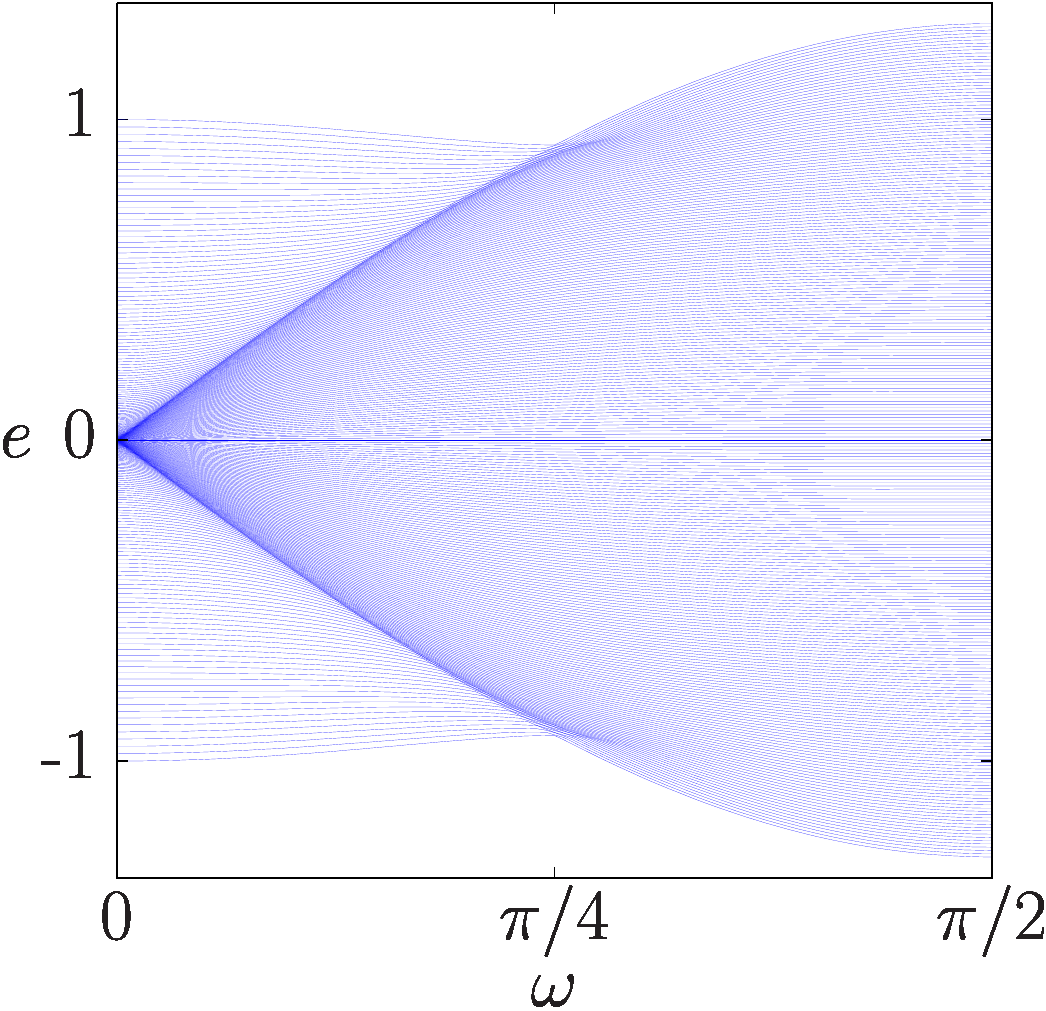}
	\includegraphics[width=0.49\columnwidth ]{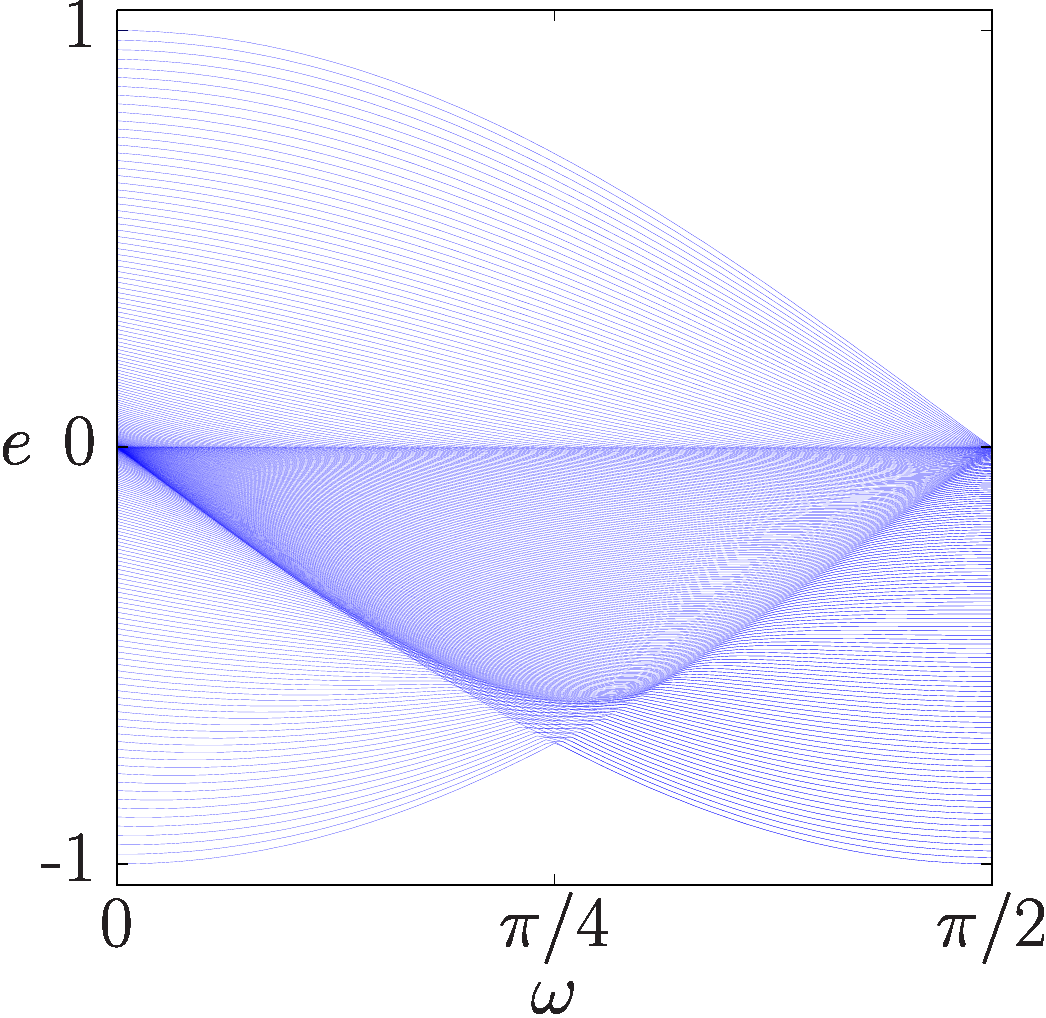}
	\vspace{0.15cm}

	\includegraphics[width=0.49\columnwidth ]{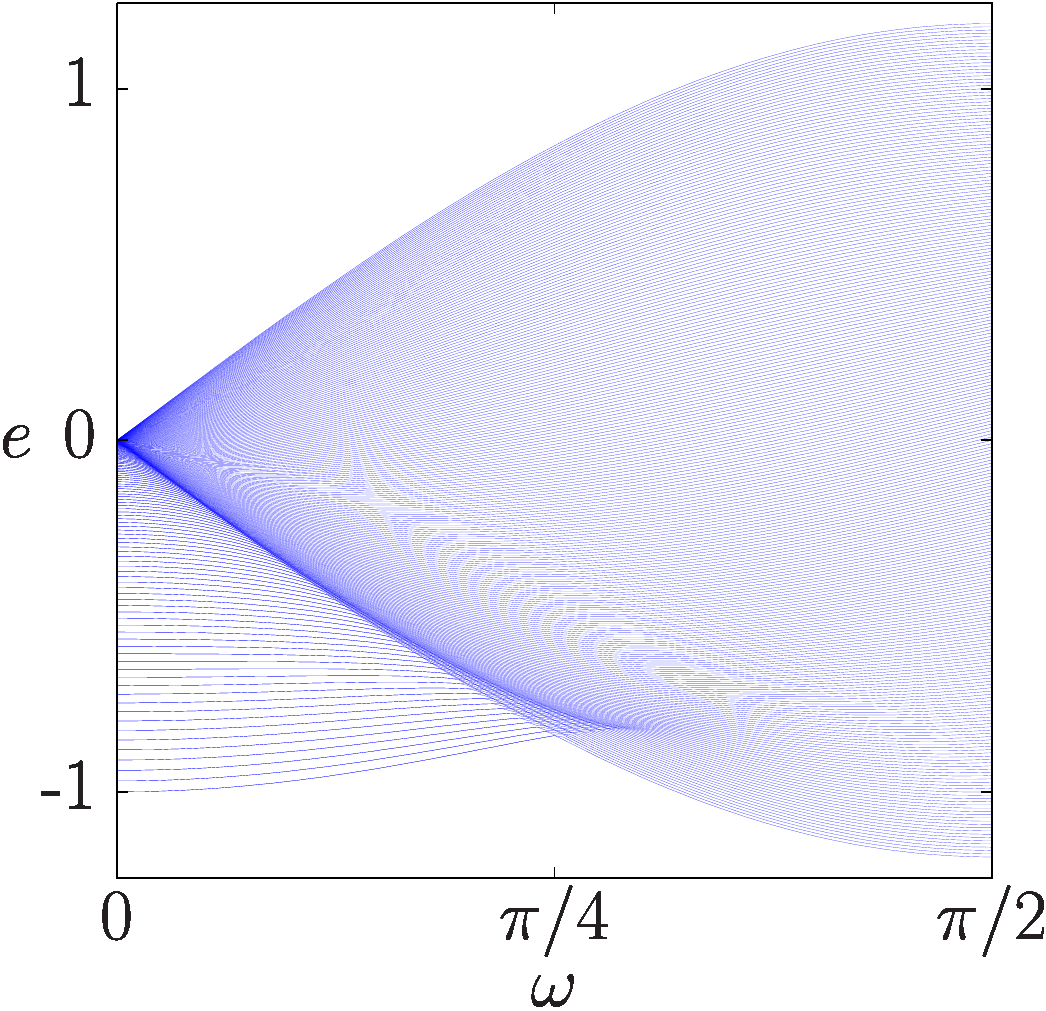}
	\includegraphics[width=0.49\columnwidth ]{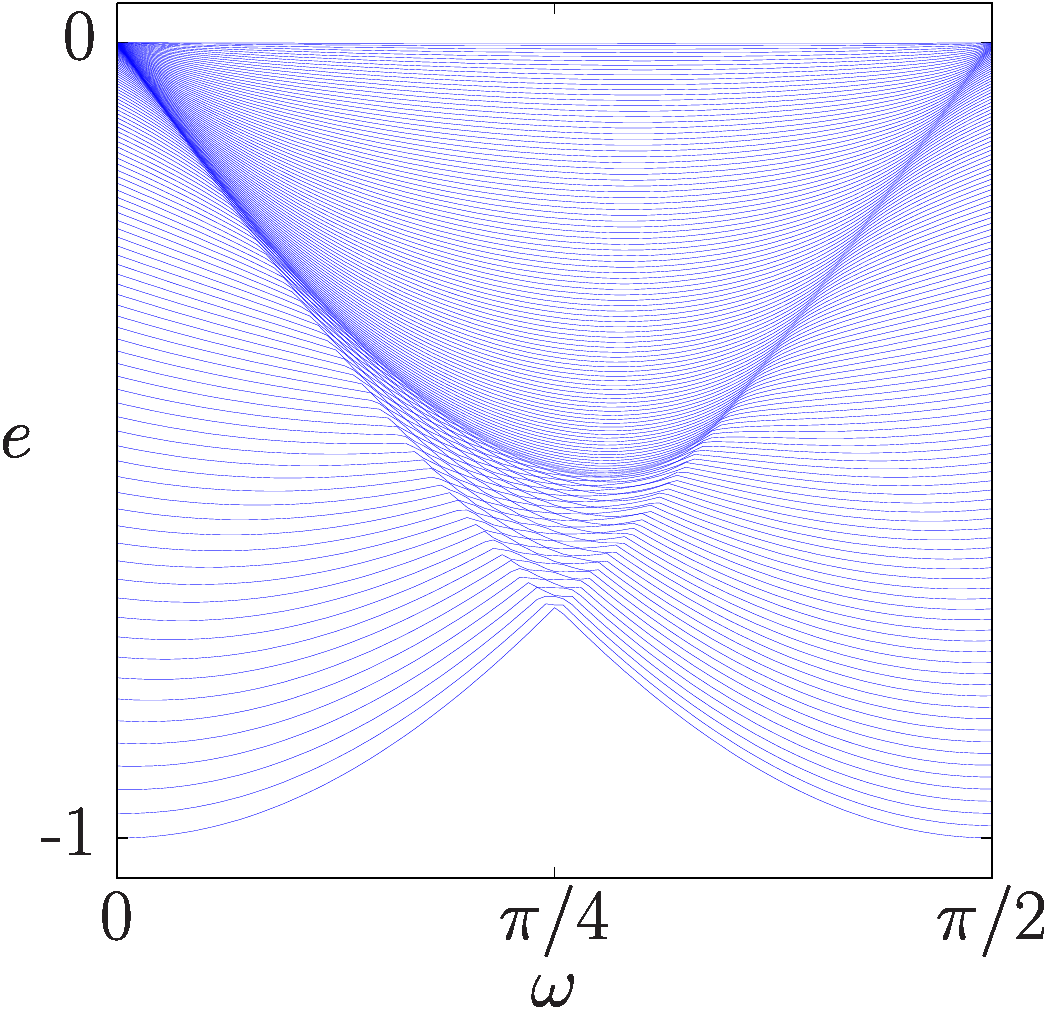}

    \caption{(Color online) Spectra (energies per spin) of six models as a
    function of the control parameter $\omega$, for a system of $N=256$ spins,
    in the maximum spin sector $S=N/2$.
    Left (right): $n=1$ ($n=2$). From top to bottom : $m=2$, $m=3$, and $m=4$.}
	\label{fig:spectrum_N256}
\end{figure}

A numerical study provides an idea about the phase transitions of the various
models. Such a study can be performed for rather large number of spins since, as
already mentioned, the Hamiltonians commute with $\boldsymbol{S}^2$. The collective
and ferromagnetic nature of the interactions implies that one can focus on the
maximum spin sector $S=N/2$ (of dimension $N+1$) where the ground state is found. 
The energies per spin $e$ of six different models are shown in
Figs.~\ref{fig:spectrum_N16} and \ref{fig:spectrum_N256}, for $N=16$ and $N=256$
respectively. From the evolution of the full spectra between these two figures,
one can infer the behavior of the various models in the thermodynamical limit.

The $(2,1)$ model displays a collapse of levels onto the ground state, at the
transition point $\omega=\pi/4$, but the ground-state energy displays no cusp.
From these features, the quantum phase transition is likely to be of second
order. All other models (except the $(2,2)$ model) have avoided level crossings,
which tend to become true level crossings in the thermodynamical limit. In
particular, the ground-state energy has a cusp, and the transition is of first
order. Note that, in Fig.~\ref{fig:spectrum_N256}, the darker regions are those
where one finds many levels (which are finite-size precursors of singularities
in the density of states in the thermodynamical limit). But, contrary to what
happens for the $(2,1)$ model \cite{Ribeiro07,Ribeiro08}, these regions do not touch the ground-state
energy at the transition.
This is, however, not true for the $(2,2)$ model, which displays both a collapse
of levels onto the ground state as well as a cusp in the ground-state energy.
From the second feature, one concludes that the transition is first order, although
the first feature is reminiscent of a second-order quantum phase transition. 
Let us stress that the $(2,2)$ model is trivially integrable at the transition point, since its Hamiltonian reads $H=-\frac{2\sqrt{2}}{N}(S^2-S_y^2)$, where $S^2=(N/2)(N/2+1)$. This additional symmetry ($[H,S_y]=0$) is responsible for the presence of non-avoided level crossings at the transition point, even at finite $N$. Actually, all these models are exactly solvable, but not in such a trivial way \cite{Pan99,Links03,Ortiz05}.

It is this variety of behaviors displayed by the different models that
motivates the study of entanglement measures and their sensitivity to the
characteristics of the quantum phase transition. Before turning to this, we
shall, however, provide analytical results for the low-energy spectrum, which
will allow us to introduce the basic techniques needed to perform analytical
computations of entanglement measures.

%
%

%
%

\subsubsection{Ground-state energy}
\label{sec:sub:sub:GSE}

Since a large spin behaves classically, a classical analysis is expected to
provide exact results for the ground-state energy of the collective models in
the thermodynamical limit. We are therefore led to substitute the spin operators by
their expectation values, namely
%
\begin{equation}
	\big(\left\langle S_x\right\rangle, \left\langle S_y\right\rangle, \left\langle S_z\right\rangle \big)
	= \frac{N}{2}\left(\sin\theta\cos\phi,\sin\theta\sin\phi,\cos\theta\right),
	\label{eq:expectation_value_Sxyz}
\end{equation}
%
where $\theta\in\left[0,\pi\right]$ and $\phi\in\left[0,2\pi\right[$ are the
usual angles of spherical coordinates. They are the variational parameters that
will be tuned to minimize the associated classical energy per spin
%
\begin{equation}
	e(\theta,\phi) = - \cos\omega (\sin\theta\cos\phi)^m
	- K_{m,n}\sin\omega (\cos\theta)^n.
	\label{eq:classical_energy}
\end{equation}
%

When $m>n=1$, the analysis follows the mean-field calculation of
Ref.~\onlinecite{Maritan84} [see also Refs.~\onlinecite{Botet83,Dusuel05_2} for the
$(2,1)$ model]. At ``large'' $\omega$ (close to $\pi/2$), the state $(\theta_0=0,\phi)$ is the
only minimum, with energy $e(0,\phi)=- K_{m,1}\sin\omega$. At ``small''
$\omega$ (close to $0$), $\phi_0=0$ ($\phi_0=0$ or $\pi$) when $m$ is odd (even), and the angle
$\theta_0$ minimizing the energy satisfies
$m\cos\theta_0\sin^{m-2}\theta_0=K_{m,1}\tan\omega$.
Requiring that the transition take place at $\omega=\pi/4$, one is led to solve
the following system of equations:
%
\begin{eqnarray}
	K_{m,1} &=& m\cos\theta_0^*\sin^{m-2}\theta_0^*, \\
	K_{m,1} &=& \sin^m\theta_0^* + K_{m,1} \cos\theta_0^*,
	\label{eq:Kmn_and_theta}
\end{eqnarray}
%
where the second equation stems from the continuity of $e$ at the transition,
and $\theta_0^*$ is the value of $\theta_0$ at the transition, in the
small-$\omega$ phase. The solution of this system yields Eqs.~(\ref{eq:K21}) and
(\ref{eq:Km1}), as well as $\cos\theta_0^*=\frac{1}{m-1}$. It is therefore clear
that, except for the $(2,1)$ model, $\theta_0$ is discontinuous at the
transition, which is thus of first order for all $(m>2,1)$ models.

\begin{figure}[t]
	\includegraphics[width=0.49\columnwidth ]{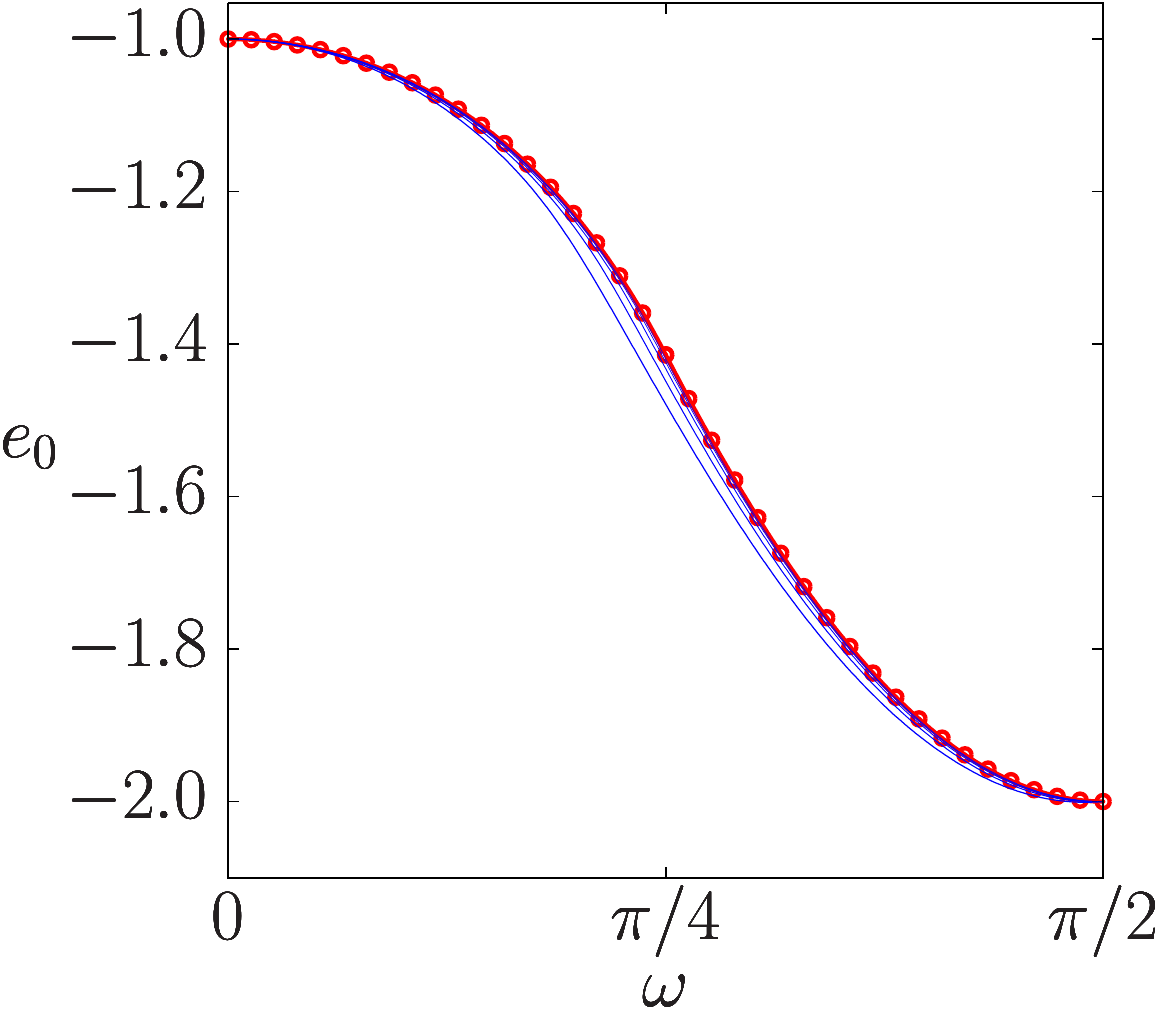}
	\includegraphics[width=0.49\columnwidth ]{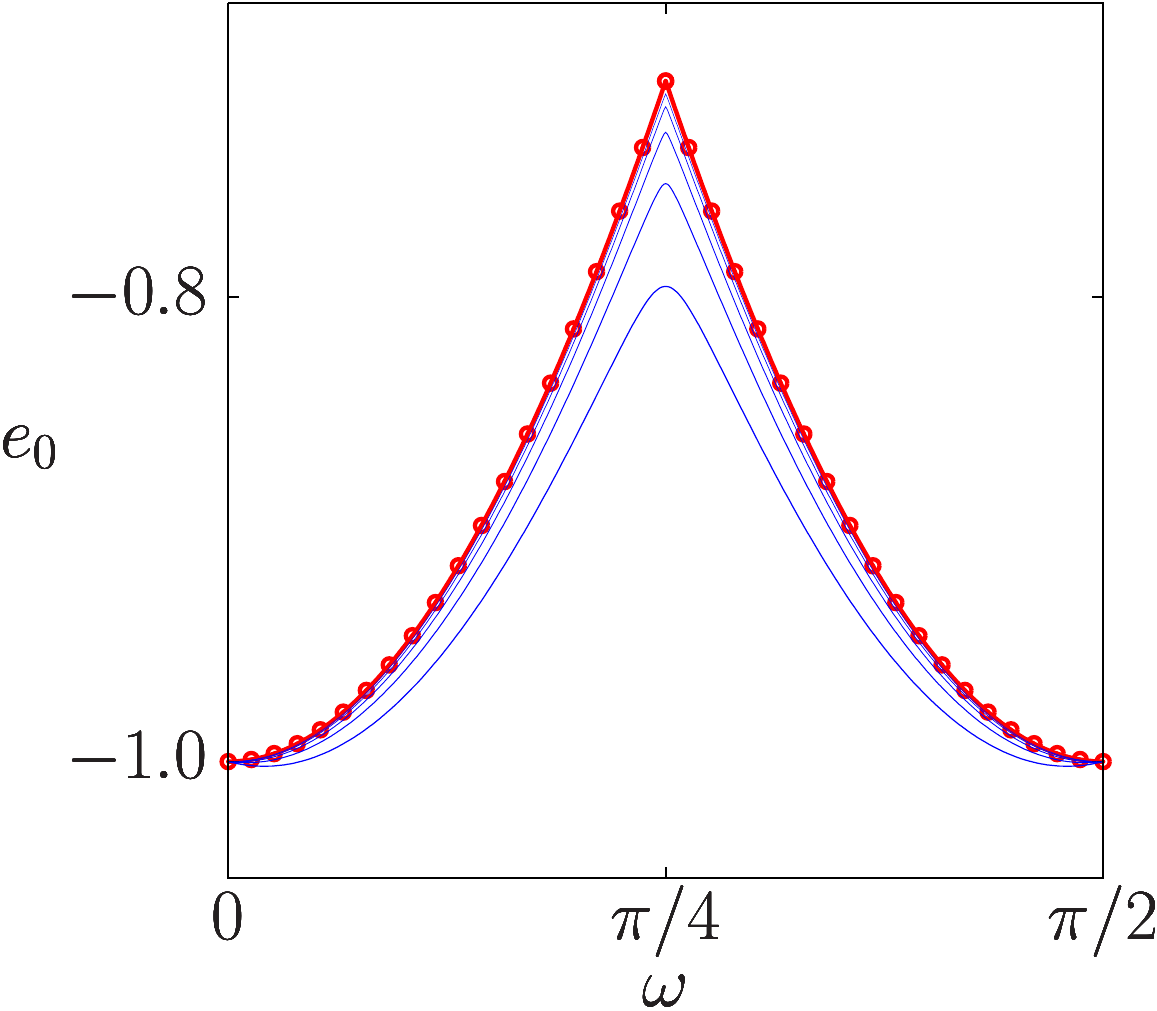}
	\vspace{0.15cm}

	\includegraphics[width=0.49\columnwidth ]{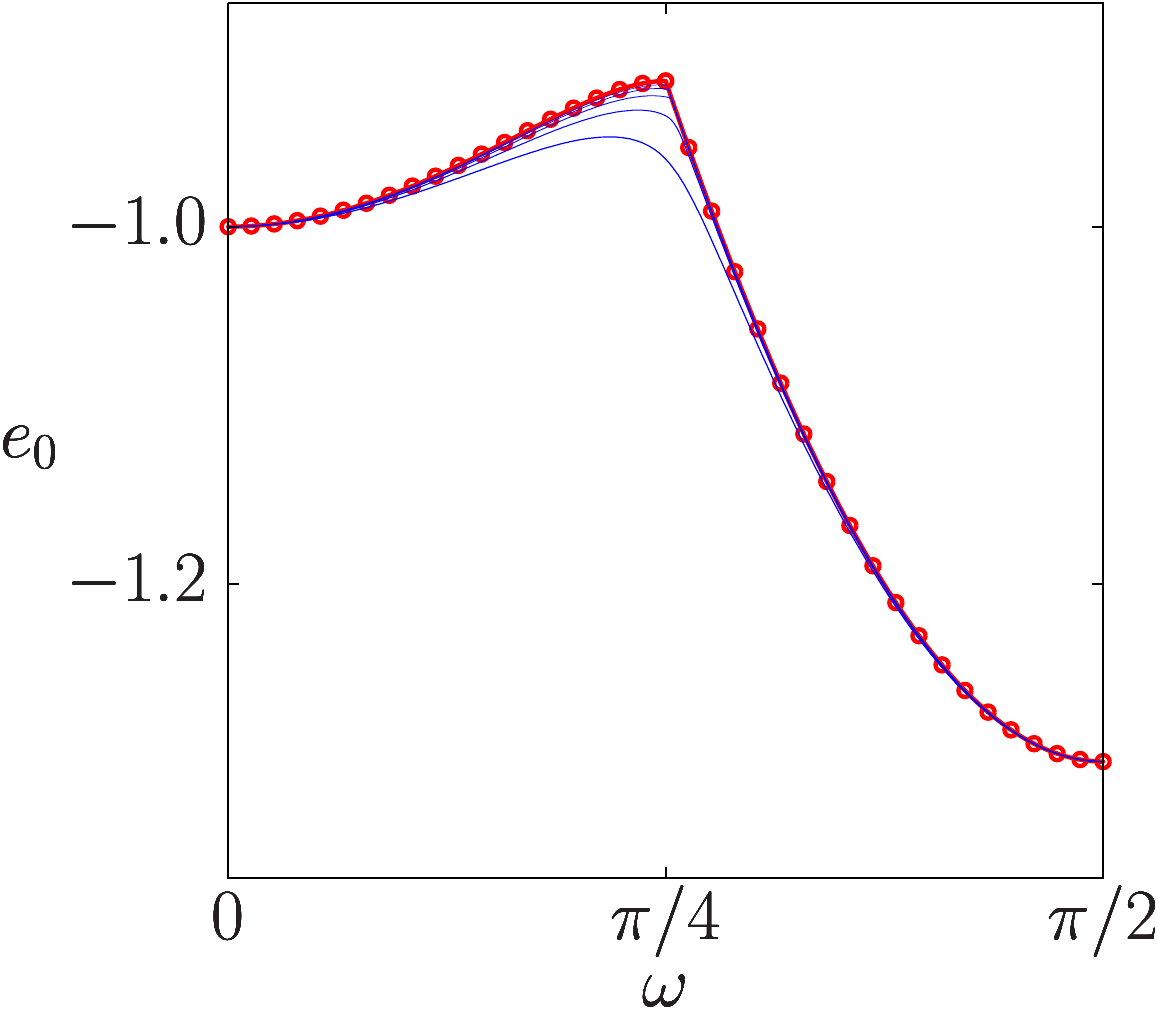}
	\includegraphics[width=0.49\columnwidth ]{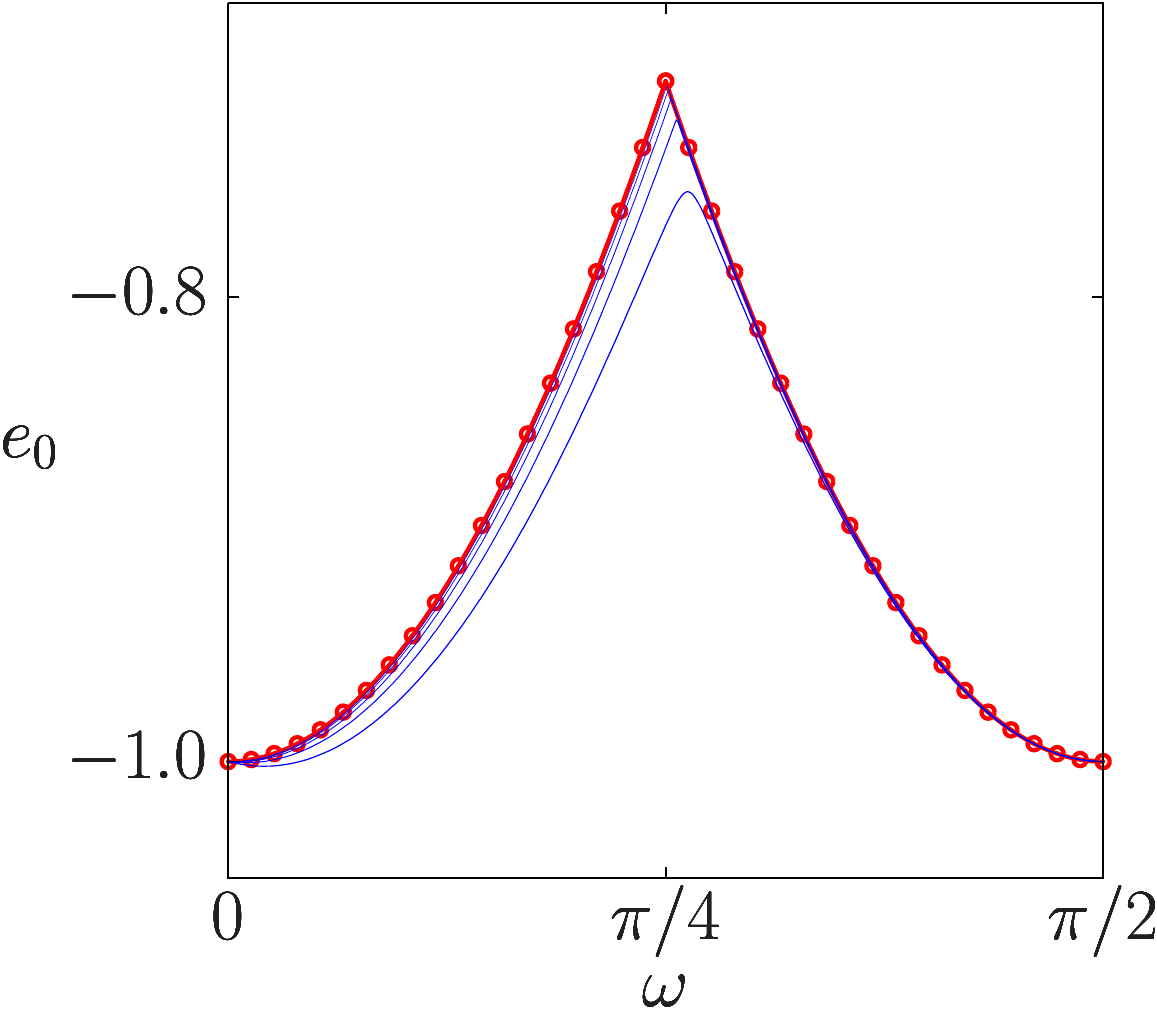}
	\vspace{0.15cm}

	\includegraphics[width=0.49\columnwidth ]{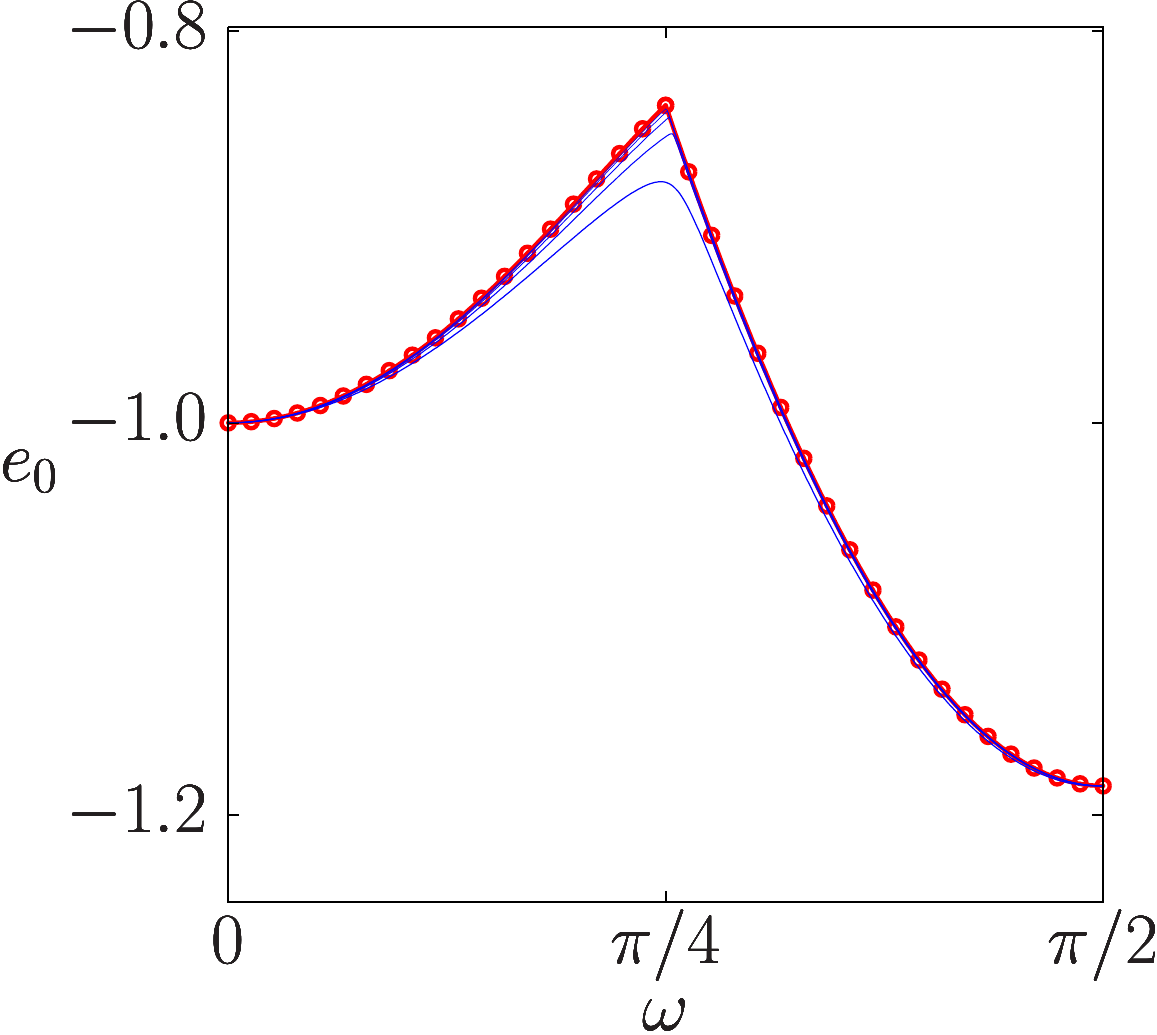}
	\includegraphics[width=0.49\columnwidth ]{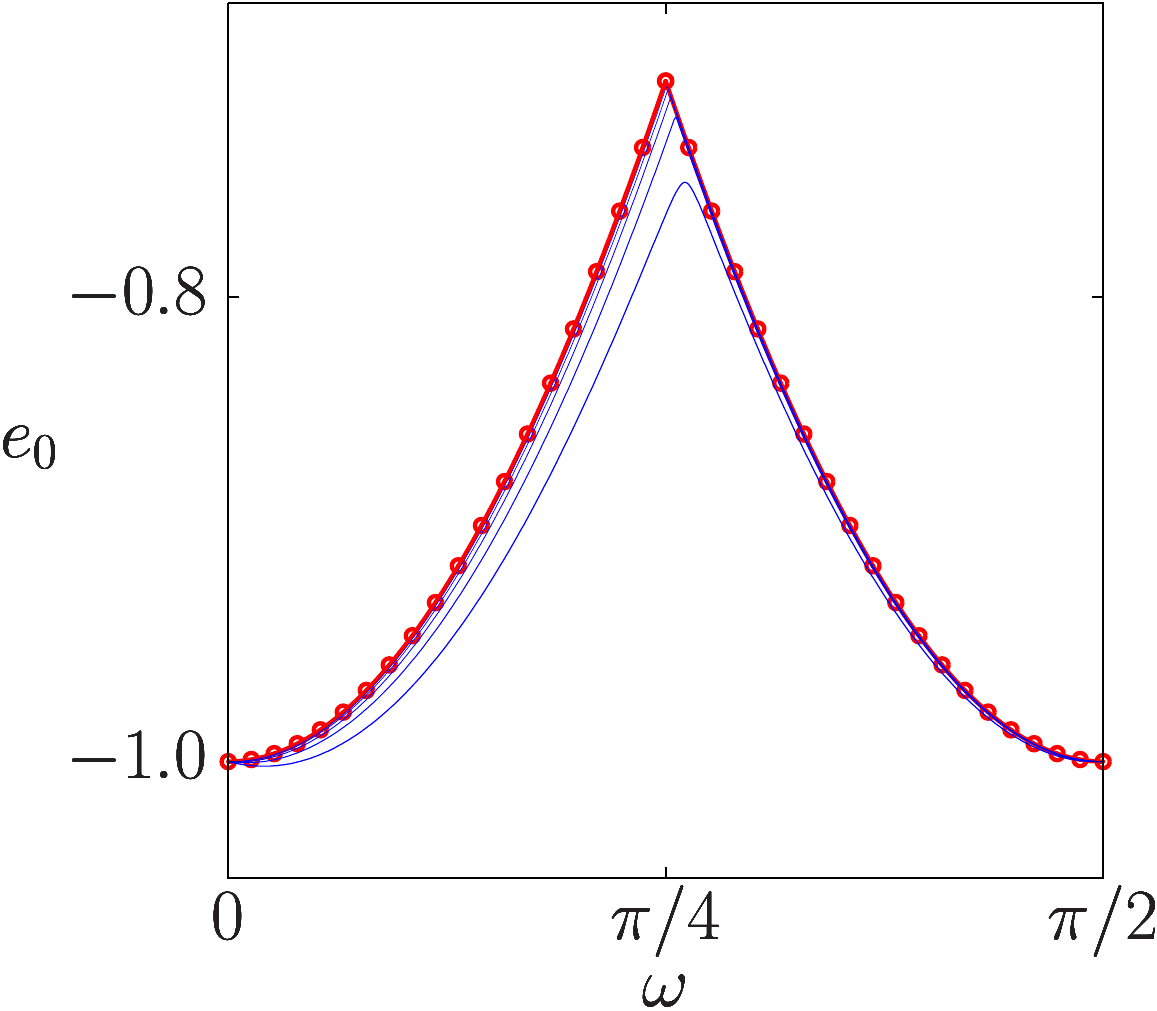}

	\caption{(Color online) Ground-state energy per spin $e_0$ of six models as
    a function of $\omega$, for $N=16$, $32$, $64$, $128$, $256$ and in the
    thermodynamical limit (red thick line with dots).
    Left (right): $n=1$ ($n=2$). From top to bottom : $m=2$, $m=3$, and $m=4$.}
	\label{fig:GSE_vary_N}
\end{figure}

For the $(2,1)$ model, the transition is of second order (see \textit{e.~g.}
Ref~\onlinecite{Botet83}), as can be seen from the discontinuity of the
second derivative $\frac{\partial^2 e}{\partial\omega^2}$ which jumps from the
value $-3\sqrt{2}$ at $\omega=(\pi/4)^-$ to $\sqrt{2}$ at $\omega=(\pi/4)^+$.
The large-$\omega$ phase is a symmetric phase with a non-degenerate ground
state, while the small-$\omega$ phase is a broken phase with a doubly-degenerate
ground state, the broken symmetry being the parity $S_x\leftrightarrow -S_{x}$.
The validity of the classical analysis can be assessed in
Fig.~\ref{fig:GSE_vary_N} where numerical data can be seen to converge to the
classical result.

When $m\geqslant n\geqslant 2$, one can proceed in the same way. One finds that
the transitions are all of first-order nature, that Eq.~(\ref{eq:Kmn}) has to
hold in order to have a transition at $\omega=\pi/4$, and that the angles
$\theta_0$ and $\phi_0$ take the following values
%
\begin{equation}
	\begin{tabular}{|c|c|}
	\hline
	$\omega\leqslant\pi/4$ & $\omega\geqslant\pi/4$ \\
	\hline
	$\theta_0=\pi/2$ & $\theta_0=0$ ($\pi$)\\
	$\phi_0=0$ ($\pi$) & any $\phi_0$\\
	\hline
	\end{tabular}
\end{equation}
%
The values in parentheses are other possible values depending on the parity of
$m$ and $n$. For $\omega\leqslant\pi/4$, the states $(\pi/2,0)$ and
$(\pi/2,\pi)$ are degenerate when $m$ is even, whereas for
$\omega\geqslant\pi/4$, the states $(0,\phi_0)$ and $(\pi,\phi_0)$ are
degenerate when $n$ is even. The degeneracies are already predictable from
Fig.~\ref{fig:spectrum_N16}, and the validity of the classical results can again
be checked in Fig.~\ref{fig:GSE_vary_N}.

%
%

\subsubsection{Gap}
\label{sec:sub:sub:gap}

In order to conclude the analysis of the spectrum and of the quantum phase
transition of these models, let us now turn to the computation of
the gap, in the maximum spin sector $S=N/2$. We follow the procedure described
in Refs.~\cite{Dusuel04_3,Dusuel05_2} and refer the reader to these references for
details. As a first step, we perform a rotation around the $y$-axis, in order to
bring the $z$-axis along the classical magnetization direction
%
\begin{equation}
	\begin{pmatrix}
		S_x \\ S_y \\ S_z
	\end{pmatrix}
	=
	\begin{pmatrix}
		\cos\theta_0 & 0 & \sin\theta_0 \\
		0 & 1 & 0 \\
		-\sin\theta_0 & 0 & \cos\theta_0
	\end{pmatrix}
	\begin{pmatrix}
		\widetilde{S}_x \\ \widetilde{S}_y \\ \widetilde{S}_z
	\end{pmatrix}.
	\label{eq:rotation}
\end{equation}
%
%
\begin{figure}[t]
	\includegraphics[width=0.49\columnwidth ]{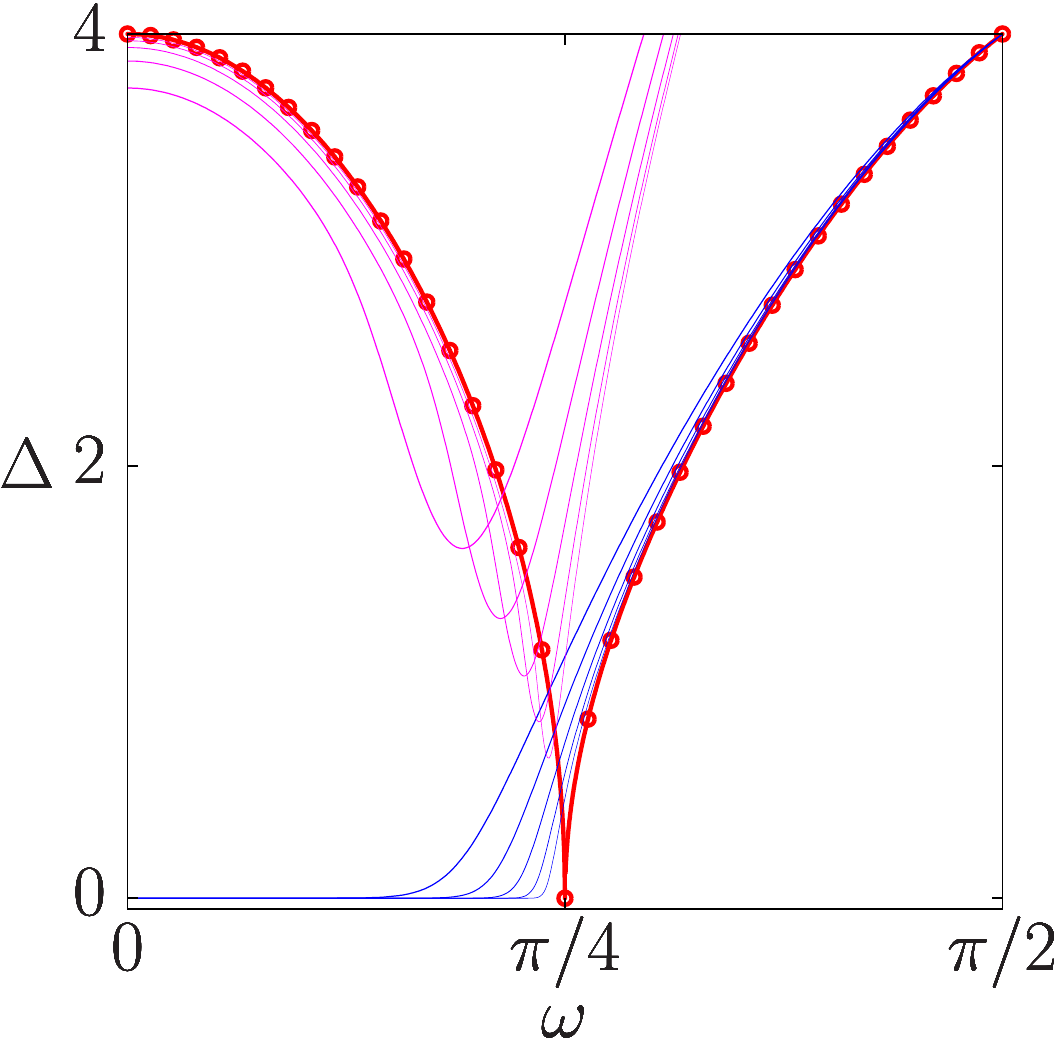}
	\includegraphics[width=0.49\columnwidth ]{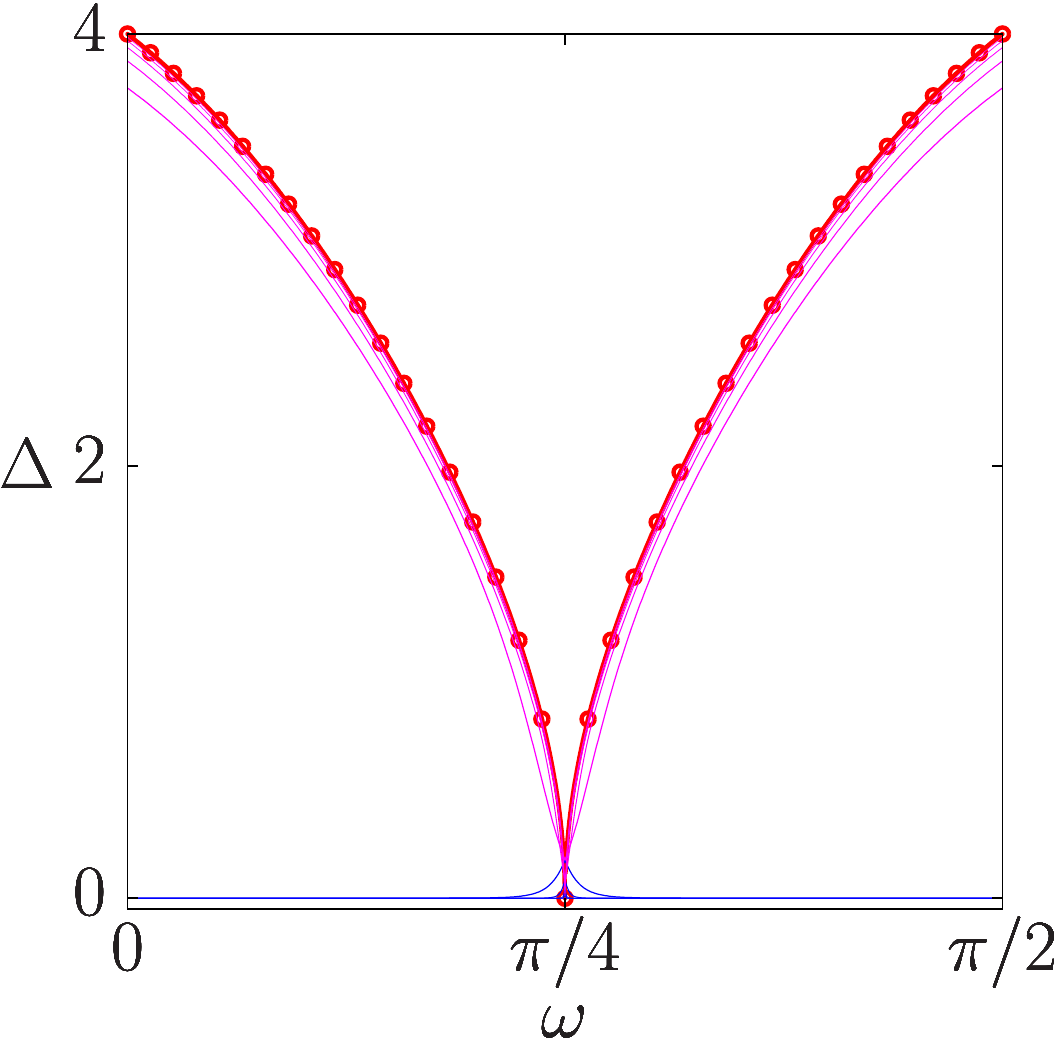}
	\vspace{0.15cm}

	\includegraphics[width=0.49\columnwidth ]{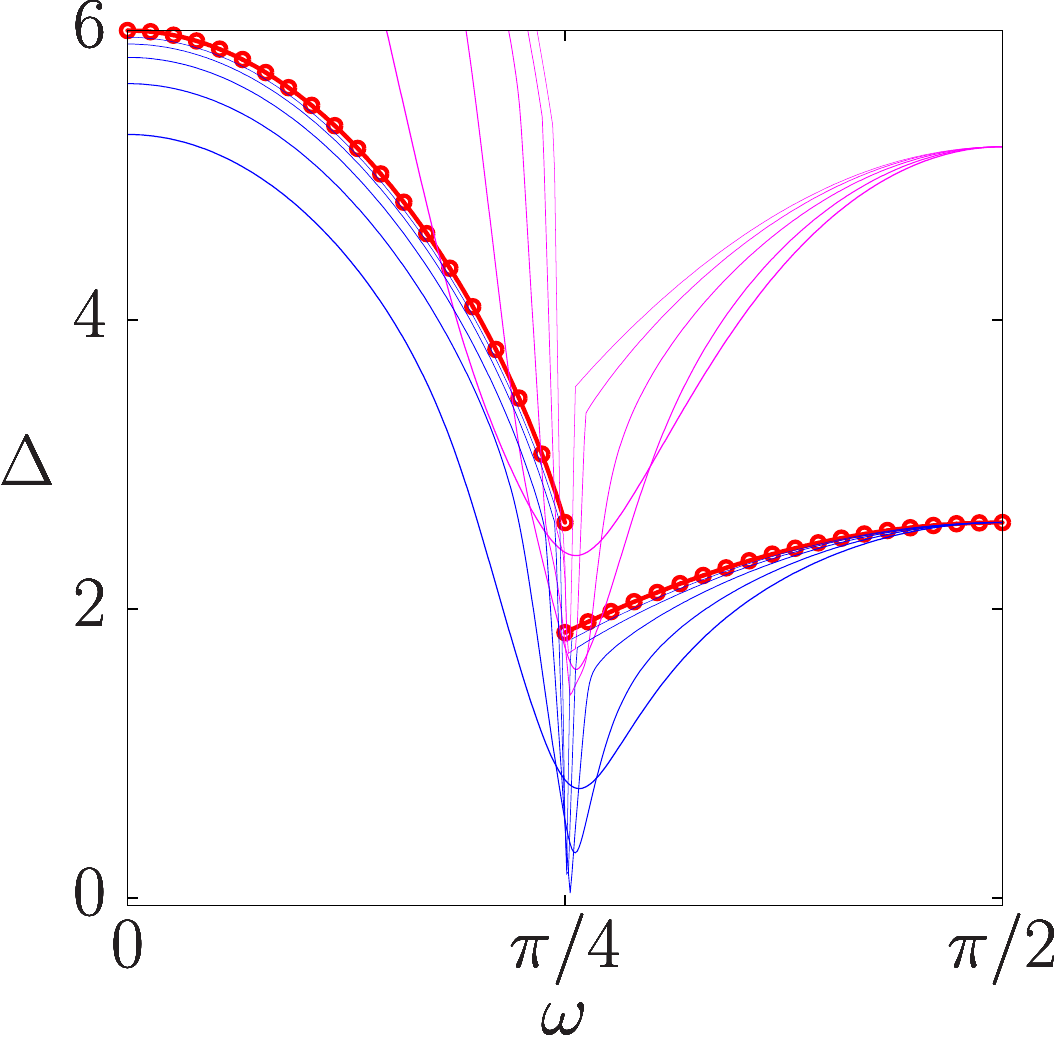}
	\includegraphics[width=0.49\columnwidth ]{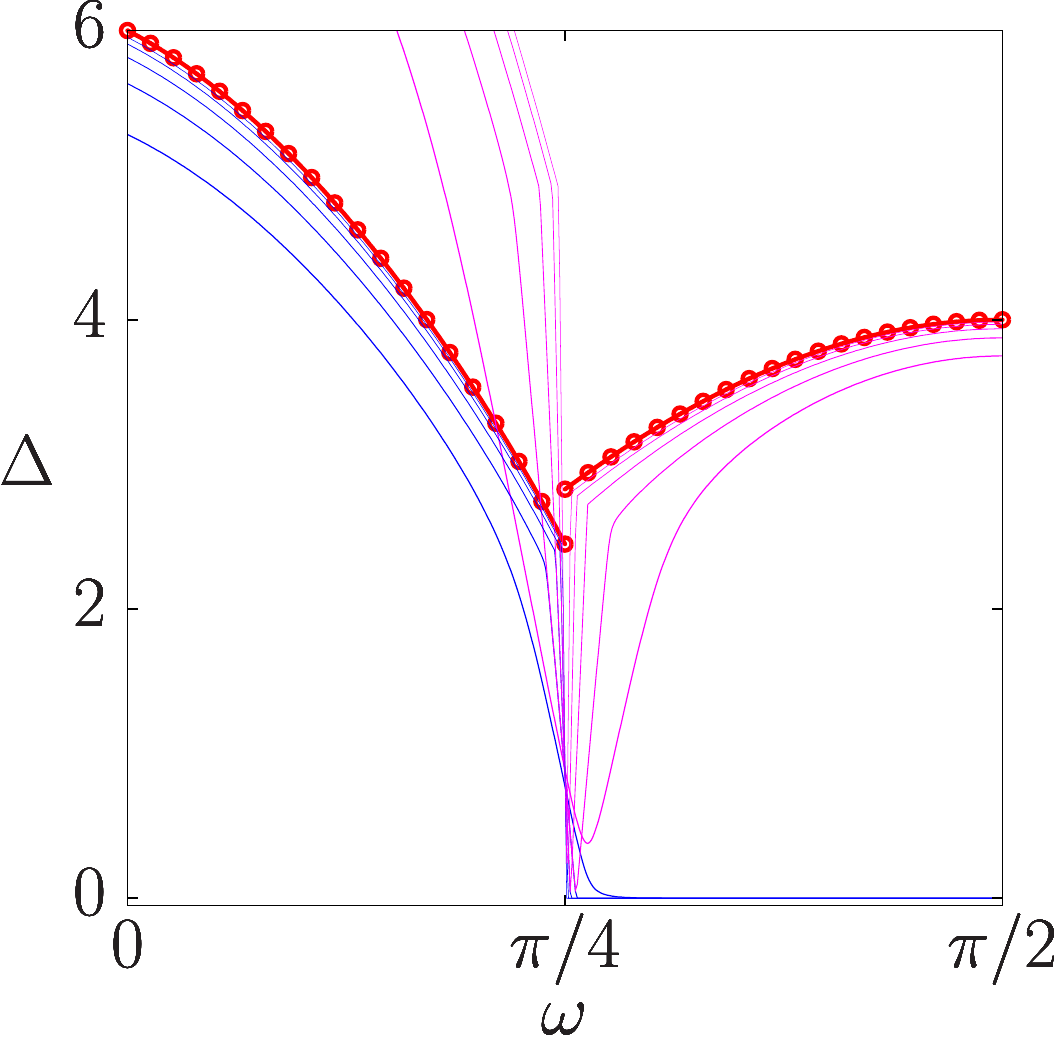}
	\vspace{0.15cm}

	\includegraphics[width=0.49\columnwidth ]{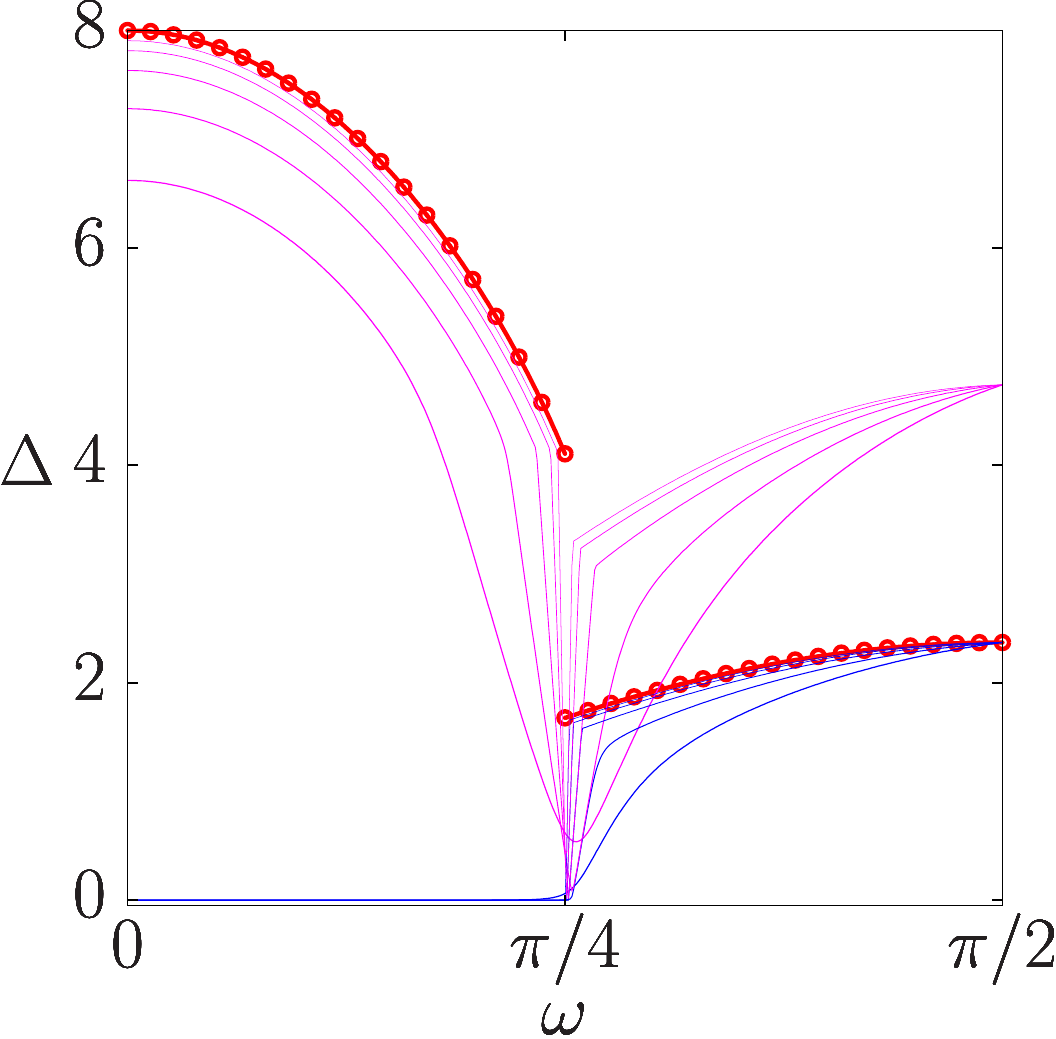}
	\includegraphics[width=0.49\columnwidth ]{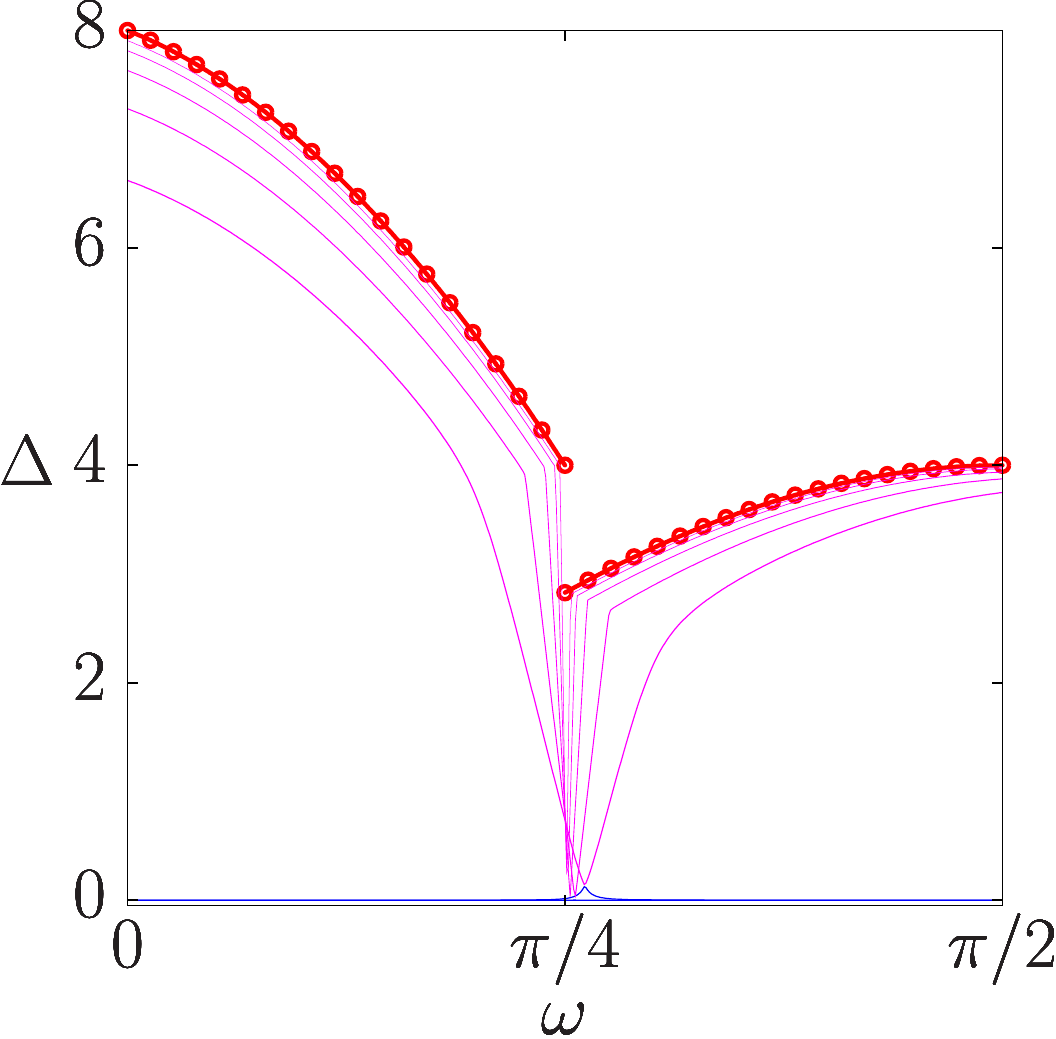}

	\caption{(Color online) Gaps to the first two excited states of six models
    as a function of $\omega$, for $N=16$, $32$, $64$, $128$, $256$ and in the
    thermodynamical limit (red thick line with dots).
    Left (right): $n=1$ ($n=2$). From top to bottom : $m=2$, $m=3$, and $m=4$.}
	\label{fig:gap_vary_N}
\end{figure}
%
Next, we make use of the bosonic Holstein-Primakoff representation
of the rotated spin operators \cite{Holstein40}
%
\begin{equation}
	\widetilde{S}_z = \frac{N}{2}-a^\dagger a \,\mbox{ and }\, \widetilde{S}_+ = \left(N-a^\dagger a\right)^{1/2} a = {\widetilde{S}_-}^\dagger,
	\label{eq:HP}
\end{equation}
%
where $\widetilde{S}_\pm=\widetilde{S}_x \pm \mathrm{i} \widetilde{S}_y$ and the
$a$ operator is a bosonic annihilation operator, satisfying
$\left[a,a^\dagger\right]=1$. As we shall only focus on the thermodynamical
limit, it will be sufficient to keep terms of order $N^1$ and $N^0$ in the
Hamiltonian, neglecting all terms that go to zero as $N\to\infty$, (assuming a
finite number $a^\dagger a$ of bosons). For example, in the large-$\omega$
phase, one can write $S_x = \widetilde{S}_x = N^{1/2}(a^\dagger+a)/2$. Note that
no term of order $\sqrt{N}$ appears, thanks to the rotation we have performed.
The Hamiltonian then reads
%
\begin{equation}
	H = N e_0 + \gamma + \delta\, a^\dagger a + \gamma\left({a^\dagger}^2 +
    {a^{\phantom{\dagger}\!\!\!}}^2\right),
	\label{eq:hamHP}
\end{equation}
%
where $e_0$, $\gamma$ and $\delta$ have the following expressions (the value
of $\theta_0$ has been given in Sec.~\ref{sec:sub:sub:GSE})
%
\begin{eqnarray}
    \label{eq:e_0}
    e_0 &=& - \cos\omega (\sin\theta_0)^m
        - K_{m,n}\sin\omega (\cos\theta_0)^n, \\
    \label{eq:gamma}
    \gamma &=& -\frac{1}{2} \big[
    m(m-1)\cos\omega (\sin\theta_0)^{m-2}(\cos\theta_0)^2\nonumber\\
    && + n(n-1) K_{m,n}\sin\omega (\cos\theta_0)^{n-2} (\sin\theta_0)^2
    \big], \\
    \label{eq:delta}
    \delta &=& 2\big[m \cos\omega (\sin\theta_0)^m
    + n K_{m,n} \sin\omega(\cos\theta_0)^n \big]\nonumber\\
    && + 2\gamma.
\end{eqnarray}
%
Note that $e_0$ is simply the minimum of the classical ground-state energy
(\ref{eq:classical_energy}).

Such a quadratic Hamiltonian is diagonalized via a Bogoliubov transformation
%
\begin{equation}
	a = \cosh(\Theta/2) b + \sinh(\Theta/2) b^\dagger,
	\label{eq:bogo}
\end{equation}
%
where $b$ is a bosonic annihilation operator, satisfying
$\left[b,b^\dagger\right]=1$. The value of $\Theta$ diagonalizing the
Hamiltonian satisfies $\tanh\Theta=\varepsilon=-2\gamma/\delta$. With these
notations,
%
\begin{equation}
	H = N e_0 + \gamma + \frac{\delta}{2}\left(\sqrt{1-\varepsilon^2}-1\right)+\Delta\, b^\dagger b,
	\label{eq:ham_bogo}
\end{equation}
%
where the gap is $\Delta=\delta\sqrt{1-\varepsilon^2}$. Let us emphasize that
$\Delta$ is the gap above the possibly-degenerate ground state, but does not
capture the energy splitting between the ground states if they are degenerate.
We compare the spectrum of Eq.~(\ref{eq:ham_bogo}) with numerics in
Fig.~\ref{fig:gap_vary_N} for the gap to the first and second excited states, so
we get at least one (and possibly two) nonzero value in the
thermodynamical limit. The relevance of this simple ``spin-wave"-like approach can be appreciated.

\section{Entanglement measures}
\label{sec:entanglement_measures}

\subsection{Technical prerequisite}
\label{sec:sub:technical_prerequisite}

We shall now compute three entanglement measures, namely, the concurrence, the
entanglement entropy, and the logarithmic negativity.  These measures have
already been computed for the $(2,1)$ model in Refs.~\onlinecite{Dusuel04_3,
Dusuel05_2, Latorre05_2, Barthel06_2, Vidal07, Wichterich09_2}. As can be
inferred from these works, the analytical computations require to write the
spin operators as the sum of one, two or three spin operators, for the
concurrence, entanglement entropy and negativity respectively. Then, one should
use the Holstein-Primakoff representation. The necessary steps for the
computation of the concurrence have already been performed in
Sec.~\ref{sec:sub:sub:gap}, but let us give the key ingredients that are useful to obtain
 the other entanglement measures.

The very first step is to perform the rotation (\ref{eq:rotation}). Then, one
splits the system into $p$ subsystems, so that the the spin operators read as
$\widetilde{S}_\alpha=\sum_{i=1}^p\widetilde{S}_\alpha^{(i)}$, where
$\alpha=x$, $y$ or $z$. Depending on the entanglement measure one wishes to
compute, one has $p=1$, $2$, or $3$. One then introduces $p$ bosonic operators
$a_i$ and their conjugates $a_i^\dagger$ for each subsystem. Denoting the
number of spins of each subsystem by $N_i$, with $\sum_{i=1}^p N_i=N$,
the $p$ Holstein-Primakoff representations read as
%
\begin{equation}
    \widetilde{S}_z^{(i)} = \frac{N_i}{2}-a_i^\dagger a_i
    \,\mbox{ and }\, \widetilde{S}_+^{(i)} =
    \left( N_i-a_i^\dagger a_i\right)^{1/2} a_i.
    \label{eq:HPp}
\end{equation}
%
One can then insert these expressions in the Hamiltonian, expand all operators
and keep terms of order $N^1$ and $N^0$, neglecting contributions that vanish in
the thermodynamical limit. After simple algebra, one gets a quadratic Hamiltonian
%
\begin{equation}
    H = N e_0 + \gamma + \delta\sum_{i=1}^pa_i^\dagger a_i +
    \gamma \sum_{k,l=1}^p \sqrt{\tau_k \tau_l}
        \big(a_k^\dagger a_l^\dagger+\mbox{H.c.}\big),
    \label{eq:hamHPp}
\end{equation}
%
where $e_0$, $\gamma$, and $\delta$ are given in
Eqs.~(\ref{eq:e_0})-(\ref{eq:delta}) and where $\tau_i=N_i/N$.
Let us note that, to obtain this precise quadratic form, with only diagonal
boson-conserving terms, one must get rid of terms of the form $a_k^\dagger a_l$
with $k\neq l$. To this end, one should use the relation
\mbox{$\sum_\alpha \widetilde{S}_\alpha^2 = S(S+1)$}, written in the bosonic language,
namely,
%
\begin{equation}
\sum_{k\neq l}\sqrt{\tau_k\tau_l}\big(a_k^\dagger a_l+\mbox{H.c.}\big)
    =2\sum_i(1-\tau_i)a_i^\dagger a_i.
\end{equation}
%
Of course, Eq.~(\ref{eq:hamHPp}) yields Eq.~(\ref{eq:hamHP}) when only one
bosonic mode is considered.

In the three subsections that follow, we shall give a minimal amount of
computational details, knowing that these can already be found in the
literature.

\subsection{Concurrence}
\label{sec:sub:concurrence}

\begin{figure}[t]
	\includegraphics[width=0.49\columnwidth ]{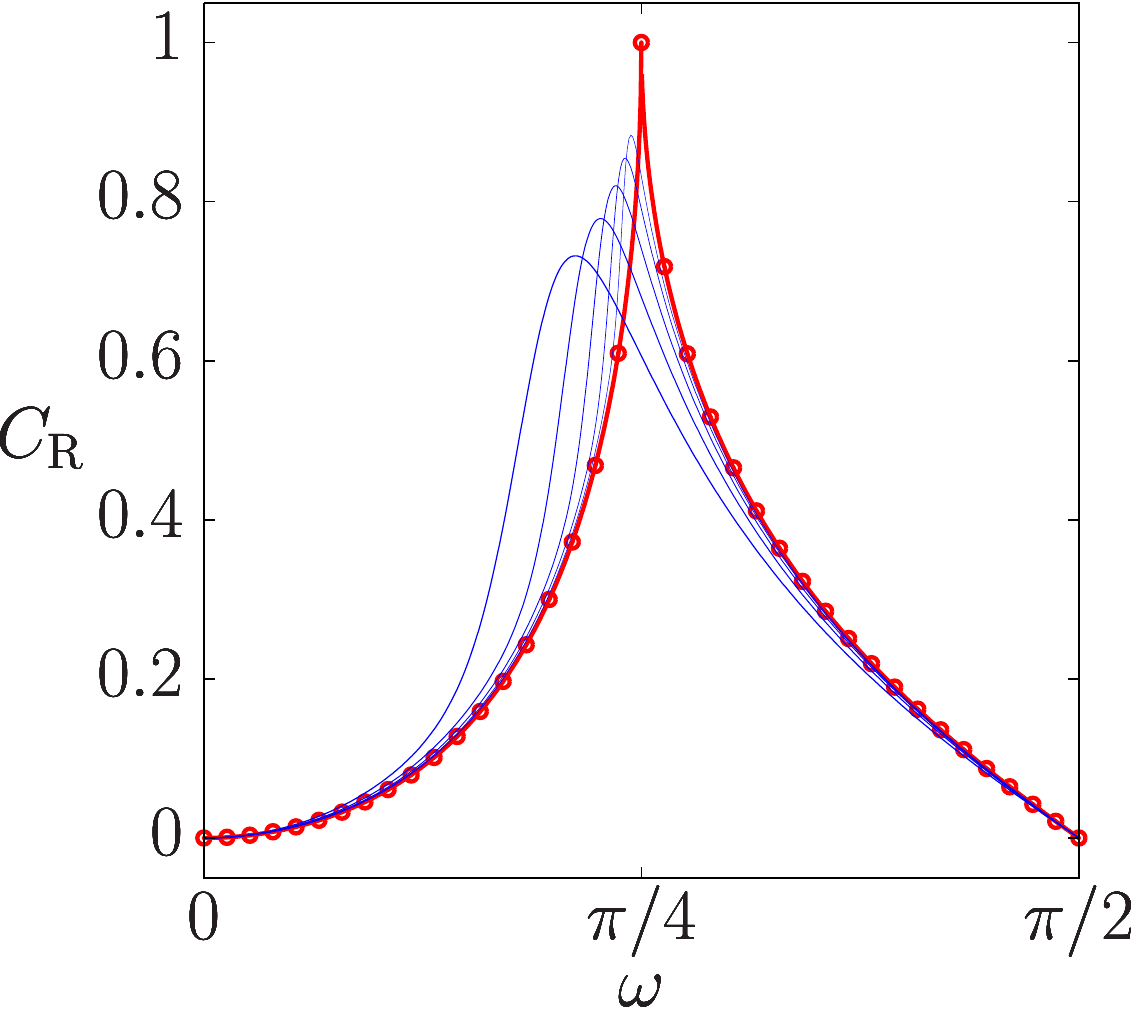}
	\includegraphics[width=0.49\columnwidth ]{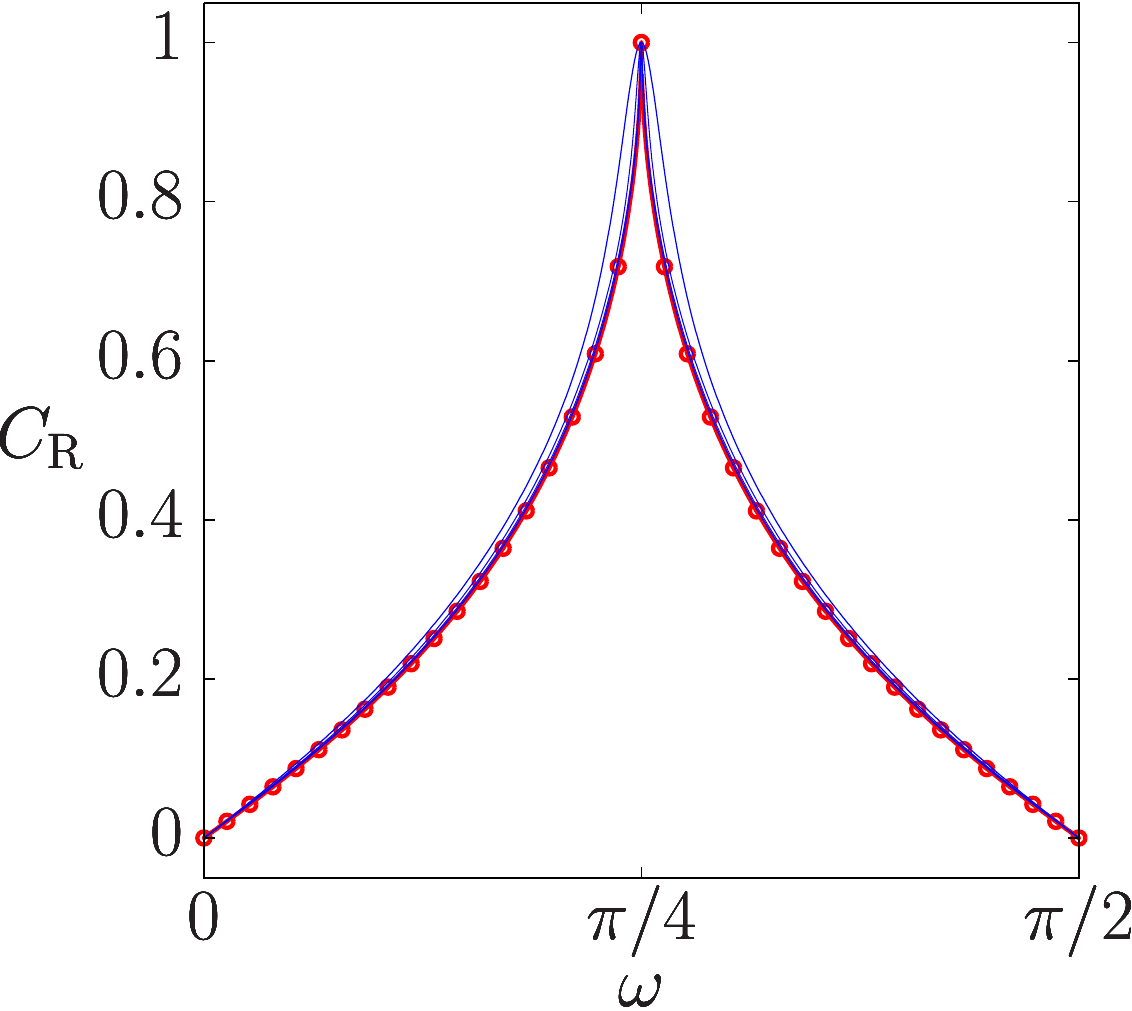}
	\vspace{0.15cm}

	\includegraphics[width=0.49\columnwidth ]{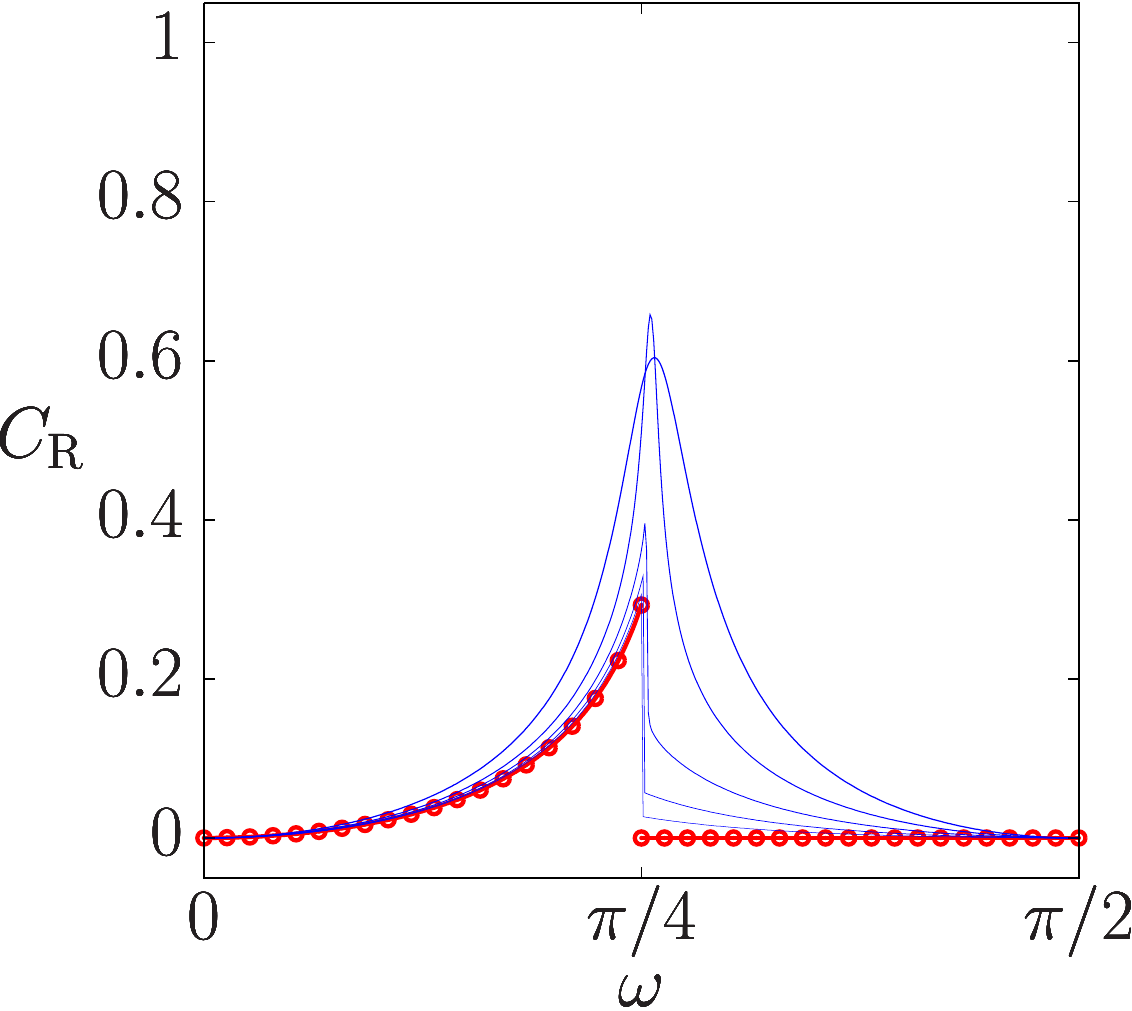}
	\includegraphics[width=0.49\columnwidth ]{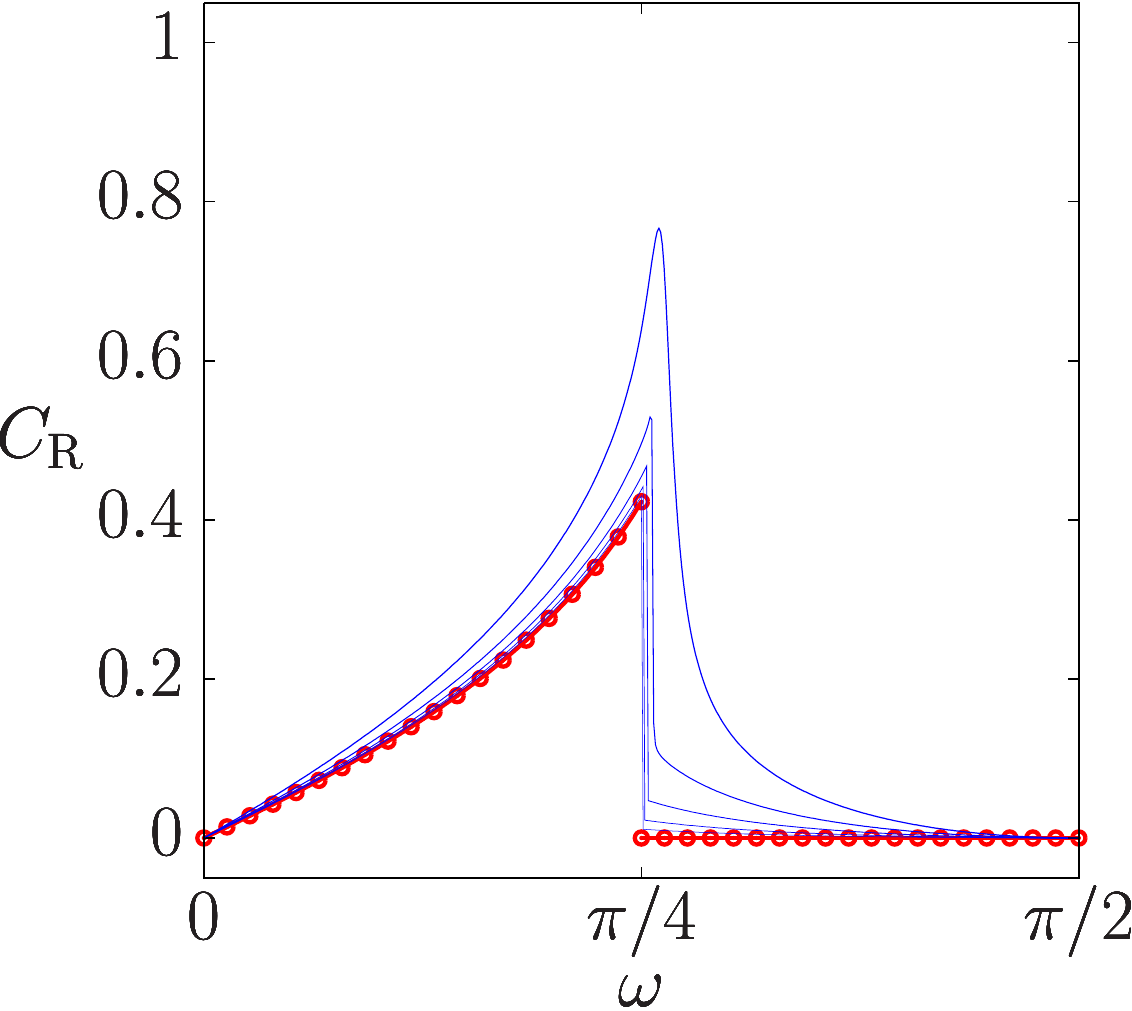}
	\vspace{0.15cm}

	\includegraphics[width=0.49\columnwidth ]{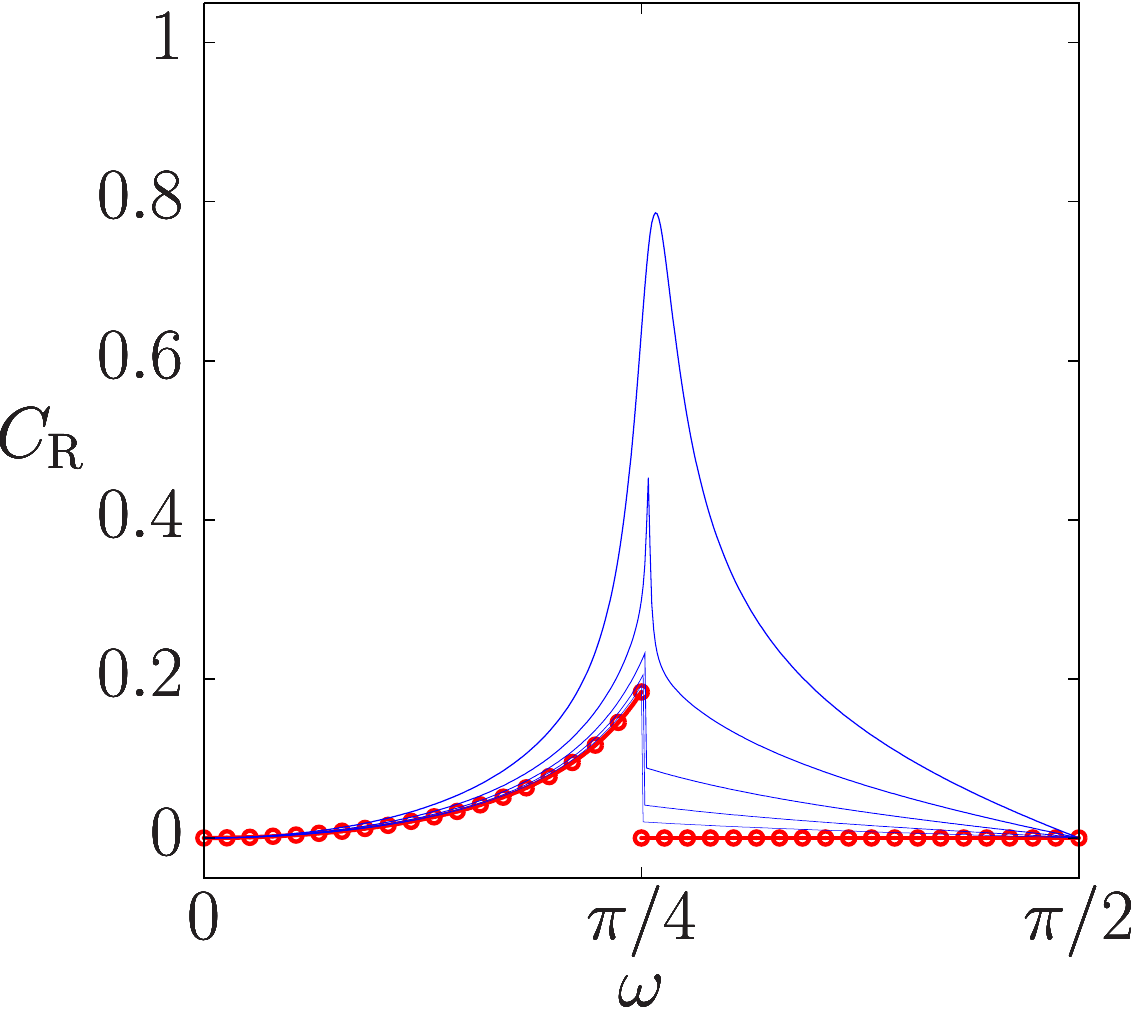}
	\includegraphics[width=0.49\columnwidth ]{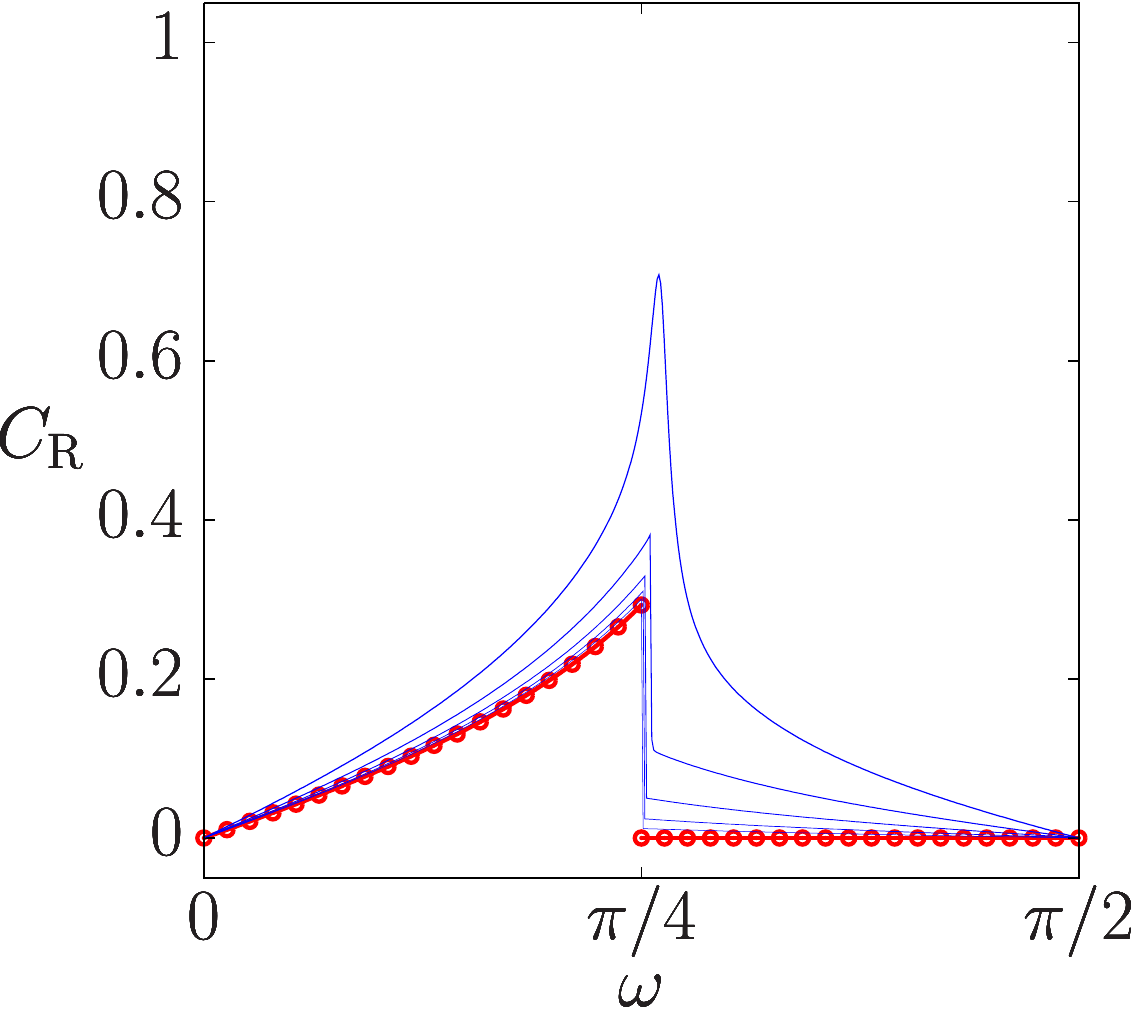}

	\caption{(Color online) Concurrence of six models as a function of
    $\omega$, for $N=16$, $32$, $64$, $128$, $256$ and in the thermodynamical
    limit (red thick line with dots).
    Left (right): $n=1$ ($n=2$). From top to bottom : $m=2$, $m=3$, and $m=4$.}
	\label{fig:concurrence_vary_N}
\end{figure}

The concurrence $C$ measures the entanglement between two spins half, these spins
being in either a pure or a mixed state \cite{Wootters98}.
Here, we are interested in quantifying the entanglement between any two spins,
the others being traced over. Finding the concurrence amounts to computing the
entries of the reduced density matrix, which can be done easily for symmetric
states \cite{Wang02}. However, except in the case of systems possessing a
spin-flip symmetry, finding a simple analytical formula for the concurrence
is not such an easy task \cite{Vidal06_3}. Although we have no formal proof,
we have checked numerically for finite-size systems and a couple of values of $m$
and $n$ (even when $m$ and $n$ are odd so that there is no spin-flip symmetry),
that the rescaled concurrence $C_\mathrm{R}$ could be simply expressed as
%
\begin{equation}
    C_\mathrm{R} = (N-1) C = 1 - \frac{4\langle S_y^2\rangle}{N}.
    \label{eq:conc_sy}
\end{equation}
%
This rescaling is needed here because each spin shares entanglement with its $N-1$ ``neighbors" \cite{Vidal04_1}.
Thanks to the Holstein-Primakoff representation
(\ref{eq:HP}) and to the Bogoliubov diagonalization of the associated
Hamiltonian (\ref{eq:bogo}), one can show \cite{Dusuel05_2} that in the
thermodynamical limit
%
\begin{equation}
    \alpha = \lim_{N\to\infty} \frac{4\langle S_y^2\rangle}{N}
    = \sqrt{\frac{1-\varepsilon}{1+\varepsilon}},
    \label{eq:alpha}
\end{equation}
%
where $\varepsilon$ is given just before Eq.~(\ref{eq:ham_bogo}). 
Figure \ref{fig:concurrence_vary_N} displays numerical results for increasing system sizes which clearly converge toward the expression computed above in the thermodynamical limit.  
%
%

As can be inferred from this figure, the concurrence is cusped but continuous at
the second-order quantum phase transition for the $(2,1)$ model, while it
displays a jump at the first-order transition of the $(m,n)$ models, except for
the $(2,2)$ model where it shows a cusp and is continuous. The spectral
peculiarities of the latter, which have been discussed in
Sec.~\ref{sec:sub:sub:numerical_spectra}, do not lead to a discontinuous
concurrence. It therefore seems, in this very special case, that an entanglement
measure such as the concurrence is more sensitive to the ``level collapse" on the ground
state (Anderson's tower structure), than to the level
crossing, when both effects are present. We shall show that this conclusion remains valid for
the other entanglement measures we have calculated, starting with the
entanglement entropy.

\subsection{Entanglement entropy}
\label{sec:sub:entropy}

The ground-state entanglement between two complementary subsystems $\mathcal{A}$
and $\mathcal{B}$ can be quantified by the R\'enyi entropy, defined by
%
\begin{equation}
    \mathcal{E}_q = \frac{1}{1-q}\ln\left[
    \mathrm{Tr}\left({\rho_\mathcal{A}}^q\right)\right].
    \label{eq:def_renyi}
\end{equation}
%
%
\begin{figure}[t]
	\includegraphics[width=0.49\columnwidth ]{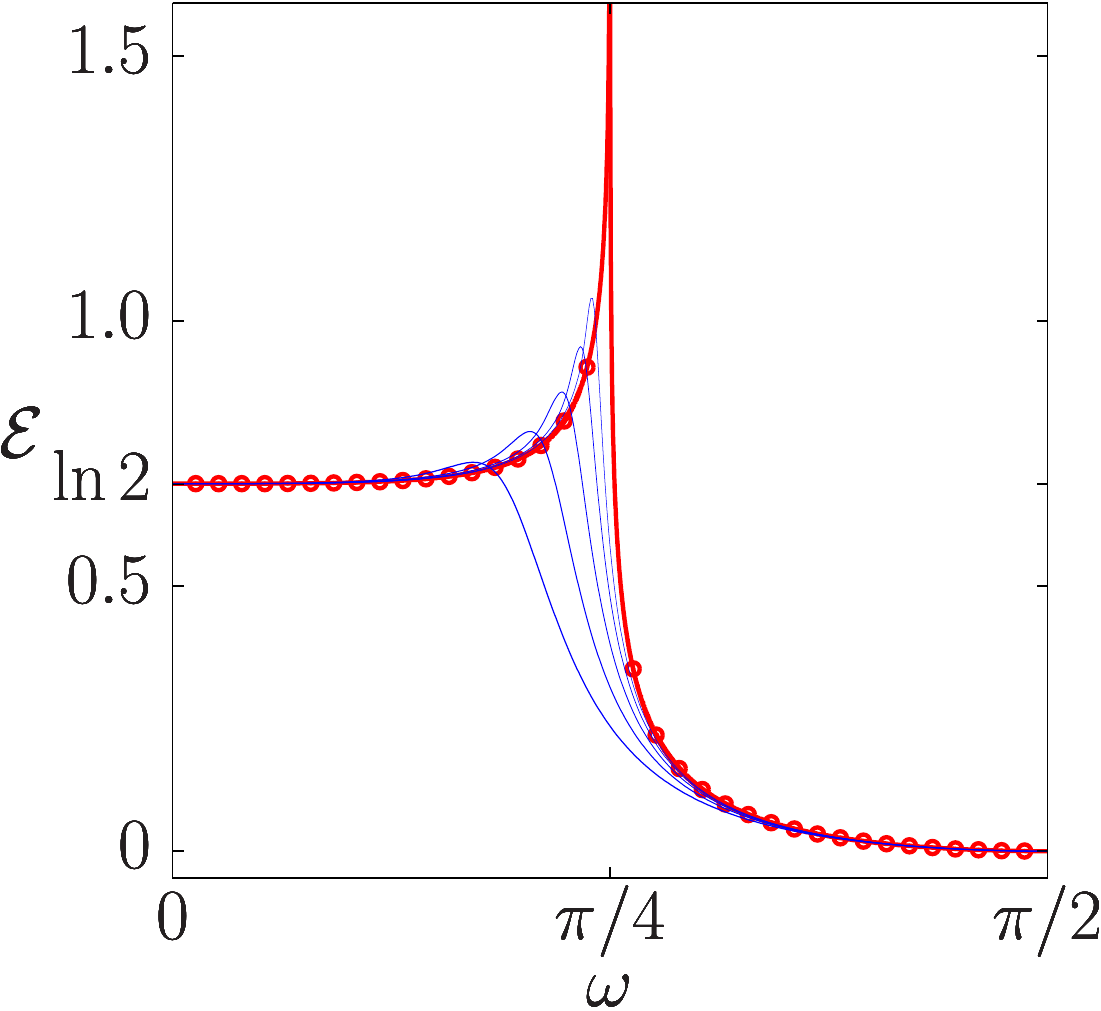}
	\includegraphics[width=0.49\columnwidth ]{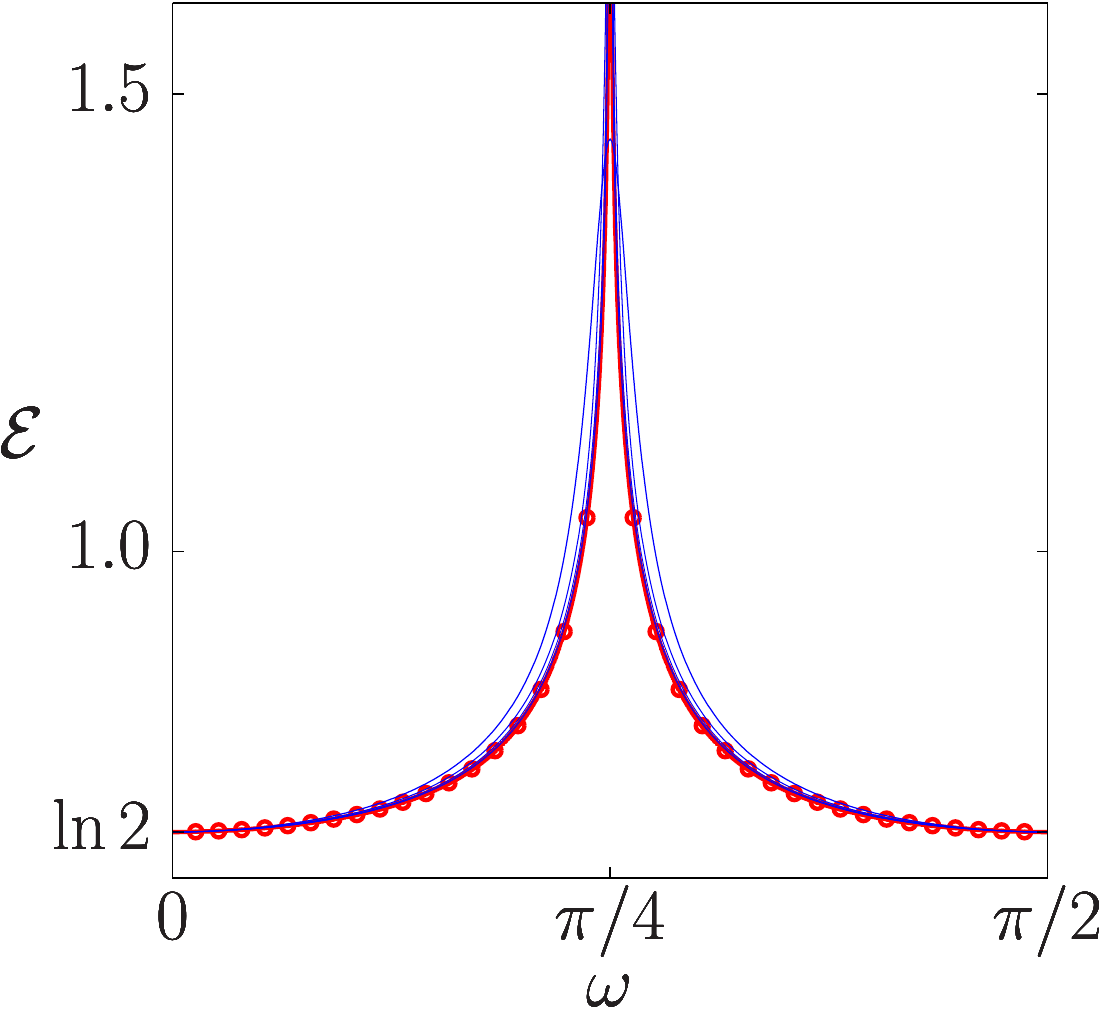}
	\vspace{0.15cm}

	\includegraphics[width=0.49\columnwidth ]{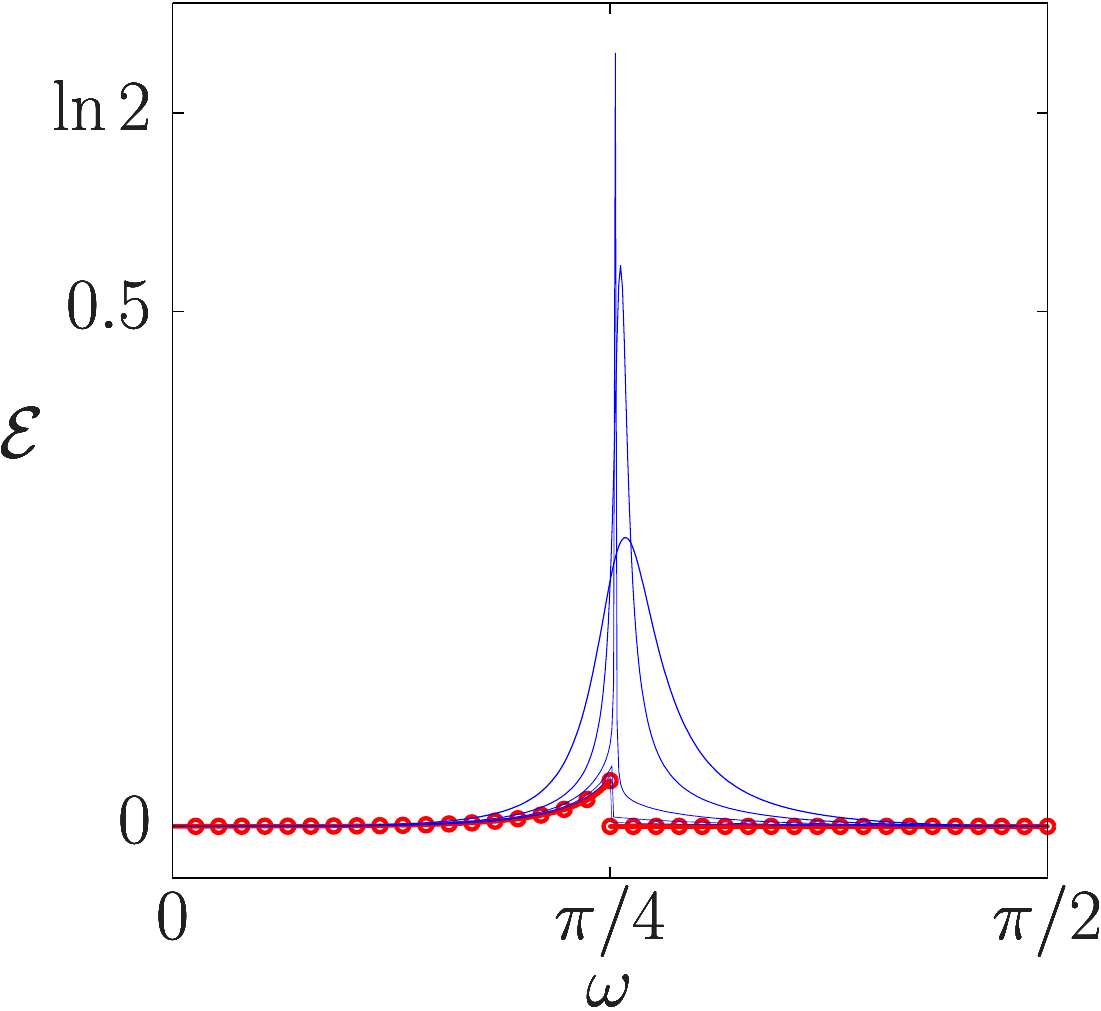}
	\includegraphics[width=0.49\columnwidth ]{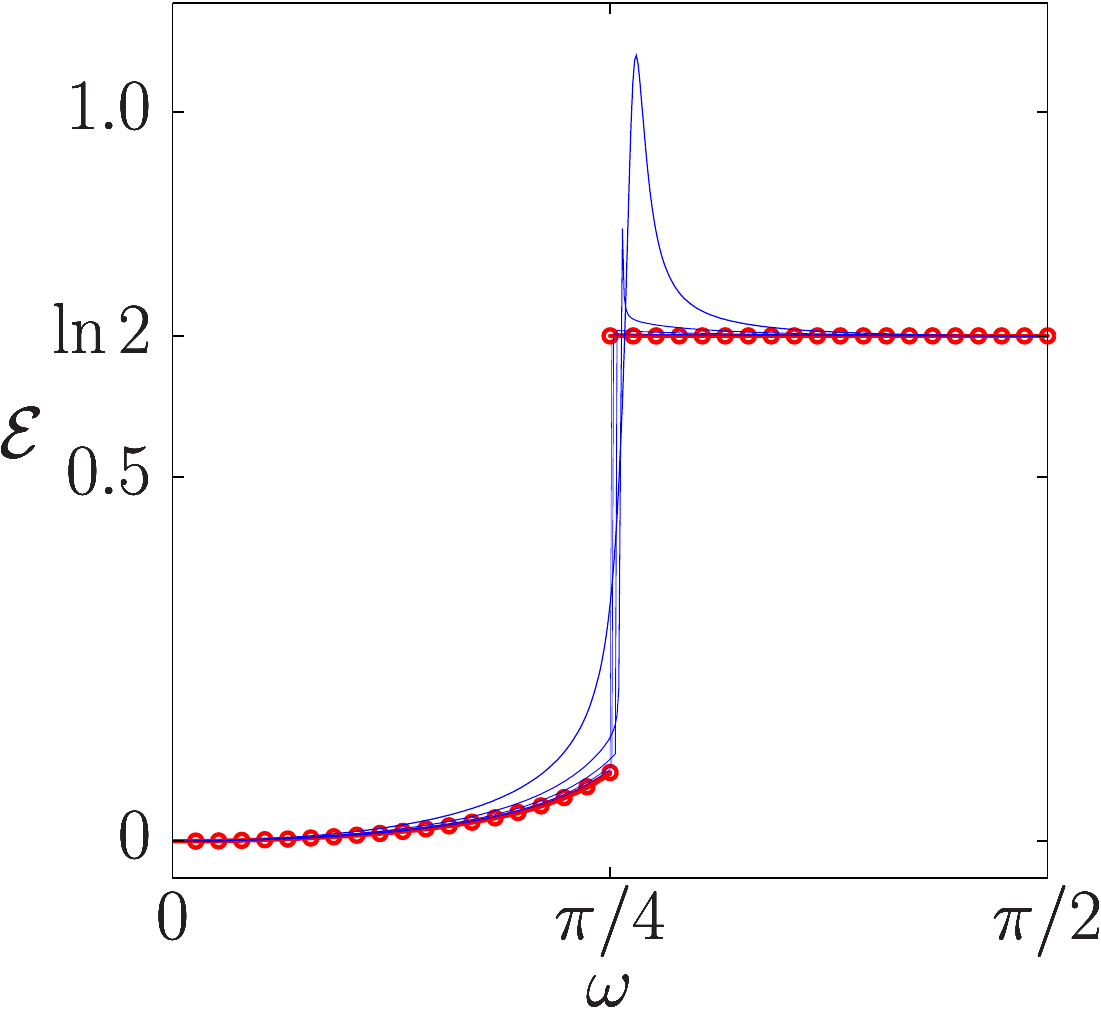}
	\vspace{0.15cm}

	\includegraphics[width=0.49\columnwidth ]{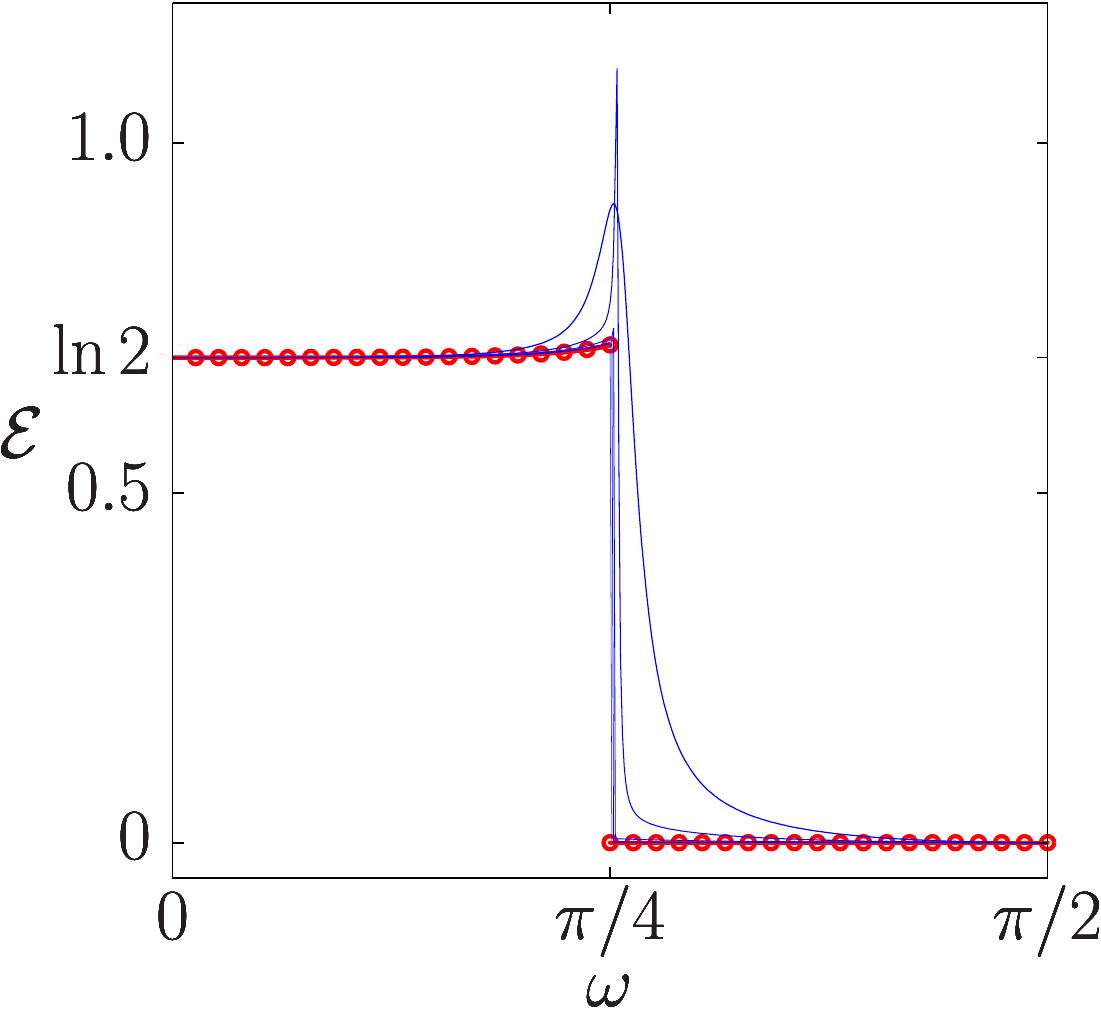}
	\includegraphics[width=0.49\columnwidth ]{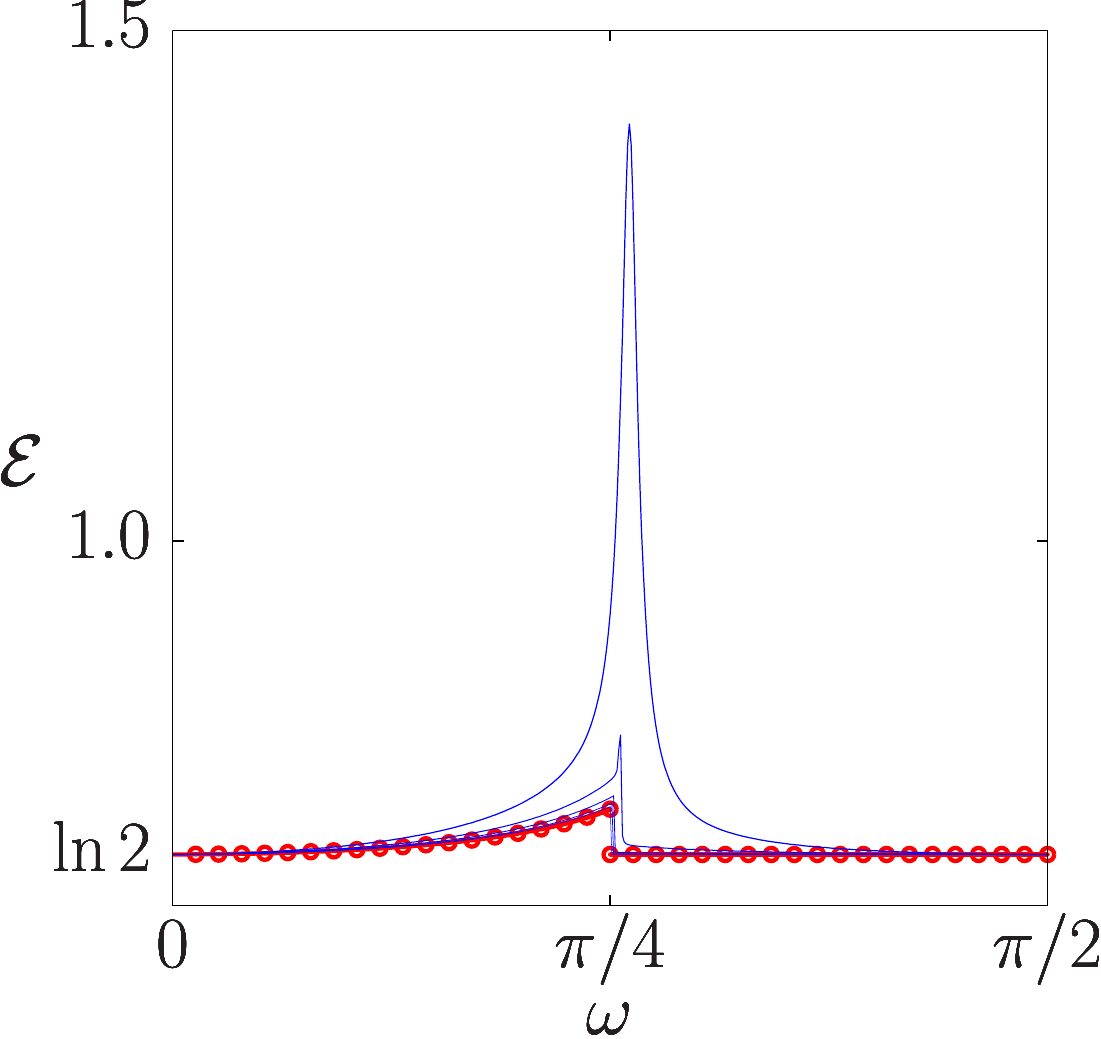}

	\caption{(Color online) Von Neumann entanglement entropy of six models
    as a function of $\omega$, for $N=16$, $32$, $64$, $128$, $256$ and in the
    thermodynamical limit (red thick line with dots). In all cases, the system is separated in two parts
    of equal sizes $N_1=N_2=N/2$.
    Left (right): $n=1$ ($n=2$). From top to bottom : $m=2$, $m=3$, and $m=4$.}
	\label{fig:entropy_vary_N}
\end{figure}
%
In the above equation, $\rho_\mathcal{A}=\mathrm{Tr}_\mathcal{B}\rho$ is the
reduced density matrix of subsystem $\mathcal{A}$ ($\rho$ is
the ground-state density matrix) and $q$ is a positive number.
In the limit $q\to 1$, one recovers the usual von Neumann entropy,
namely \mbox{$\mathcal{E}=\lim_{q\to 1}\mathcal{E}_q =
-\mathrm{Tr}[\rho_\mathcal{A}\ln\rho_\mathcal{A}]$}.
The technique for computing $\rho_\mathcal{A}$ has been exposed in
Refs.~\onlinecite{Barthel06_2, Vidal07}. To use this method one simply needs the Bogoliubov transformation which diagonalizes the Hamiltonian (\ref{eq:hamHPp}) for $p=2$ given in Appendix \ref{appendix:diag2}. In summary, for subsystems
$\mathcal{A}$ and $\mathcal{B}$ of sizes $N_\mathcal{A}=\tau N$ and
$N_\mathcal{B}=(1-\tau) N$, one has (in the thermodynamical limit and in an
appropriate basis)
%
\begin{eqnarray}
    \rho_\mathcal{A} &=& \frac{2}{\mu+1}
    \exp\left[-\ln\left(\frac{\mu+1}{\mu-1}\right)c^\dagger c\right],
    \quad\mbox{with}\\
    \label{eq:rho_A}
    \mu &=& \sqrt{[\tau+(1-\tau)/\alpha][(1-\tau)+\tau/\alpha]},
\end{eqnarray}
%
where $c$ and $c^\dagger$ are bosonic annihilation and creation operators and
where $\alpha$ has been defined in Eq.~(\ref{eq:alpha}). It is then
straightforward to compute the R\'enyi entropy
%
\begin{equation}
    \mathcal{E}_q = \frac{1}{1-q}\Big\{
    q\ln 2-\ln\big[(\mu+1)^q-(\mu-1)^q\big]\Big\},
    \label{eq:result_renyi}
\end{equation}
%
as well as the von Neumann entropy
%
\begin{equation}
    \mathcal{E} = \frac{\mu+1}{2}\ln\left(\frac{\mu+1}{2}\right)
    - \frac{\mu-1}{2}\ln\left(\frac{\mu-1}{2}\right).
    \label{eq:result_von_neumann}
\end{equation}
%
We have computed the latter numerically. As can be seen in
Fig.~\ref{fig:entropy_vary_N}, when the system size grows, the numerical
results converge to the analytical expressions obtained above (to which one
must in fact add a term $\ln 2$ when the ground state is two-fold degenerate).

One can furthermore see that similar conclusions to those for the concurrence
can be drawn here. Indeed, the von Neumann entropy of the $(2,2)$ model diverges
at the first-order transition point, like the entropy of the $(2,1)$ model but
contrary to the entropy of all other models which is finite but discontinuous
at the transition. In fact, the entropies of the $(2,2)$ model and of the
$(2,1)$ model diverge logarithmically at the transition, as $(1/2)\ln N$ and
$(1/6)\ln N$, respectively (see Refs.~\onlinecite{Latorre05_2,Barthel06_2}). So,
once again, from an entanglement perspective, the peculiar first-order
transition of the $(2,2)$ model looks like a second-order transition.

\subsection{Logarithmic negativity}
\label{sec:sub:negativity}

As a final study of the ground-state entanglement properties of our class of
models, let us compute the logarithmic negativity \cite{Vidal02}.  This
quantity, which quantifies the entanglement between any two subsystems (in a mixed or in a pure state), was
already worked out for the $(2,1)$ model \cite{Wichterich09_2} and is obtained
as follows. The system is divided into three subsystems $\mathcal{A}$,
$\mathcal{B}$ and $\mathcal{C}$ of respective sizes $N_1$, $N_2$ and $N_3$. One
then traces the ground-state density matrix over one of the subsystems, say
$\mathcal{B}$, to obtain the reduced density matrix
$\rho_{\mathcal{AC}}=\mathrm{Tr}_\mathcal{B}\rho$.
The logarithmic negativity $\mathcal{L}$ is defined as
%
\begin{equation}
    \mathcal{L}=\ln\mathrm{Tr}\left[\sqrt{
    \left(\rho_{\!\mathcal{AC}}^{\,\,\mathrm{T}_{\!\mathcal{A}}}\right)^\dagger
    \rho_{\!\mathcal{AC}}^{\,\,\mathrm{T}_{\!\mathcal{A}}}}\right],
    \label{eq:def_negativity}
\end{equation}
%
where $\mathrm{T}_{\!\mathcal{A}}$ denotes the partial transposition with respect to
subsystem $\mathcal{A}$. In a basis of states
$|\phi,\psi\rangle=|\phi\rangle_\mathcal{A}\otimes|\psi\rangle_\mathcal{C}$,
this operation reads as
$\langle\phi',\psi'|
\rho_{\!\mathcal{AC}}^{\,\,\mathrm{T}_{\!\mathcal{A}}}
|\phi,\psi\rangle = \langle\phi,\psi'|
\rho_{\!\mathcal{AC}}|\phi',\psi\rangle$.
Since $\mathcal{L}$ measures the entanglement between subsystems $\mathcal{A}$ and $\mathcal{C}$, we would obtain the same result by considering $\mathrm{T}_{\mathcal{C}}$ in Eq.~(\ref{eq:def_negativity}).

%
\begin{figure}[t]
	\includegraphics[width=0.49\columnwidth ]{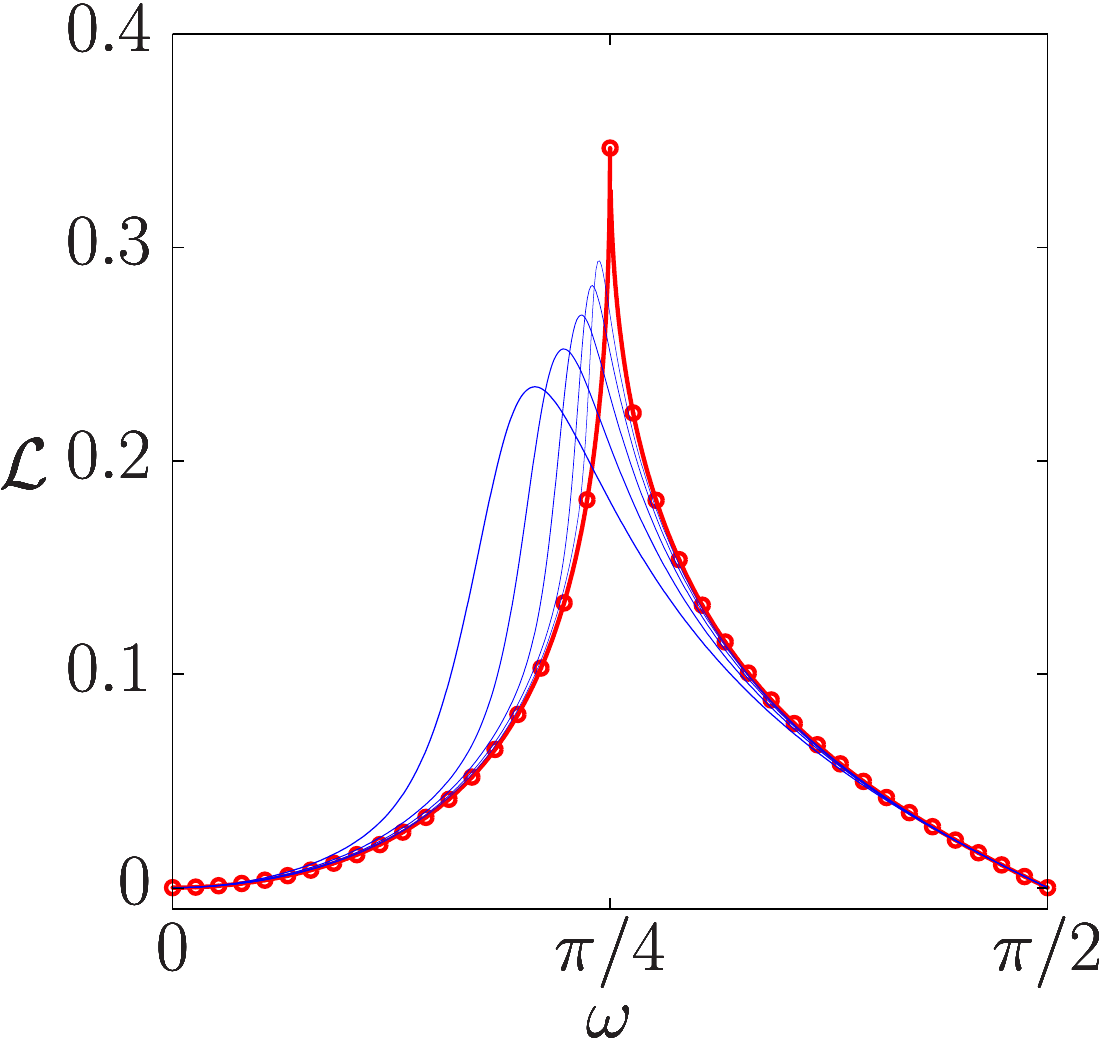}
	\includegraphics[width=0.49\columnwidth ]{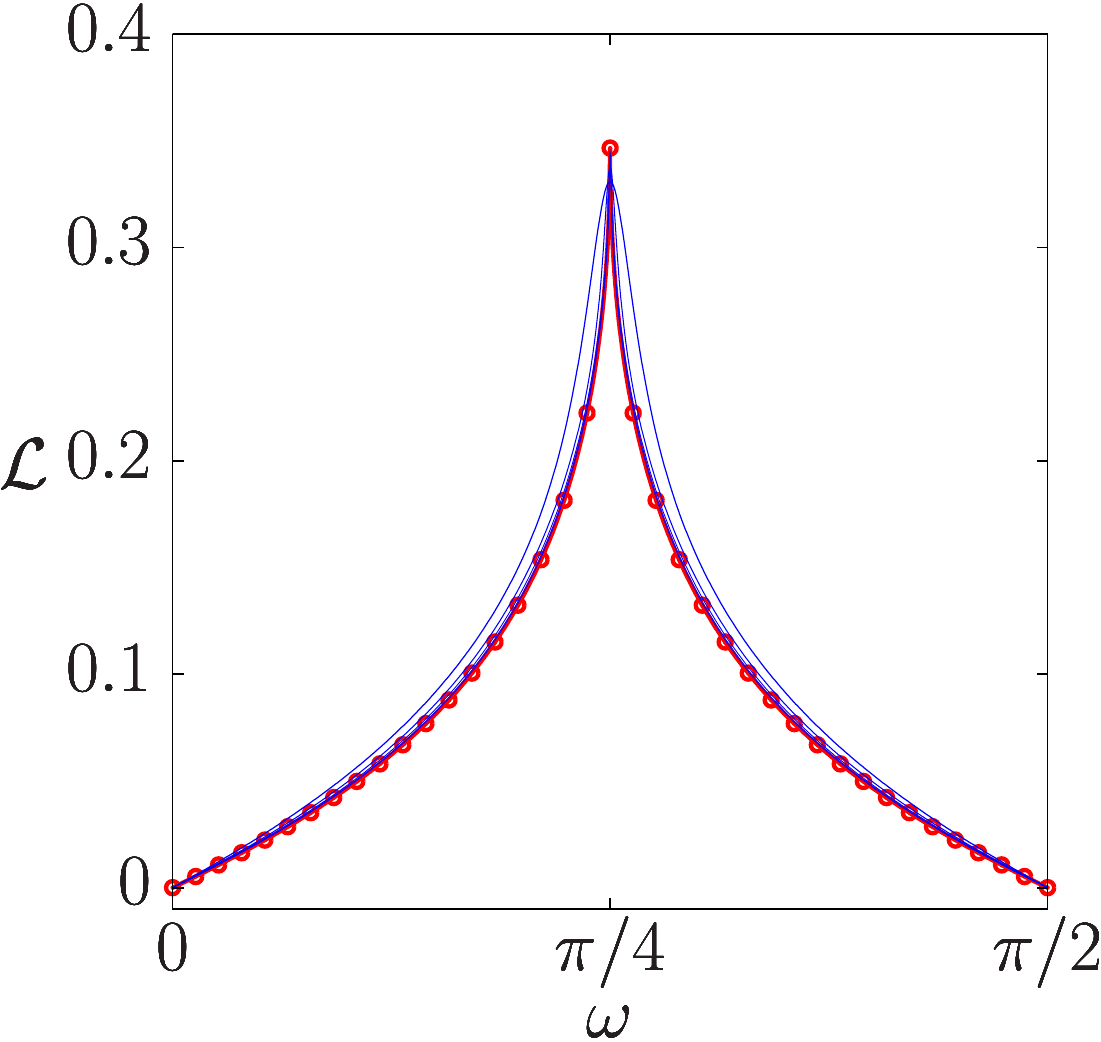}
	\vspace{0.15cm}

	\includegraphics[width=0.49\columnwidth ]{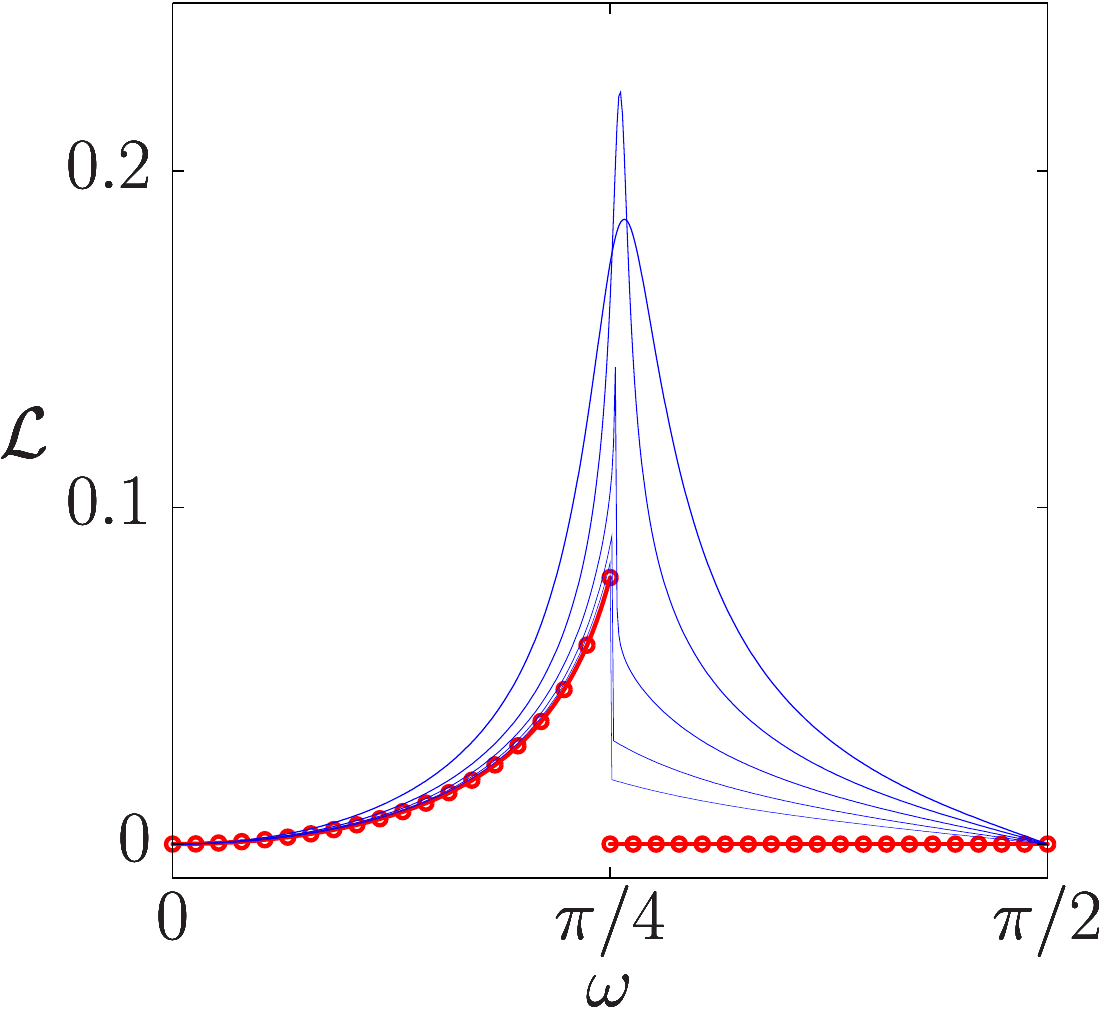}
	\includegraphics[width=0.49\columnwidth ]{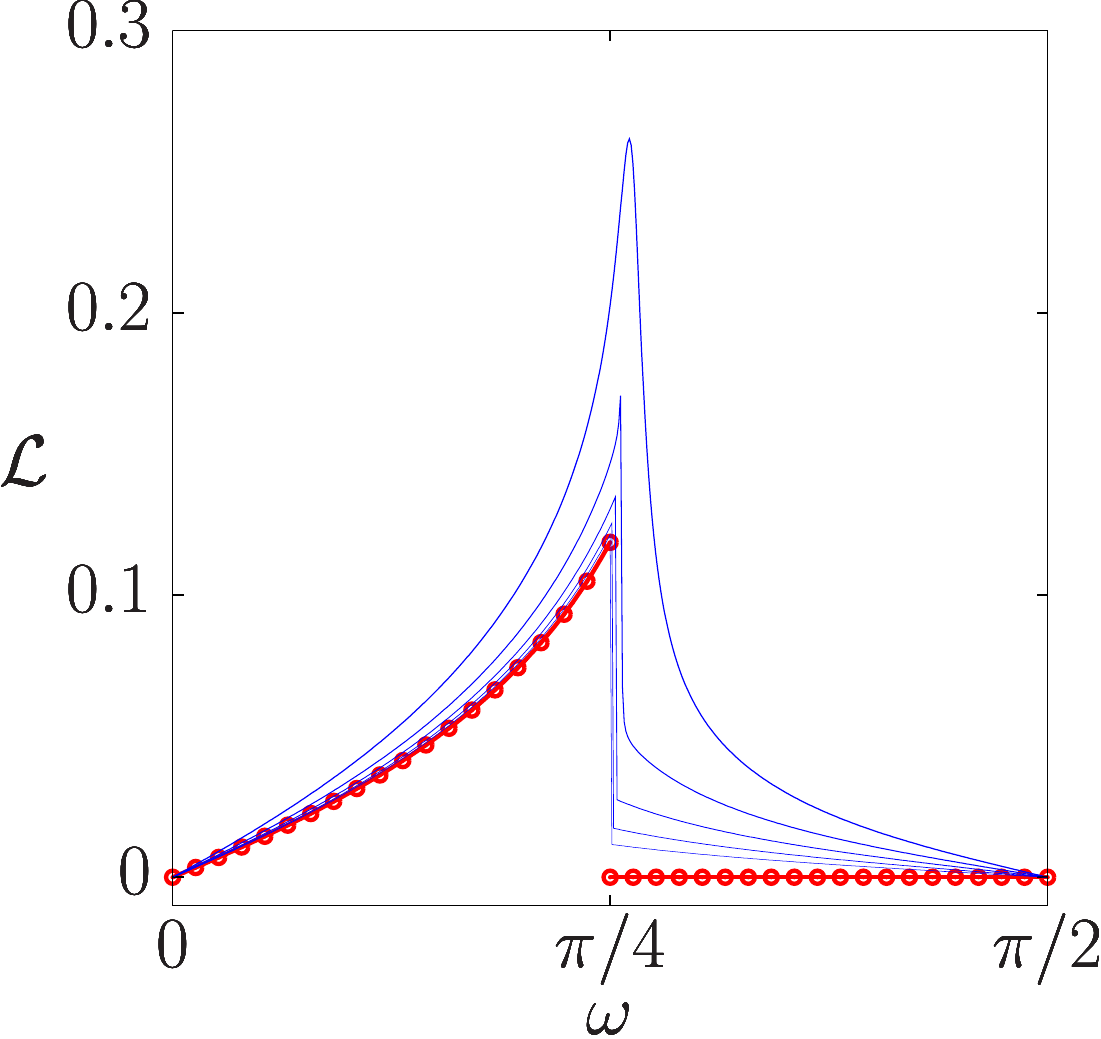}
	\vspace{0.15cm}

	\includegraphics[width=0.49\columnwidth ]{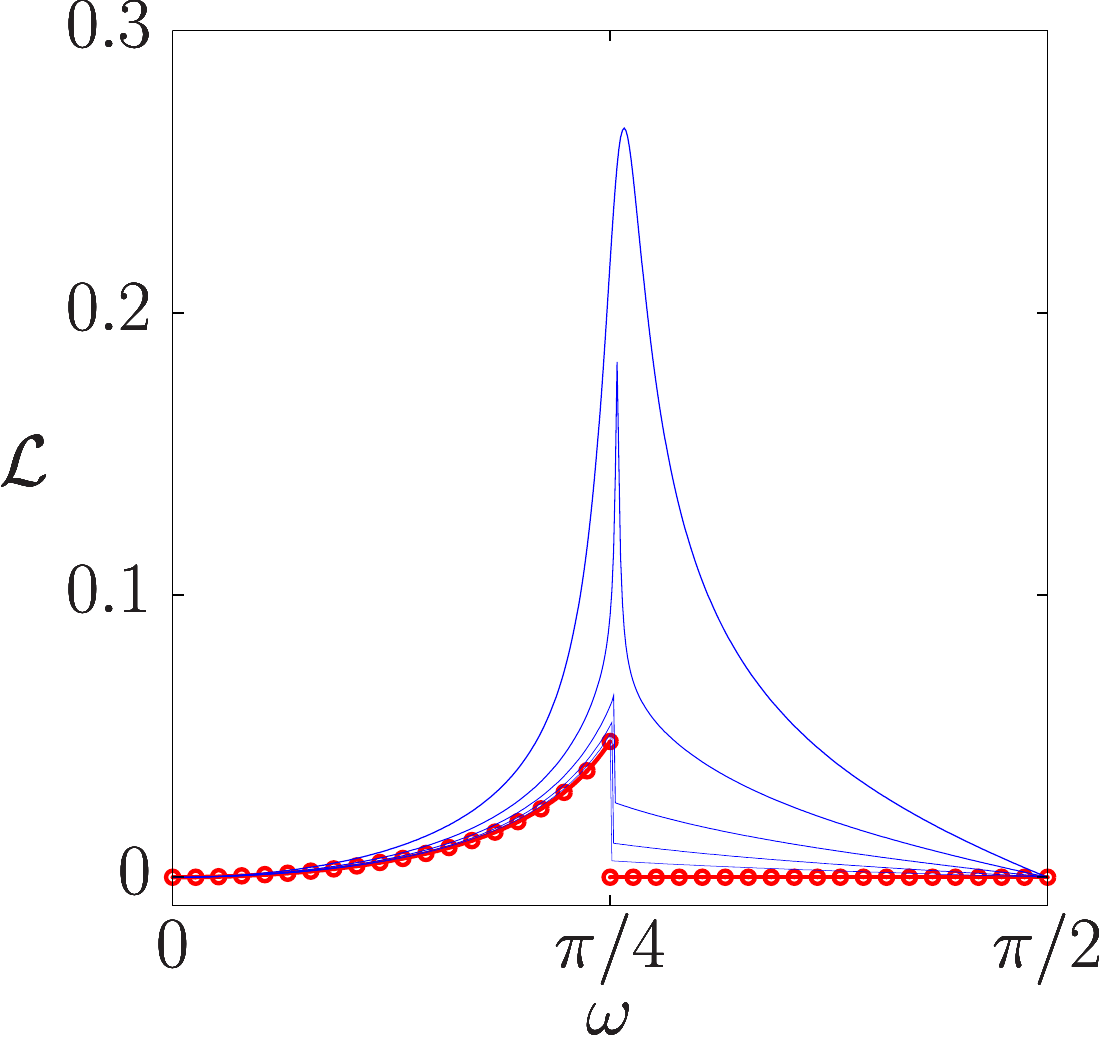}
	\includegraphics[width=0.49\columnwidth ]{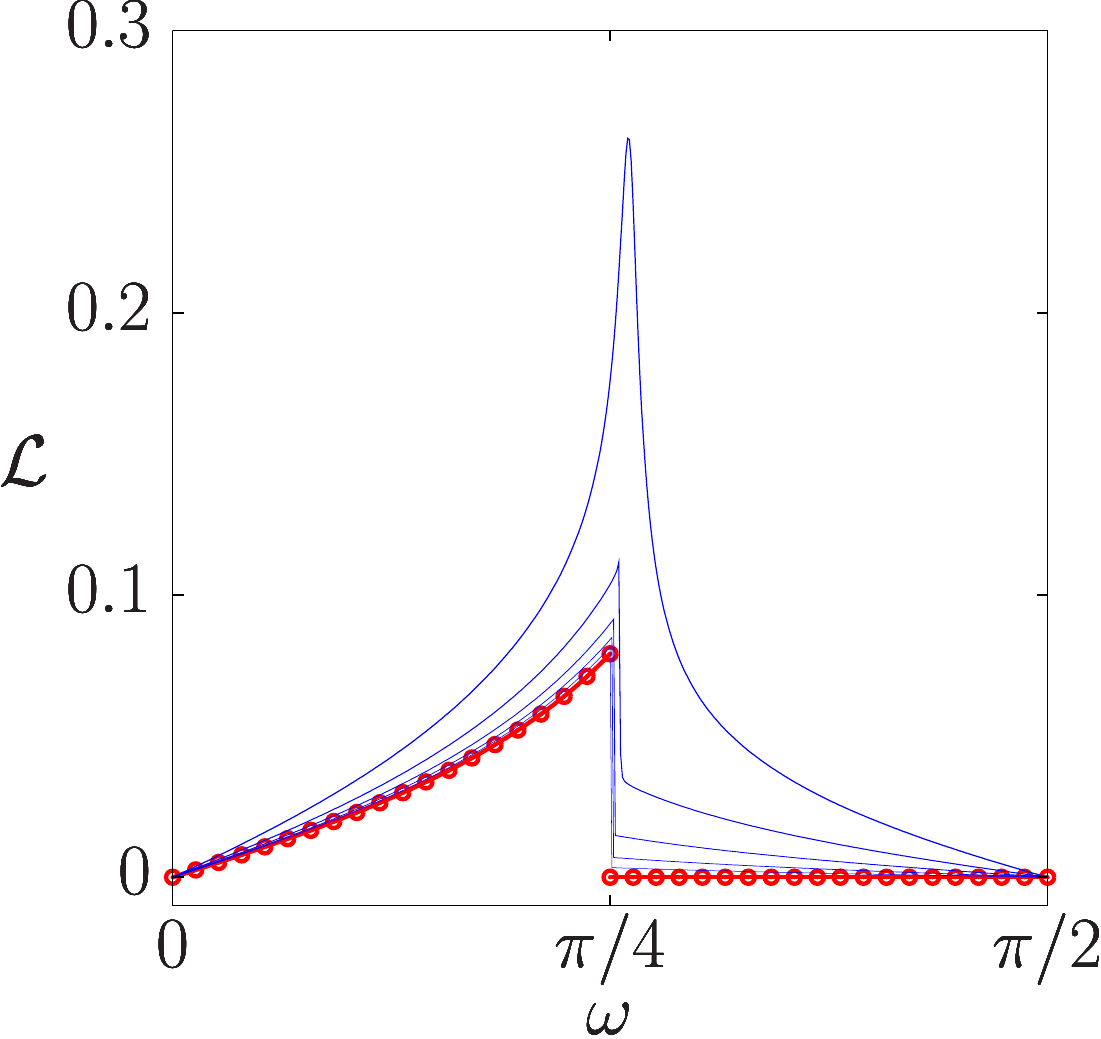}

	\caption{(Color online) Logarithmic negativity of six models
    as a function of $\omega$, for $N=16$, $32$, $64$, $128$, $256$ and in the
    thermodynamical limit (red thick line with dots). In all cases, one first traces over $N_2=N/2$ spins,
    and the logarithmic negativity is computed for the remaining spins which
    are partitioned in two subsystems of equal sizes $N_1=N_3=N/4$.
    Left (right): $n=1$ ($n=2$). From top to bottom : $m=2$, $m=3$, and $m=4$.}
	\label{fig:negativity_vary_N}
\end{figure}
%
In Ref.~\onlinecite{Wichterich09_2}, Wichterich \textit{et al.} have shown that
once the Hamiltonian is written as a three-boson Hamiltonian, that is
(\ref{eq:hamHPp}) with $p=3$, the logarithmic negativity is given by
%
\begin{equation}
    \mathcal{L} = -\frac{1}{2}\ln\left[1+g
        -\sqrt{g^2+4\tau_1\tau_3(\alpha+1/\alpha-2)}\right],
\end{equation}
with
 %
\begin{equation}
          g = \left[\tau_1+\tau_3-(\tau_1-\tau_3)^2\right]
        (\alpha+1/\alpha-2)/2,
    \label{eq:negativity}
\end{equation}
%
where $\alpha$ is given in Eq.~(\ref{eq:alpha}).
This result, which is valid in the thermodynamical limit, is plotted in
Fig.~\ref{fig:negativity_vary_N}. One can furthermore see in this figure that
the finite-size data from exact diagonalizations converge to the value
(\ref{eq:negativity}) when the system size $N$ grows.
In addition, all that was said for the behavior of the concurrence of the
various models under investigation holds again here for the logarithmic
negativity.

\section{Conclusion}
\label{sec:conclusion}

The concurrence, the entanglement entropy, and the logarithmic negativity, although
different entanglement measures, show similar features when used to
characterize the quantum phase transitions of the class of collective models we
have introduced in this paper. However, when the transition is of
first-order nature but accompanied by a collapse of levels on the ground state [see model $(2, 2)$],
as is usually characteristic of second-order transitions, the entanglement of
the ground state does not show any discontinuity at the transition, but
behaves exactly as in a usual second-order transition.
In such a situation, one may wonder whether other ``intrinsic measures"  would be more sensitive to this discontinuous transition. One may think about studying the fidelity that has already been analyzed for the (2,1) (Lipkin-Meshkov-Glick) model at zero \cite{Kwok08,Ma08,Ma09,Gu10} and at finite temperature \cite{Quan09,Scherer09} or to the geometric entanglement, which is also known for the (2,1) case \cite{Orus08_2}. 
We have computed these quantities for the (2,2) model and found that their behavior is similar in the (2,1) and the (2,2) cases although, as already underlined, finite-size scalings are different. 
We wish to underline that it this not an isolated case since, for all $(m,m)$ with $m\geqslant2$, one has a first-order transition and the symmetry of the Hamiltonian under the exchange $(x\leftrightarrow z)$ implies that entanglement measures must be continuous. However, as can be checked from the exact formulas given in this paper, there is no entanglement for $(m,n)$ models when $m\geqslant n\geqslant 3$. Thus, from this perpspective the $(2,2)$ model is a bit singular.

To conclude, let us emphasize that we focused here on the ground-state entanglement. Nevertheless, it would be worth considering the full spectrum of these models to investigate finite-temperature entanglement, which may unveil interesting properties \cite{Canosa07,Matera08}. This is beyond the scope of this paper but it will be the topic of a forthcoming publication \cite{Wilms11}.

\appendix

\section{Diagonalization of the two-mode Hamiltonian (\ref{eq:hamHPp})}
\label{appendix:diag2}

To compute the entanglement entropy, one needs to diagonalize the Hamiltonian (\ref{eq:hamHPp}) for $p=2$. This is done by performing the following Bogoliubov transformation~:
%
\begin{eqnarray}
    a_1 &=& \big[\cosh(\Theta/2) b_1 + \sinh(\Theta/2) b_1^\dagger \big]\sqrt{\tau_1}+b_2 \sqrt{\tau_2}, \qquad\\
    a_2 &=& \big[\cosh(\Theta/2) b_1 + \sinh(\Theta/2) b_1^\dagger \big] \sqrt{\tau_2}-b_2 \sqrt{\tau_1}, \qquad
\end{eqnarray}
%
\mbox{}\\
where  $\tanh\Theta=\varepsilon=-2\gamma/\delta$. New bosonic operators mutually commutes and satisfy $\left[b_1,b_1^\dagger\right]=\left[b_2,b_2^\dagger\right]=1$.

Inserting these relations in Eq.~(\ref{eq:hamHPp}) for $p=2$, one gets~:
%
\begin{equation}
H=N e_0 + \gamma + \frac{\delta}{2}\left(\sqrt{1-\varepsilon^2}-1\right)+\Delta_1 b_1^\dagger b_1 +\Delta_2 b_2^\dagger b_2,\\
\end{equation}
%
where $\Delta_1=\delta\sqrt{1-\varepsilon^2}$ and $\Delta_2=\delta$.


\begin{thebibliography}{46}
\expandafter\ifx\csname natexlab\endcsname\relax\def\natexlab#1{#1}\fi
\expandafter\ifx\csname bibnamefont\endcsname\relax
  \def\bibnamefont#1{#1}\fi
\expandafter\ifx\csname bibfnamefont\endcsname\relax
  \def\bibfnamefont#1{#1}\fi
\expandafter\ifx\csname citenamefont\endcsname\relax
  \def\citenamefont#1{#1}\fi
\expandafter\ifx\csname url\endcsname\relax
  \def\url#1{\texttt{#1}}\fi
\expandafter\ifx\csname urlprefix\endcsname\relax\def\urlprefix{URL }\fi
\providecommand{\bibinfo}[2]{#2}
\providecommand{\eprint}[2][]{\url{#2}}

\bibitem[{\citenamefont{Amico et~al.}(2008)\citenamefont{Amico, Fazio,
  Osterloh, and Vedral}}]{Amico08}
\bibinfo{author}{\bibfnamefont{L.}~\bibnamefont{Amico}},
  \bibinfo{author}{\bibfnamefont{R.}~\bibnamefont{Fazio}},
  \bibinfo{author}{\bibfnamefont{A.}~\bibnamefont{Osterloh}}, \bibnamefont{and}
  \bibinfo{author}{\bibfnamefont{V.}~\bibnamefont{Vedral}},
  \bibinfo{journal}{Rev. Mod. Phys.} \textbf{\bibinfo{volume}{80}},
  \bibinfo{pages}{517} (\bibinfo{year}{2008}).

\bibitem[{\citenamefont{Osborne and Nielsen}(2002)}]{Osborne02}
\bibinfo{author}{\bibfnamefont{T.~J.} \bibnamefont{Osborne}} \bibnamefont{and}
  \bibinfo{author}{\bibfnamefont{M.~A.} \bibnamefont{Nielsen}},
  \bibinfo{journal}{Phys. Rev. A} \textbf{\bibinfo{volume}{66}},
  \bibinfo{pages}{032110} (\bibinfo{year}{2002}).

\bibitem[{\citenamefont{Osterloh et~al.}(2002)\citenamefont{Osterloh, Amico,
  Falci, and Fazio}}]{Osterloh02}
\bibinfo{author}{\bibfnamefont{A.}~\bibnamefont{Osterloh}},
  \bibinfo{author}{\bibfnamefont{L.}~\bibnamefont{Amico}},
  \bibinfo{author}{\bibfnamefont{G.}~\bibnamefont{Falci}}, \bibnamefont{and}
  \bibinfo{author}{\bibfnamefont{R.}~\bibnamefont{Fazio}},
  \bibinfo{journal}{Nature (London)} \textbf{\bibinfo{volume}{416}},
  \bibinfo{pages}{608} (\bibinfo{year}{2002}).

\bibitem[{\citenamefont{Vidal et~al.}(2003)\citenamefont{Vidal, Latorre, Rico,
  and Kitaev}}]{Vidal03_1}
\bibinfo{author}{\bibfnamefont{G.}~\bibnamefont{Vidal}},
  \bibinfo{author}{\bibfnamefont{J.~I.} \bibnamefont{Latorre}},
  \bibinfo{author}{\bibfnamefont{E.}~\bibnamefont{Rico}}, \bibnamefont{and}
  \bibinfo{author}{\bibfnamefont{A.}~\bibnamefont{Kitaev}},
  \bibinfo{journal}{Phys. Rev. Lett.} \textbf{\bibinfo{volume}{90}},
  \bibinfo{pages}{227902} (\bibinfo{year}{2003}).

\bibitem[{\citenamefont{Lipkin et~al.}(1965)\citenamefont{Lipkin, Meshkov, and
  Glick}}]{Lipkin65}
\bibinfo{author}{\bibfnamefont{H.~J.} \bibnamefont{Lipkin}},
  \bibinfo{author}{\bibfnamefont{N.}~\bibnamefont{Meshkov}}, \bibnamefont{and}
  \bibinfo{author}{\bibfnamefont{A.~J.} \bibnamefont{Glick}},
  \bibinfo{journal}{Nucl. Phys.} \textbf{\bibinfo{volume}{62}},
  \bibinfo{pages}{188} (\bibinfo{year}{1965}).

\bibitem[{\citenamefont{Meshkov et~al.}(1965)\citenamefont{Meshkov, Glick, and
  Lipkin}}]{Meshkov65}
\bibinfo{author}{\bibfnamefont{N.}~\bibnamefont{Meshkov}},
  \bibinfo{author}{\bibfnamefont{A.~J.} \bibnamefont{Glick}}, \bibnamefont{and}
  \bibinfo{author}{\bibfnamefont{H.~J.} \bibnamefont{Lipkin}},
  \bibinfo{journal}{Nucl. Phys.} \textbf{\bibinfo{volume}{62}},
  \bibinfo{pages}{199} (\bibinfo{year}{1965}).

\bibitem[{\citenamefont{Glick et~al.}(1965)\citenamefont{Glick, Lipkin, and
  Meshkov}}]{Glick65}
\bibinfo{author}{\bibfnamefont{A.~J.} \bibnamefont{Glick}},
  \bibinfo{author}{\bibfnamefont{H.~J.} \bibnamefont{Lipkin}},
  \bibnamefont{and} \bibinfo{author}{\bibfnamefont{N.}~\bibnamefont{Meshkov}},
  \bibinfo{journal}{Nucl. Phys.} \textbf{\bibinfo{volume}{62}},
  \bibinfo{pages}{211} (\bibinfo{year}{1965}).

\bibitem[{\citenamefont{Vidal et~al.}(2004{\natexlab{a}})\citenamefont{Vidal,
  Palacios, and Mosseri}}]{Vidal04_1}
\bibinfo{author}{\bibfnamefont{J.}~\bibnamefont{Vidal}},
  \bibinfo{author}{\bibfnamefont{G.}~\bibnamefont{Palacios}}, \bibnamefont{and}
  \bibinfo{author}{\bibfnamefont{R.}~\bibnamefont{Mosseri}},
  \bibinfo{journal}{Phys. Rev. A} \textbf{\bibinfo{volume}{69}},
  \bibinfo{pages}{022107} (\bibinfo{year}{2004}{\natexlab{a}}).

\bibitem[{\citenamefont{Vidal et~al.}(2004{\natexlab{b}})\citenamefont{Vidal,
  Mosseri, and Dukelsky}}]{Vidal04_2}
\bibinfo{author}{\bibfnamefont{J.}~\bibnamefont{Vidal}},
  \bibinfo{author}{\bibfnamefont{R.}~\bibnamefont{Mosseri}}, \bibnamefont{and}
  \bibinfo{author}{\bibfnamefont{J.}~\bibnamefont{Dukelsky}},
  \bibinfo{journal}{Phys. Rev. A} \textbf{\bibinfo{volume}{69}},
  \bibinfo{pages}{054101} (\bibinfo{year}{2004}{\natexlab{b}}).

\bibitem[{\citenamefont{Vidal et~al.}(2004{\natexlab{c}})\citenamefont{Vidal,
  Palacios, and Aslangul}}]{Vidal04_3}
\bibinfo{author}{\bibfnamefont{J.}~\bibnamefont{Vidal}},
  \bibinfo{author}{\bibfnamefont{G.}~\bibnamefont{Palacios}}, \bibnamefont{and}
  \bibinfo{author}{\bibfnamefont{C.}~\bibnamefont{Aslangul}},
  \bibinfo{journal}{Phys. Rev. A} \textbf{\bibinfo{volume}{70}},
  \bibinfo{pages}{062304} (\bibinfo{year}{2004}{\natexlab{c}}).

\bibitem[{\citenamefont{Dusuel and Vidal}(2004)}]{Dusuel04_3}
\bibinfo{author}{\bibfnamefont{S.}~\bibnamefont{Dusuel}} \bibnamefont{and}
  \bibinfo{author}{\bibfnamefont{J.}~\bibnamefont{Vidal}},
  \bibinfo{journal}{Phys. Rev. Lett.} \textbf{\bibinfo{volume}{93}},
  \bibinfo{pages}{237204} (\bibinfo{year}{2004}).

\bibitem[{\citenamefont{Latorre et~al.}(2005)\citenamefont{Latorre, Or\'us,
  Rico, and Vidal}}]{Latorre05_2}
\bibinfo{author}{\bibfnamefont{J.~I.} \bibnamefont{Latorre}},
  \bibinfo{author}{\bibfnamefont{R.}~\bibnamefont{Or\'us}},
  \bibinfo{author}{\bibfnamefont{E.}~\bibnamefont{Rico}}, \bibnamefont{and}
  \bibinfo{author}{\bibfnamefont{J.}~\bibnamefont{Vidal}},
  \bibinfo{journal}{Phys. Rev. A} \textbf{\bibinfo{volume}{71}},
  \bibinfo{pages}{064101} (\bibinfo{year}{2005}).

\bibitem[{\citenamefont{Dusuel and Vidal}(2005)}]{Dusuel05_2}
\bibinfo{author}{\bibfnamefont{S.}~\bibnamefont{Dusuel}} \bibnamefont{and}
  \bibinfo{author}{\bibfnamefont{J.}~\bibnamefont{Vidal}},
  \bibinfo{journal}{Phys. Rev. B} \textbf{\bibinfo{volume}{71}},
  \bibinfo{pages}{224420} (\bibinfo{year}{2005}).

\bibitem[{\citenamefont{Unanyan et~al.}(2005)\citenamefont{Unanyan, Ionescu,
  and Fleischhauer}}]{Unanyan05_1}
\bibinfo{author}{\bibfnamefont{R.~G.} \bibnamefont{Unanyan}},
  \bibinfo{author}{\bibfnamefont{C.}~\bibnamefont{Ionescu}}, \bibnamefont{and}
  \bibinfo{author}{\bibfnamefont{M.}~\bibnamefont{Fleischhauer}},
  \bibinfo{journal}{Phys. Rev. A} \textbf{\bibinfo{volume}{72}},
  \bibinfo{pages}{022326} (\bibinfo{year}{2005}).

\bibitem[{\citenamefont{Barthel et~al.}(2006)\citenamefont{Barthel, Dusuel, and
  Vidal}}]{Barthel06_2}
\bibinfo{author}{\bibfnamefont{T.}~\bibnamefont{Barthel}},
  \bibinfo{author}{\bibfnamefont{S.}~\bibnamefont{Dusuel}}, \bibnamefont{and}
  \bibinfo{author}{\bibfnamefont{J.}~\bibnamefont{Vidal}},
  \bibinfo{journal}{Phys. Rev. Lett.} \textbf{\bibinfo{volume}{97}},
  \bibinfo{pages}{220402} (\bibinfo{year}{2006}).

\bibitem[{\citenamefont{Vidal}(2006)}]{Vidal06_3}
\bibinfo{author}{\bibfnamefont{J.}~\bibnamefont{Vidal}},
  \bibinfo{journal}{Phys. Rev. A} \textbf{\bibinfo{volume}{73}},
  \bibinfo{pages}{062318} (\bibinfo{year}{2006}).

\bibitem[{\citenamefont{Vidal et~al.}(2007)\citenamefont{Vidal, Dusuel, and
  Barthel}}]{Vidal07}
\bibinfo{author}{\bibfnamefont{J.}~\bibnamefont{Vidal}},
  \bibinfo{author}{\bibfnamefont{S.}~\bibnamefont{Dusuel}}, \bibnamefont{and}
  \bibinfo{author}{\bibfnamefont{T.}~\bibnamefont{Barthel}},
  \bibinfo{journal}{J. Stat. Mech.: Theory Exp.}
  \textbf{\bibinfo{volume}{P01015}} (\bibinfo{year}{2007}).

\bibitem[{\citenamefont{Morrison and Parkins}(2008)}]{Morrison08_2}
\bibinfo{author}{\bibfnamefont{S.}~\bibnamefont{Morrison}} \bibnamefont{and}
  \bibinfo{author}{\bibfnamefont{A.~S.} \bibnamefont{Parkins}},
  \bibinfo{journal}{Phys. Rev. A} \textbf{\bibinfo{volume}{77}},
  \bibinfo{pages}{043810} (\bibinfo{year}{2008}).

\bibitem[{\citenamefont{Cui}(2008)}]{Cui08}
\bibinfo{author}{\bibfnamefont{H.~T.} \bibnamefont{Cui}},
  \bibinfo{journal}{Phys. Rev. A} \textbf{\bibinfo{volume}{77}},
  \bibinfo{pages}{052105} (\bibinfo{year}{2008}).

\bibitem[{\citenamefont{Or\'us et~al.}(2008)\citenamefont{Or\'us, Dusuel, and
  Vidal}}]{Orus08_2}
\bibinfo{author}{\bibfnamefont{R.}~\bibnamefont{Or\'us}},
  \bibinfo{author}{\bibfnamefont{S.}~\bibnamefont{Dusuel}}, \bibnamefont{and}
  \bibinfo{author}{\bibfnamefont{J.}~\bibnamefont{Vidal}},
  \bibinfo{journal}{Phys. Rev. Lett.} \textbf{\bibinfo{volume}{101}},
  \bibinfo{pages}{025701} (\bibinfo{year}{2008}).

\bibitem[{\citenamefont{Caneva et~al.}(2008)\citenamefont{Caneva, Fazio, and
  Santoro}}]{Caneva08}
\bibinfo{author}{\bibfnamefont{T.}~\bibnamefont{Caneva}},
  \bibinfo{author}{\bibfnamefont{R.}~\bibnamefont{Fazio}}, \bibnamefont{and}
  \bibinfo{author}{\bibfnamefont{G.~E.} \bibnamefont{Santoro}},
  \bibinfo{journal}{Phys. Rev. B} \textbf{\bibinfo{volume}{78}},
  \bibinfo{pages}{104426} (\bibinfo{year}{2008}).

\bibitem[{\citenamefont{Ma et~al.}(2009)\citenamefont{Ma, Wang, and Gu}}]{Ma09}
\bibinfo{author}{\bibfnamefont{J.}~\bibnamefont{Ma}},
  \bibinfo{author}{\bibfnamefont{X.}~\bibnamefont{Wang}}, \bibnamefont{and}
  \bibinfo{author}{\bibfnamefont{S.-J.} \bibnamefont{Gu}},
  \bibinfo{journal}{Phys. Rev. E} \textbf{\bibinfo{volume}{80}},
  \bibinfo{pages}{021124} (\bibinfo{year}{2009}).

\bibitem[{\citenamefont{Wichterich et~al.}(2010)\citenamefont{Wichterich,
  Vidal, and Bose}}]{Wichterich09_2}
\bibinfo{author}{\bibfnamefont{H.}~\bibnamefont{Wichterich}},
  \bibinfo{author}{\bibfnamefont{J.}~\bibnamefont{Vidal}}, \bibnamefont{and}
  \bibinfo{author}{\bibfnamefont{S.}~\bibnamefont{Bose}},
  \bibinfo{journal}{Phys. Rev. A} \textbf{\bibinfo{volume}{81}},
  \bibinfo{pages}{032311} (\bibinfo{year}{2010}).

\bibitem[{\citenamefont{Wootters}(1998)}]{Wootters98}
\bibinfo{author}{\bibfnamefont{W.~K.} \bibnamefont{Wootters}},
  \bibinfo{journal}{Phys. Rev. Lett.} \textbf{\bibinfo{volume}{80}},
  \bibinfo{pages}{2245} (\bibinfo{year}{1998}).

\bibitem[{\citenamefont{Vidal and Werner}(2002)}]{Vidal02}
\bibinfo{author}{\bibfnamefont{G.}~\bibnamefont{Vidal}} \bibnamefont{and}
  \bibinfo{author}{\bibfnamefont{R.~F.} \bibnamefont{Werner}},
  \bibinfo{journal}{Phys. Rev. A} \textbf{\bibinfo{volume}{65}},
  \bibinfo{pages}{032314} (\bibinfo{year}{2002}).

\bibitem[{\citenamefont{Kugel and Khomskii}(1982)}]{Kugel82}
\bibinfo{author}{\bibfnamefont{K.~I.} \bibnamefont{Kugel}} \bibnamefont{and}
  \bibinfo{author}{\bibfnamefont{D.~I.} \bibnamefont{Khomskii}},
  \bibinfo{journal}{Sov. Phys. Usp.} \textbf{\bibinfo{volume}{25}},
  \bibinfo{pages}{231} (\bibinfo{year}{1982}).

\bibitem[{\citenamefont{Xu and Moore}(2004)}]{Xu04}
\bibinfo{author}{\bibfnamefont{C.}~\bibnamefont{Xu}} \bibnamefont{and}
  \bibinfo{author}{\bibfnamefont{J.~E.} \bibnamefont{Moore}},
  \bibinfo{journal}{Phys. Rev. Lett.} \textbf{\bibinfo{volume}{93}},
  \bibinfo{pages}{047003} (\bibinfo{year}{2004}).

\bibitem[{\citenamefont{Xu and Moore}(2005)}]{Xu05}
\bibinfo{author}{\bibfnamefont{C.}~\bibnamefont{Xu}} \bibnamefont{and}
  \bibinfo{author}{\bibfnamefont{J.~E.} \bibnamefont{Moore}},
  \bibinfo{journal}{Nucl. Phys. B} \textbf{\bibinfo{volume}{716}},
  \bibinfo{pages}{487} (\bibinfo{year}{2005}).

\bibitem[{\citenamefont{Nussinov and Fradkin}(2005)}]{Nussinov05_1}
\bibinfo{author}{\bibfnamefont{Z.}~\bibnamefont{Nussinov}} \bibnamefont{and}
  \bibinfo{author}{\bibfnamefont{E.}~\bibnamefont{Fradkin}},
  \bibinfo{journal}{Phys. Rev. B} \textbf{\bibinfo{volume}{71}},
  \bibinfo{pages}{195120} (\bibinfo{year}{2005}).

\bibitem[{\citenamefont{Maritan et~al.}(1984)\citenamefont{Maritan, Stella, and
  Vanderzande}}]{Maritan84}
\bibinfo{author}{\bibfnamefont{A.}~\bibnamefont{Maritan}},
  \bibinfo{author}{\bibfnamefont{A.}~\bibnamefont{Stella}}, \bibnamefont{and}
  \bibinfo{author}{\bibfnamefont{C.}~\bibnamefont{Vanderzande}},
  \bibinfo{journal}{Phys. Rev. B} \textbf{\bibinfo{volume}{29}},
  \bibinfo{pages}{519} (\bibinfo{year}{1984}).

\bibitem[{\citenamefont{Ribeiro et~al.}(2007)\citenamefont{Ribeiro, Vidal, and
  Mosseri}}]{Ribeiro07}
\bibinfo{author}{\bibfnamefont{P.}~\bibnamefont{Ribeiro}},
  \bibinfo{author}{\bibfnamefont{J.}~\bibnamefont{Vidal}}, \bibnamefont{and}
  \bibinfo{author}{\bibfnamefont{R.}~\bibnamefont{Mosseri}},
  \bibinfo{journal}{Phys. Rev. Lett.} \textbf{\bibinfo{volume}{99}},
  \bibinfo{pages}{050402} (\bibinfo{year}{2007}).

\bibitem[{\citenamefont{Ribeiro et~al.}(2008)\citenamefont{Ribeiro, Vidal, and
  Mosseri}}]{Ribeiro08}
\bibinfo{author}{\bibfnamefont{P.}~\bibnamefont{Ribeiro}},
  \bibinfo{author}{\bibfnamefont{J.}~\bibnamefont{Vidal}}, \bibnamefont{and}
  \bibinfo{author}{\bibfnamefont{R.}~\bibnamefont{Mosseri}},
  \bibinfo{journal}{Phys. Rev. E} \textbf{\bibinfo{volume}{78}},
  \bibinfo{pages}{021106} (\bibinfo{year}{2008}).

\bibitem[{\citenamefont{Pan and Draayer}(1999)}]{Pan99}
\bibinfo{author}{\bibfnamefont{F.}~\bibnamefont{Pan}} \bibnamefont{and}
  \bibinfo{author}{\bibfnamefont{J.~P.} \bibnamefont{Draayer}},
  \bibinfo{journal}{Phys. Lett. B} \textbf{\bibinfo{volume}{451}},
  \bibinfo{pages}{1} (\bibinfo{year}{1999}).

\bibitem[{\citenamefont{Links et~al.}(2003)\citenamefont{Links, Zhou, McKenzie,
  and Gould}}]{Links03}
\bibinfo{author}{\bibfnamefont{J.}~\bibnamefont{Links}},
  \bibinfo{author}{\bibfnamefont{H.-Q.} \bibnamefont{Zhou}},
  \bibinfo{author}{\bibfnamefont{R.~H.} \bibnamefont{McKenzie}},
  \bibnamefont{and} \bibinfo{author}{\bibfnamefont{M.~D.} \bibnamefont{Gould}},
  \bibinfo{journal}{J. Phys. A} \textbf{\bibinfo{volume}{36}},
  \bibinfo{pages}{R63} (\bibinfo{year}{2003}).

\bibitem[{\citenamefont{Ortiz et~al.}(2005)\citenamefont{Ortiz, Somma,
  Dukelsky, and Rombouts}}]{Ortiz05}
\bibinfo{author}{\bibfnamefont{G.}~\bibnamefont{Ortiz}},
  \bibinfo{author}{\bibfnamefont{R.}~\bibnamefont{Somma}},
  \bibinfo{author}{\bibfnamefont{J.}~\bibnamefont{Dukelsky}}, \bibnamefont{and}
  \bibinfo{author}{\bibfnamefont{S.}~\bibnamefont{Rombouts}},
  \bibinfo{journal}{Nucl. Phys. B} \textbf{\bibinfo{volume}{707}},
  \bibinfo{pages}{421} (\bibinfo{year}{2005}).

\bibitem[{\citenamefont{Botet and Jullien}(1983)}]{Botet83}
\bibinfo{author}{\bibfnamefont{R.}~\bibnamefont{Botet}} \bibnamefont{and}
  \bibinfo{author}{\bibfnamefont{R.}~\bibnamefont{Jullien}},
  \bibinfo{journal}{Phys. Rev. B} \textbf{\bibinfo{volume}{28}},
  \bibinfo{pages}{3955} (\bibinfo{year}{1983}).

\bibitem[{\citenamefont{Holstein and Primakoff}(1940)}]{Holstein40}
\bibinfo{author}{\bibfnamefont{T.}~\bibnamefont{Holstein}} \bibnamefont{and}
  \bibinfo{author}{\bibfnamefont{H.}~\bibnamefont{Primakoff}},
  \bibinfo{journal}{Phys. Rev.} \textbf{\bibinfo{volume}{58}},
  \bibinfo{pages}{1098} (\bibinfo{year}{1940}).

\bibitem[{\citenamefont{Wang and M{\o}lmer}(2002)}]{Wang02}
\bibinfo{author}{\bibfnamefont{X.}~\bibnamefont{Wang}} \bibnamefont{and}
  \bibinfo{author}{\bibfnamefont{K.}~\bibnamefont{M{\o}lmer}},
  \bibinfo{journal}{Eur. Phys. J. D} \textbf{\bibinfo{volume}{18}},
  \bibinfo{pages}{385} (\bibinfo{year}{2002}).

\bibitem[{\citenamefont{Kwok et~al.}(2008)\citenamefont{Kwok, Ning, Gu, and
  Lin}}]{Kwok08}
\bibinfo{author}{\bibfnamefont{H.-M.} \bibnamefont{Kwok}},
  \bibinfo{author}{\bibfnamefont{W.-Q.} \bibnamefont{Ning}},
  \bibinfo{author}{\bibfnamefont{S.-J.} \bibnamefont{Gu}}, \bibnamefont{and}
  \bibinfo{author}{\bibfnamefont{H.-Q.} \bibnamefont{Lin}},
  \bibinfo{journal}{Phys. Rev. E} \textbf{\bibinfo{volume}{78}},
  \bibinfo{pages}{032103} (\bibinfo{year}{2008}).

\bibitem[{\citenamefont{Ma et~al.}(2008)\citenamefont{Ma, Xu, Xiong, and
  Wang}}]{Ma08}
\bibinfo{author}{\bibfnamefont{J.}~\bibnamefont{Ma}},
  \bibinfo{author}{\bibfnamefont{L.}~\bibnamefont{Xu}},
  \bibinfo{author}{\bibfnamefont{H.-N.} \bibnamefont{Xiong}}, \bibnamefont{and}
  \bibinfo{author}{\bibfnamefont{X.}~\bibnamefont{Wang}},
  \bibinfo{journal}{Phys. Rev. E} \textbf{\bibinfo{volume}{78}},
  \bibinfo{pages}{051126} (\bibinfo{year}{2008}).

\bibitem[{\citenamefont{Gu}(2010)}]{Gu10}
\bibinfo{author}{\bibfnamefont{S.-J.} \bibnamefont{Gu}}, \bibinfo{journal}{Int.
  J. Mod. Phys. B} \textbf{\bibinfo{volume}{24}}, \bibinfo{pages}{4371}
  (\bibinfo{year}{2010}).

\bibitem[{\citenamefont{Quan and Cucchietti}(2009)}]{Quan09}
\bibinfo{author}{\bibfnamefont{H.~T.} \bibnamefont{Quan}} \bibnamefont{and}
  \bibinfo{author}{\bibfnamefont{F.~M.} \bibnamefont{Cucchietti}},
  \bibinfo{journal}{Phys. Rev. E} \textbf{\bibinfo{volume}{79}},
  \bibinfo{pages}{031101} (\bibinfo{year}{2009}).

\bibitem[{\citenamefont{Scherer et~al.}(2009)\citenamefont{Scherer, M\"{u}ller,
  and Kastner}}]{Scherer09}
\bibinfo{author}{\bibfnamefont{D.~D.} \bibnamefont{Scherer}},
  \bibinfo{author}{\bibfnamefont{C.~A.} \bibnamefont{M\"{u}ller}},
  \bibnamefont{and} \bibinfo{author}{\bibfnamefont{M.}~\bibnamefont{Kastner}},
  \bibinfo{journal}{J. Phys. A} \textbf{\bibinfo{volume}{42}},
  \bibinfo{pages}{465304} (\bibinfo{year}{2009}).

\bibitem[{\citenamefont{Canosa et~al.}(2007)\citenamefont{Canosa, Matera, and
  Rossignoli}}]{Canosa07}
\bibinfo{author}{\bibfnamefont{N.}~\bibnamefont{Canosa}},
  \bibinfo{author}{\bibfnamefont{J.~M.} \bibnamefont{Matera}},
  \bibnamefont{and}
  \bibinfo{author}{\bibfnamefont{R.}~\bibnamefont{Rossignoli}},
  \bibinfo{journal}{Phys. Rev. A} \textbf{\bibinfo{volume}{76}},
  \bibinfo{pages}{022310} (\bibinfo{year}{2007}).

\bibitem[{\citenamefont{Matera et~al.}(2008)\citenamefont{Matera, Rossignoli,
  and Canosa}}]{Matera08}
\bibinfo{author}{\bibfnamefont{J.~M.} \bibnamefont{Matera}},
  \bibinfo{author}{\bibfnamefont{R.}~\bibnamefont{Rossignoli}},
  \bibnamefont{and} \bibinfo{author}{\bibfnamefont{N.}~\bibnamefont{Canosa}},
  \bibinfo{journal}{Phys. Rev. A} \textbf{\bibinfo{volume}{78}},
  \bibinfo{pages}{012316} (\bibinfo{year}{2008}).

\bibitem[{\citenamefont{{J. Wilms {\it et al.}}}()}]{Wilms11}
\bibinfo{author}{\bibnamefont{{J. Wilms {\it et al.}}}}, \bibinfo{note}{in
  preparation}.

\end{thebibliography}

\end{document}